\documentclass[floats,floatfix,showpacs,amssymb,prd,superscriptaddress,twocolumn,aps,nofootinbib]{revtex4-1}
\usepackage[dvipdfm]{graphicx} %include figure files
\usepackage{epsfig, amssymb} %include figure files
\usepackage{amsmath, amsfonts}
\usepackage{bm} %include bold math: \bm{} creates bold letters in math mode

\usepackage[caption=false]{subfig}
\usepackage[usenames]{color}

\DeclareMathOperator\erf{erf}

\newcommand{\dif}{{\rm d}}
\newcommand{\tdif}{\tilde{D}}
\newcommand{\tgam}{\tilde{\gamma}}

\newcommand{\tA}{\tilde{A}}
\newcommand{\tR}{\tilde{R}}

\def\Lie{\mathcal{L}}
\def\A{\mathcal{A}}

\def\H{\mathcal{H}}
\def\M{\mathcal{M}}

\def\beq{\begin{equation}}
\def\eeq{\end{equation}}
\def\beqa{\begin{eqnarray}}
\def\eeqa{\end{eqnarray}}

\def\p{\partial}

\def\tg{\tilde{\gamma}}
\def\tA{\tilde{A}}

\def\tG{\tilde{\Gamma}}
\def\tD{\tilde{D}}

\begin{document}

\title{\large Black holes and fundamental fields in Numerical Relativity:\\
initial data construction and evolution of bound states%\\
}

\author{Hirotada Okawa}
\email{hirotada.okawa@ist.utl.pt}
\affiliation{CENTRA, Departamento de F\'{\i}sica, Instituto Superior T\'ecnico, Universidade T\'ecnica de Lisboa - UTL,
Avenida Rovisco Pais 1, 1049 Lisboa, Portugal.}
\author{Helvi Witek} 
\email{h.witek@damtp.cam.ac.uk}
\affiliation{Department of Applied Mathematics and Theoretical Physics,
Centre for Mathematical Sciences, University of Cambridge,
Wilberforce Road, Cambridge CB3 0WA, UK}
\author{Vitor Cardoso} 
\email{vitor.cardoso@ist.utl.pt}
\affiliation{CENTRA, Departamento de F\'{\i}sica, Instituto Superior T\'ecnico, Universidade T\'ecnica de Lisboa - UTL,
Avenida Rovisco Pais 1, 1049 Lisboa, Portugal.}
\affiliation{Perimeter Institute for Theoretical Physics, Waterloo, Ontario N2L 2Y5, Canada.}
\affiliation{Department of Physics and Astronomy, The University of Mississippi, University, MS 38677, USA.}

\date{\today} 

\begin{abstract} 
Fundamental fields are a natural outcome in cosmology and particle physics and 
might therefore serve as a proxy for more complex interactions.
The equivalence principle implies that all forms of matter gravitate, and one therefore expects relevant, universal imprints of new physics
in strong field gravity, such as that encountered close to black holes. Fundamental fields in the vicinities
of supermassive black holes give rise to extremely long-lived, or even unstable, configurations
which slowly extract angular momentum from the black hole or simply evolve non-linearly over long timescales,
with important implications for particle physics and gravitational-wave physics.

Here, we perform a fully non-linear study of scalar-field condensates around rotating black holes.
We provide novel ways to specify initial data for the Einstein-Klein-Gordon system, with potential applications in a variety of scenarios.
Our numerical results confirm the existence of long-lived bar-modes which act as lighthouses for gravitational
wave emission:
the scalar field condenses outside the black hole geometry and acts as a constant frequency gravitational-wave source
for very long timescales. This effect could turn out to be a potential signature of beyond standard model physics and also a promising
source of gravitational waves for future gravitational wave detectors.

\end{abstract}


\maketitle
%\tableofcontents
%\newpage

%-----------------------------------------------------------------------------
%%%%%%%%%%%%%%%%%%%%%%%%%%%%%%%%%%%%%%%%%%%%%%%%%%%%%%%%%%%%%%%%%%%%%%%%%%%%%%
\section{Introduction}
%%%%%%%%%%%%%%%%%%%%%%%%%%%%%%%%%%%%%%%%%%%%%%%%%%%%%%%%%%%%%%%%%%%%%%%%%%%%%%
Black holes (BHs) are among the most fascinating and numerous inhabitants of our universe,
a remarkable consequence of General Relativity (GR) or extensions thereof. 
Stellar mass BHs of a few solar masses ($3M_{\odot}\lesssim M_{\rm BH} \lesssim 30M_{\odot}$) are expected to be the end-state 
of massive stars; compelling evidence for their existence is provided by
observations of ultra-compact binaries made up of pulsars and BHs as well as 
observations of BHs accreting matter from surrounding disks~\cite{Belczynski:2006br,McClintock:2009as,Ivanova:2010ia,Nissanke:2012eh,Belczynski:2008nh,Heinke:2013ela,Seoane:2013qna,Reynolds:2013qqa}.
On the other end of the mass spectrum we expect supermassive BHs (SMBHs) with 
$10^{6}M_{\odot}\lesssim M_{\rm BH} \lesssim 10^{9}M_{\odot}$ to be hosted at the
centre of most galaxies and, in fact, observations of trajectories of stars close to the centre of the
Milky Way hint at ``our very own'' SMBH with mass $M_{\rm BH}\sim4.2\cdot10^{6}M_{\odot}$~\cite{Begelman:1980vb,Rees:1984si,
Ferrarese:2004qr,Alexander:2005jz,Ferrarese:2006fd,Denney:2010cn,Volonteri:2012yn,Wang:2013oga,Reynolds:2013rva}.
Many of the aforementioned observations ranging from the radio to the X-ray frequency band
allow for accurate estimates of the mass and spin of these BHs, thus making precision BH physics possible; this in turn may allow a
mapping of compact objects across the visible universe with second generation ground-based gravitational wave (GW) 
observatories such as 
the advanced LIGO / VIRGO network~\cite{Abbott:2007kv, Acernese:2008zzf,Abadie:2011kd,Aasi:2013wya,
aLIGO,aLIGOind}, the KAGRA detector~\cite{Aso:2013eba,Somiya:2011np}
currently under construction or future space-based LISA-like missions~\cite{AmaroSeoane:2012km,Seoane:2013qna}.

Fundamental fields -- either constituents of dark matter or other types of stable fundamental fields --
may play a crucial role in the context of GW emission and detection. They may influence how 
a SMBH is bound to its host galaxy or affect the very properties of both stellar-mass or supermassive BH systems and, 
consequently, their GW emission~\cite{Eda:2013gg,Macedo:2013qea,Arvanitaki:2010sy}.
One example of such drastic effects concerns the existence of very long lived massive states
of fundamental fields around BHs and their possible growth via superradiance~\cite{zeldovich1,zeldovich2,East:2013mfa}.
We remind the reader that (rotational-induced) superradiance of a wavepacket of (real) frequency $\omega_R$ requires the 
condition
\begin{align}
0 < &\, \omega_{R} < m \Omega_{H} = \omega_C\,,\label{eq:SRcond}
\end{align}
where $m$ is an azimuthal number and $\Omega_H$ is the BH's angular frequency.
This mechanism to amplify waves scattering off Kerr BHs allows for exciting phenomena: one example is the 
Gedankenexperiment suggested in Refs.~\cite{zeldovich1,zeldovich2,Press:1972zz}, and consists of enclosing 
the system by a perfectly reflecting mirror. The bouncing on the mirror and amplification in the ergoregion
will result in an exponential growth of the field and an increasing ``radiation'' pressure will eventually render the system 
unstable. This so-called ``black-hole bomb'' mechanism~\cite{Press:1972zz} or ``superradiant instability'' has attracted some attention over the years~\cite{zeldovich1,zeldovich2,Press:1972zz,Cardoso:2004nk,Hod:2009cp,Rosa:2009ei,Witek:2010qc,Dolan:2012yt}.

Astrophysical ``BH bombs'' may exist in the presence of ultralight degrees of freedom, thought to arise in a variety of 
scenarios~\cite{Peccei:1977hh,Arvanitaki:2009fg,Arvanitaki:2010sy}~\footnote{A very natural ``BH bomb'' also 
arises in asymptotically anti-de Sitter (AdS) spacetimes, for which the timelike boundary takes on 
the role of the reflecting cavity. Indeed, it has been shown that small Kerr-AdS BHs do suffer from 
the superradiant instability~\cite{Cardoso:2004nk,Hawking:1999dp,Cardoso:2004hs,Cardoso:2006wa,Cardoso:2013pza}. 
Although they are not relevant for astrophysical scenarios, BHs in AdS play an important role
in high energy physics, in particular in the context of the gauge/gravity duality.}.
Massive fields in general can create a ``trapping well'' some distance away from the BH, with a tail that extends
into the ergoregion, effectively working as a confining box at low frequencies~\cite{Damour:1976,Detweiler:1980uk,Zouros:1979iw}.
An exploration of the behavior of massive fields in the vicinities of BHs has uncovered remarkable results, ranging from floating orbits to
the possibility of constraining fundamental fields from the observation of SMBHs~\cite{Cardoso:2011xi,Pani:2012bp}.
This has led to a vast number of investigations of massive fields around rotating BHs in the frequency-~\cite{Cardoso:2005vk,Dolan:2007mj,Berti:2009kk,Pani:2012bp,Cardoso:2011xi,Burt:2011pv,Barranco:2013rua}
and in the time-domain~\cite{Strafuss:2004qc,Barranco:2012qs,Yoshino:2012kn,Witek:2012tr,Dolan:2012yt}.

These incursions are also motivated by extensions of General Relativity such as scalar-tensor 
theories~\cite{Damour:1993hw,Yunes:2013dva,Cardoso:2013fwa,Stein:2013wza}
but also by the ``axiverse'' scenario proposed in Refs.~\cite{Arvanitaki:2009fg, Arvanitaki:2010sy,Kodama:2011zc}.
Inspired by the QCD axion~\cite{Peccei:1977hh} Arvanitaki et al~\cite{Arvanitaki:2009fg, Arvanitaki:2010sy}
suggested the existence of a plethora of ultra-light bosonic degrees of freedom which could play an important 
role in BH astrophysics if their masses range from $10^{-21}{\rm eV}\lesssim \mu_S \lesssim 10^{-8}{\rm eV}$.
Conversely, the observation of BHs in a certain mass--spin %$(M,a/M)$ 
parameter range can put stringent constraints
on the existence of these particles as suggested in Refs.~\cite{Arvanitaki:2010sy,Pani:2012vp}.
In other words, one can use SMBHs to explore beyond-standard model physics.

Within the past year, these studies 
inspired a ``gold-rush'' of investigations going beyond the simplest model
considering scalar fields in GR~\cite{Brito:2013wya,
Brito:2013yxa,Pani:2013hpa,Cardoso:2013opa,Yoshino:2012kn,Degollado:2013eqa,Herdeiro:2013pia,Degollado:2013bha,Hod:2013fvl,Pani:2013ija,Pani:2013wsa,Rosa:2011my,Pani:2012bp,Pani:2012vp,Witek:2012tr} (see also, e.g., Refs.~\cite{Cardoso:2013krh,Berti:2013uda} for recent reviews). Recently, it was shown that the superradiant instability can, at least for complex scalars, drive the system to a new, truly stationary solution describing a hairy BH~\cite{Herdeiro:2014goa}.

Despite this immense progress in exploring the superradiant mechanism in various configurations
most dynamical studies so far have been restricted to the \textit{linearized} regime, i.e. fixing the BH spacetime as a background
over which matter fields evolve (but see Ref.~\cite{East:2013mfa} for recent fully non-linear investigations).
While this assumption is valid as long as the amplified field is small, it is inevitable 
to break down eventually. Then, back-reaction onto the spacetime
will become important. Depending on the frequency composition of the field this might yield
spin-down and mass-loss of the central BH in the superradiant regime
or accretion, i.e., increase of the BH spin and mass in the ``normal'' regime.

Thus, the question about the end-state of the instability remains; both scenarios, 
that of a true non-linear instability as well as a 
quasi-equilibrium configuration, might be possible.
In fact, this is exactly the regime with potentially exciting new physical signatures:
The formation of a bosonic cloud around BHs might produce a ``gravitational atom'',
where the name is adopted due to the hydrogen-like spectrum of the field~\cite{Dolan:2007mj,Pani:2012bp},
possibly observable through scalar and (modified) GW emission.
Depending on the efficiency with which the bosonic cloud can be accreted, 
we might even observe GW pulsars or ``light-houses'' or find gaps in the Regge plane, i.e., 
the mass-spin phase-space of BHs~\cite{Arvanitaki:2009fg, Arvanitaki:2010sy,Kodama:2011zc,Yoshino:2013ofa,
Cardoso:2011xi,Berti:2013uda,Cardoso:2013krh,Pani:2012vp,Mocanu:2012fd}.
The evolution may also drive the system to a new stationary state such as the hairy BH solutions recently uncovered~\cite{Herdeiro:2014goa},
although these solutions are also likely to be unstable, at least in part of the solution parameter space~\cite{Pani:2010jz}.
The understanding of all these possibilities require nonlinear evolutions of the equations of motion.

In the present study we explore the effects of the back-reaction and nonlinearities
of the field equations by performing full-blown numerical simulations of GR coupled to a minimally coupled massive scalar field.

As guideline for our investigations let us briefly review 
the key results of linearized studies concerning massive scalar fields surrounding 
Kerr BHs~\cite{Cardoso:2005vk,Dolan:2007mj,Berti:2009kk,Pani:2012bp,Cardoso:2011xi,Burt:2011pv,
Yoshino:2012kn,Witek:2012tr,Dolan:2012yt}.
First of all, because of the presence of the potential well due to the mass term, 
two types of modes are possible -- quasi-normal modes (QNMs) which decay
at infinity and quasi-bound state modes (BSMs) which are localized 
in the dip of the potential around the BH.
The latter class can yield exponentially growing modes if the BH is spinning.
Because these are dynamical, non-stationary spacetimes, the no-hair theorem is {\it not} violated.
The strongest instability growth rate of $\omega_{I}\equiv \tfrac{1}{\tau}\sim1.5\cdot10^{-7}\left(\tfrac{GM}{c^3}\right)^{-1}$
was found for the $l=m=1$ mode of a massive scalar with mass coupling 
$M_{\rm BH}\mu=0.42$ evolving around a rotating BH with spin $a/M_{\rm BH}=0.99$~\cite{Cardoso:2005vk,Dolan:2007mj}.
Here $\tau$ is the typical instability timescale.
Recent simulations in the time domain~\cite{Yoshino:2012kn,Witek:2012tr,Dolan:2012yt}
revealed beating phenomena for generic Gaussian wave packets 
due to the presence of various overtone modes and space dependent excitation of modes; very much like the excitation and modulation
of tones as a guitar string is excited. 
The latter effect has been further investigated in Ref.~\cite{Zhang:2013ksa}.
Guided by these results, we will numerically evolve complex, massive scalar fields
initialized both as generic Gaussian pulse as well as pseudo-bound states with a mass coupling
close to the one expected for strong instability growth rates
around Kerr BHs. Following the approach in Liu et al~\cite{Liu:2009al} we have been able to accurately set up puncture initial data representing Kerr BHs with an initial spin up to 
$95\%$ of the Kerr bound.

This paper is organized as follows:
In Sec.~\ref{sec:setup} we will describe the setup of our model and its formulation
as Cauchy problem. We complement the setup by constructing constraint-preserving initial data in Sec.~\ref{sec:initdata}.
In particular, we develop analytic and numerical solutions for a number of different configurations involving 
Schwarzschild or Kerr BHs surrounded by ``scalar clouds''.
In Secs.~\ref{sec:SFresultsII}-\ref{sec:SFresultsIII}
we present the results of our non-linear time evolutions starting, respectively, with
non-rotating or highly rotating BHs.
We finalize the paper with some conclusions and prospects for future work in Sec.~\ref{sec:conclusions}.
Further checks of our numerical simulations including error estimates and
benchmark tests can be found in Apendices.~\ref{app:convergence}
and~\ref{app:SFresultsIIIpureKerr}. Finally, illustrative snapshots of the evolution are
shown in Apendix~\ref{app:snapshots}.
Unless stated otherwise we work in natural units, i.e., $G=1=c$.

%%%%%%%%%%%%%%%%%%%%%%%%%%%%%%%%%%%%%%%%%%%%%%%%%%%%%%%%%%%%%%%%%%%%%%%%%%%%%%
\section{Setup}
\label{sec:setup}
%%%%%%%%%%%%%%%%%%%%%%%%%%%%%%%%%%%%%%%%%%%%%%%%%%%%%%%%%%%%%%%%%%%%%%%%%%%%%%

We wish to explore the non-linear dynamics of scalar clouds in the environment of rotating BHs.
This system is modelled by the Einstein--Hilbert action 
(in $4$-dimensional asymptotically flat spacetimes)
minimally coupled to a complex, massive scalar field
$\Phi$ with mass parameter $\mu_S=m_{S}/\hbar$ and described by the action~\cite{1975problembook,Wald:1984rg}
\begin{subequations}
\label{eq:action}
\begin{align}
S = & \int \dif^{4}x \sqrt{-g}\left( \frac{\,^{(4)}R}{16\pi} +\Lie_{\Phi} \right)
\,,\quad{\rm{with}}\\
\Lie_{\Phi} = & -\tfrac{1}{2} g^{\mu\nu}\p_{\mu}\Phi^{\ast}{}\p_{\nu}\Phi
                -\tfrac{1}{2}\mu_{S}^{2} \Phi^{\ast}{}\Phi - V(\Phi)
\,,
\end{align}
\end{subequations}
where $\,^{(4)}R$ is the $4$-dimensional Ricci scalar and $V(\Phi)$ is the scalar field potential.
The variation of the action~\eqref{eq:action} with respect to the metric and scalar field yields,
respectively, the tensor and scalar equations of motion (EoMs)
\begin{subequations}
\label{eq:EoM4D}
\begin{align}
\label{eq:EoMTen4D}
\,^{(4)}R_{\mu\nu} - \tfrac{1}{2}g_{\mu\nu} \,^{(4)}R - 8\pi T_{\mu\nu} = & 0
\,,\\
\label{eq:EoMSca4D}
\left( \nabla^{\mu}\nabla_{\mu} - \mu_{S}^2 \right) \Phi - V^{'}(\Phi) = & 0
\,,
\end{align}
\end{subequations}
where the energy-momentum tensor of the scalar field is determined by
\begin{align}
\label{eq:TmnScalar}
T_{\mu\nu} = & -\tfrac{1}{2}g_{\mu\nu}\left( \p_{\lambda}\Phi^{\ast}{}\p^{\lambda}\Phi + \mu_{S}^2 \Phi^{\ast}{}\Phi \right)
               - g_{\mu\nu} V(\Phi) 
\nonumber\\ &
        + \tfrac{1}{2}\left(\p_{\mu}\Phi^{\ast}{}\p_{\nu}\Phi + \p_{\mu}\Phi\p_{\nu}\Phi^{\ast}{}\right)
\,.
\end{align}
%

%%%%%%%%%%%%%%%%%%%%%%%%%%%%%%%%%%%%%%%%%%%%%%%%%%%%%%%%%%%%%%%%%%%%%%%%%%%%%%
%\subsection{Time evolution formulation}
%\label{ssec:ADMsplit}
%%%%%%%%%%%%%%%%%%%%%%%%%%%%%%%%%%%%%%%%%%%%%%%%%%%%%%%%%%%%%%%%%%%%%%%%%%%%%%
\noindent{{\bf{Time evolution formulation:}}}
Because we intend to explore the non-linear
GR-scalar field system in the highly dynamical, strong curvature regime
we need to solve the EoMs~\eqref{eq:EoM4D} numerically. 
Therefore, we employ standard Numerical Relativity (NR) techniques~\cite{Alcubierre:2008,Baumgarte2010,
York:1979,Gourgoulhon:2007ue,Centrella:2010mx,Hinder:2010vn,Baumgarte:2002jm}
based on the $3+1$ decomposition. 
Within this approach the $4$-dimensional spacetime manifold $(\M,g_{\mu\nu})$ is foliated into
$3$-dimensional spatial hypersurfaces $(\Sigma_{t},\gamma_{ij})$ 
which are parametrized by the time coordinate $t$.
The spatial metric $\gamma_{ij}$ is related to the spacetime metric $g_{\mu\nu}$via $\gamma_{\mu\nu} = g_{\mu\nu} + n_{\mu} n_{\nu}$ and $n^{\mu}$ is the unit vector normal to the hypersurfaces.
The spacetime line element writes
\begin{align}
\label{eq:lineelement}
\dif s^2 = & g_{\mu\nu} \dif x^{\mu}\dif x^{\nu}
\\
         = & - \left( \alpha^2 - \beta_{i}\beta^{i} \right) \dif t^{2}
             + 2\gamma_{ij}\beta^{i} \dif t \dif x^{j}
             + \gamma_{ij} \dif x^{i} \dif x^{j}
\,,\nonumber
\end{align}
where 
the lapse function $\alpha$ and shift vector $\beta^{i}$ encode
the coordinate degrees of freedom.
We proceed by casting Eqs.~\eqref{eq:EoM4D} into a Cauchy problem 
such that they are ``digestible'' by computers.
In a nutshell, we rewrite them as a time evolution problem with 
constraints along the lines of the ADM-York decomposition~\cite{Arnowitt:1962hi,York:1979}.
To this end, we introduce the conjugated momenta
to the scalar field, $\Pi$, and to the metric, $K_{ij}$. The extrinsic curvature $K_{ij}$
describes how the spatial hypersurfaces are embedded into the full spacetime. 
These quantities are defined by,
\begin{align}
\label{eq:defKijKPhi}
K_{ij} = - \tfrac{1}{2\alpha} \left(\p_{t} - \Lie_{\beta} \right) \gamma_{ij}
\,,\quad&
\Pi =  - \tfrac{1}{\alpha} \left(\p_{t} - \Lie_{\beta} \right) \Phi
\,.
\end{align}
Here, $\Lie_{\beta}$ denotes the Lie derivative along the shift vector $\beta^i$.
The definitions~\eqref{eq:defKijKPhi} immediately provide a prescription for the 
time evolution of the $3$-metric and scalar field
\begin{align}
\label{eq:evolGamPhi}
\left(\p_{t}-\Lie_{\beta}\right) \gamma_{ij} = - 2\alpha K_{ij}
\,,\quad&
\left(\p_{t}-\Lie_{\beta}\right) \Phi = - \alpha \Pi
\,,
\end{align}
capturing the kinematical degrees of freedom.
Performing the $3+1$ decomposition of Eqs.~\eqref{eq:EoM4D}
yields the Hamiltonian and momentum constraints
\begin{subequations}
\label{eq:constraintsADM}
\begin{align}
\label{eq:HamiltonianADM}
\H = & R +K^2 - K_{ij} K^{ij} - 16\pi \rho = 0 
\,,\\
\label{eq:MomentumADM}
\M_{i} = & D_{j} K^{j}{}_{i} - D_{i} K - 8 \pi j_{i} = 0
\,,
\end{align}
\end{subequations}
as well as time evolution equations for the extrinsic curvature and scalar field momentum
\begin{subequations}
\label{eq:EvolEqADM}
\begin{align}
\label{eq:EvolKADM}
\left(\p_{t}-\Lie_{\beta} \right) K_{ij} = &
        - D_{i} D_{j} \alpha + \alpha\left( R_{ij} -2 K^{k}{}_{i} K_{jk} + K K_{ij} \right) 
\nonumber \\ &
        + 4\pi\alpha\left(\gamma_{ij}(S-\rho) - 2S_{ij} \right)
\,,\\
\label{eq:EvolKphiADM}
\left(\p_{t}-\Lie_{\beta} \right)\Pi = & 
        \alpha\left( -D^{i}D_{i}\Phi + K \Pi + \mu_S^2 \Phi + V^{'}(\Phi) \right)
\nonumber \\ &
        - D^{i}\alpha D_{i} \Phi 
\,,
\end{align}
\end{subequations}
where $R_{ij}$ and $R$ refer to the $3$-dimensional Ricci tensor and scalar
associated with the spatial metric $\gamma_{ij}$.
The energy density $\rho$, energy-momentum flux $j_{i}$ and spatial components $S_{ij}$ of the energy momentum tensor~\eqref{eq:TmnScalar}
are given by
\begin{subequations}
\label{eq:TmnProj}
\begin{align}
\label{eq:rho}
\rho = & \tfrac{1}{2} \Pi^{\ast}{} \Pi + \tfrac{1}{2} \mu_{S}^{2} \Phi^{\ast}{}\Phi + \tfrac{1}{2} D^{i}\Phi^{\ast}{} D_{i}\Phi
        + V(\Phi)
\,,\\
\label{eq:jj}
j_i = & \tfrac{1}{2}\left( \Pi^{\ast}{} D_{i}\Phi + \Pi D_{i} \Phi^{\ast}{}\right)
\,,\\
\label{eq:Sij}
S_{ij} = & \tfrac{1}{2}\left( D_{i}\Phi^{\ast}{} D_{j}\Phi + D_{i}\Phi D_{j}\Phi^{\ast}\right)
           - \gamma_{ij} V(\Phi)
\nonumber\\ &
        + \tfrac{1}{2}\gamma_{ij}\left( \Pi^{\ast}{} \Pi - \mu_S^2\Phi^{\ast}{}\Phi - D^{k}\Phi^{\ast}{}D_{k}\Phi \right)
\,.
\end{align}
\end{subequations}
We now have all necessary ingredients at hand to simulate scalar fields in GR.
Unfortunately, the evolution equations ~\eqref{eq:evolGamPhi} and~\eqref{eq:EvolEqADM} are known to pose only a weakly hyperbolic set of 
PDEs~\cite{Alcubierre:2008,Sarbach:2012pr,Hilditch:2013sba} and are, thus, prone to numerical instabilities.

%%%%%%%%%%%%%%%%%%%%%%%%%%%%%%%%%%%%%%%%%%%%%%%%%%%%%%%%%%%%%%%%%%%%%%%%%%%%%%
%\subsection{BSSN formulation}
%\label{ssec:BSSN}
%%%%%%%%%%%%%%%%%%%%%%%%%%%%%%%%%%%%%%%%%%%%%%%%%%%%%%%%%%%%%%%%%%%%%%%%%%%%%%
\noindent{{\bf{BSSN formulation:}}}
In order to circumvent the ill-posedness of Eqs.~\eqref{eq:evolGamPhi} and~\eqref{eq:EvolEqADM}
and to obtain a stable numerical formulation we need to modify the evolution PDEs.
Specifically, we employ the strongly hyperbolic, well-posed scheme introduced by
Baumgarte \& Shapiro~\cite{Baumgarte:1998te} and Shibata \& Nakamura~\cite{Shibata:1995we}
(BSSN).
The key idea is to add the constraints~\eqref{eq:constraintsADM} to the ADM-York like Eqs.~\eqref{eq:evolGamPhi} 
and~\eqref{eq:EvolEqADM} in a specific manner,
thus changing the character of the PDEs.
Additionally, it has been found convenient to introduce a set of conformal variables as
dynamical quantities. The BSSN variables are given by
\begin{subequations}
\label{eq:gBSSNvars}
\begin{align}
\chi = \gamma^{-\tfrac{1}{3}} \,,\quad&
\tg_{ij} = \gamma^{-\tfrac{1}{3}}\gamma_{ij} = \chi \gamma_{ij} 
\,,\\
K = \gamma^{ij}K_{ij} \,,\quad & 
\tA_{ij} = \chi A_{ij} = \chi \left(K_{ij} - \tfrac{1}{3}\gamma_{ij} K\right)
\,,\\
\tG^{i} = \tg^{jk}\tG^{i}{}_{jk} = & - \p_{j} \tg^{ij}
\,,
\end{align}
\end{subequations}
where $\chi$ and $\tg_{ij}$ are the conformal factor and metric, $K$ and $\tA_{ij}$ are the trace and conformal tracefree
part of the extrinsic curvature and $\tG^{i}$ is the conformal connection function.
We denote $\gamma=\det\gamma_{ij}$, 
while $\tg=\det\tg_{ij}=1$ holds by construction.
We will not discuss the derivation of the BSSN equations 
and instead only present the final expressions
for a vanishing scalar field potential $V(\Phi)=0$
\begin{subequations}
\label{eq:evolBSSN}
\begin{align}
\label{eq:evolBSSNchi}
\p_t \chi = &\, [{\rm{BSSN}}] \,,\\
\label{eq:evolBSSNgamma}
\p_t \tg_{ij} = & \,[{\rm{BSSN}}]\,,\\
\label{eq:evolBSSNK}
\p_t K = & \,[{\rm{BSSN}}] + 8\pi\alpha\left( \Pi^{\ast}{} \Pi - \tfrac{\mu_S^2}{2}\Phi^{\ast}{}\Phi \right) 
\,,\\
\label{eq:evolBSSNAij}
\p_t \tA_{ij} = & \,[{\rm{BSSN}}]
        - 4\pi\alpha\chi\left( \tD_{i}\Phi^{\ast}{}\tD_{j}\Phi + \tD_{i}\Phi\tD_{j}\Phi^{\ast}{}
\right. \nonumber \\ & \left.\qquad\qquad\qquad\qquad
                -\tfrac{2}{3}\tg_{ij} \tD^{k}\Phi^{\ast}{}\tD_{k}\Phi \right)
\,,\\
\label{eq:evolBSSNGam}
\p_t \tG^{i} = & \,[{\rm{BSSN}}]
        - 8\pi\alpha \tg^{ij}\left( \Pi^{\ast}{} \tD_{j}\Phi + \Pi \tD_{j}\Phi^{\ast}{}\right)
\,,\\
\label{eq:evolBSSNPhi}
\p_t \Phi = & -\alpha \Pi + \Lie_{\beta} \Phi
\,,\\
\label{eq:evolBSSNKPhi}
\p_t \Pi = & \alpha\left( -\chi \tD^{i}\tD_{i}\Phi + \tfrac{1}{2}\tD^{i}\Phi\tD_{i}\chi + K \Pi + \mu_S^2 \Phi\right)
\nonumber \\ &
        - \chi \tD^{i}\alpha \tD_{i}\Phi
        + \Lie_{\beta}\Pi
\,,
\end{align} 
\end{subequations}
where ``[BSSN]'' denotes the vacuum BSSN equations given, e.g., in Refs.~\cite{Alcubierre:2008,
Baumgarte2010,Gourgoulhon:2007ue,Centrella:2010mx,Hinder:2010vn,Baumgarte:2002jm,
Witek:2013ora}
and $\tD_{i}$ is the covariant derivative with respect to the conformal metric $\tg_{ij}$.

In order to close the PDE system~\eqref{eq:evolBSSN} we additionally have to specify the 
coordinates. In particular, we choose the moving puncture gauge~\cite{Campanelli:2005dd,Baker:2005vv,
Alcubierre:2008,Baumgarte2010,vanMeter:2006vi}, i.e., the $1+\log$-slicing for the lapse function $\alpha$ 
and the $\Gamma$-driver shift condition for $\beta^{i}$ 
\begin{subequations}
\label{eq:PunctureGauge}
\begin{align}
\label{eq:1+log}
\p_t\alpha = & \beta^k\p_k\alpha - 2 \alpha K
\,,\\
\label{eq:GammaDriver}
\p_t \beta^i = & \beta^k\p_k\beta^i - \eta_{\beta} \beta^i + \zeta_{\Gamma} \tilde{\Gamma}^i
\,.
\end{align}
\end{subequations}

This system of equations was supplemented with Sommerfeld boundary conditions far away. To ensure that the boundary conditions
are not contaminating our results, the outer boundary of the numerical domain is placed sufficiently far away as to be causally disconnected from the region under study, and  each of the simulations we discuss was stopped before spurious reflections from the outer boundary can contaminate the results. 
We have verified {\it a posteriori} that the grid size is also larger than any characteristic wavelength showing up in our results.
Our results are convergent and stable when the grid size is varied.
%%%%%%%%%%%%%%%%%%%%%%%%%%%%%%%%%%%%%%%%%%%%%%%%%%%%%%%%%%%%%%%%%%%%%%%%%%%%%%
%\subsection{Extraction of physical information}
%\label{ssec:analysis}
%%%%%%%%%%%%%%%%%%%%%%%%%%%%%%%%%%%%%%%%%%%%%%%%%%%%%%%%%%%%%%%%%%%%%%%%%%%%%%

\noindent{{\bf{Extraction of physical information:}}}
Finally, we give a brief summary of the tools that we will employ to analyze our numerical data.
These include in particular
(i) the extraction of scalar and GWs,
(ii) the computation of the energy radiated in these signals,
and (iii) the estimation of the BH mass and spin using information about the apparent horizon (AH).
An extended discussion of either method can be found, e.g., in Refs.~\cite{Alcubierre:2008,Baumgarte2010,
Witek:2010qc}
and references therein. 

As a measure for the scalar and gravitational radiation we extract the scalar field $\Phi$ 
and the Newman-Penrose scalar $\Psi_4$, which encodes the outgoing gravitational 
radiation~\cite{Friedrich:1996hq,Teukolsky:1973ha},
at coordinate spheres of fixed radius $r_{\rm{ex}}$. 
Because we are interested in the multipolar structure of the radiated signals we project 
$\Phi(t,r=r_{\rm{ex}},\theta,\varphi)$ and $\Psi_{4}(t,r=r_{\rm{ex}},\theta,\varphi)$ with, respectively,
spherical and $s=-2$ spin-weighted spherical harmonics
\label{eq:sYlmProj}
\begin{align}
\Phi_{lm}(t,r_{\rm{ex}}) = & \int \dif\Omega\, \Phi(t,r_{\rm{ex}},\theta,\varphi) Y^{\ast}{}_{lm}(\theta,\varphi)
\,,\\
\Psi_{4,lm}(t,r_{\rm{ex}}) = & \int \dif\Omega\, \Psi_{4}(t,r_{\rm{ex}},\theta,\varphi) \,_{-2}Y^{\ast}{}_{lm}(\theta,\varphi)
\,.\nonumber
\end{align}
We estimate the radiated energy and angular momentum content in the GWs $\Psi_{4}$ 
by (see Eqs.~(22)-(24) in Ref.~\cite{Witek:2010qc} and references therein)
\begin{subequations}
\label{eq:FluxFromPsi4}
\begin{align}
\label{eq:calcEflux}
\frac{\dif E}{\dif t} = & \lim_{r\rightarrow\infty} \frac{r^2}{16\pi}\int \dif\Omega\, \left| \int_{-\infty}^{t} \Psi_{4} \dif\tilde{t}\,\right|^2
\,,\\
\label{eq:calcPflux}
\frac{\dif P_{i}}{\dif t} = & -\lim_{r\rightarrow\infty} \frac{r^2}{16\pi}\int \dif\Omega\,
        \ell_{i}\left| \int_{-\infty}^{t} \Psi_{4} \dif\tilde{t}\,\right|^2
\,,\\
\label{eq:calcJzflux}
\frac{\dif J_{z}}{\dif t} = &  -\lim_{r\rightarrow\infty} \frac{r^2}{16\pi} \times
\Re\left[
        \int\dif\Omega\, \left(\int_{-\infty}^{t} \Psi_{4} \dif\tilde{t}\,\right)
\right.\nonumber\\ & \left.\qquad\qquad\qquad
        \p_{\varphi}\left( \int_{-\infty}^{t}\int_{-\infty}^{\hat{t}} \Psi^{\ast}{}_{4} \dif\hat{t}\dif\tilde{t}\,\right)
\right]
\,,
\end{align}
\end{subequations}
with $\ell=(-\sin\theta \cos\varphi, -\sin\theta \sin\varphi, -\cos\theta)$.
In order to access the properties and evolution of the BH itself we characterize its AH (by using for example
the \textsc{AHFinderDirect}~\cite{Thornburg:1995cp,Thornburg:2003sf}), in terms of its
area $A_{\rm{AH}}$, equatorial circumference $C_{\rm{e}}$, 
and irreducible mass $M_{\rm{irr}}=\sqrt{A_{\rm{AH}}/(16\pi)}$, among other data.
We use this information to compute the BH dimensionless spin
\begin{align}
\label{eq:spin}
j_{\rm{AH}}\equiv  \tfrac{J}{M_{\rm BH}^2}= \sqrt{1-\left(\frac{2\pi A_{\rm{AH}}}{C_{\rm{e}}^2} - 1\right)^2}\,,
\end{align}
and mass according to Christodoulou's formula~\cite{Christodoulou:1970wf}
\begin{align}
\label{eq:MBH}
M_{\rm BH}^2 = & M_{\rm{irr}}^2 + \frac{J^2}{4M_{\rm{irr}}^2}\,.
\end{align}

\noindent{{\bf{Code description:}}}
We simulate the GR -- Klein--Gordon system using the \textsc{Cosmos} and \textsc{Lean--SR}
codes.
Both codes have been designed to solve Einstein's equations as an initial value
problem. Computationally this results in solving (i) a set of coupled elliptic
partial differential equations (PDEs) to provide initial data for BH spacetimes and
(ii) a set of coupled hyperbolic-type PDEs for the time evolution.
The evolution PDEs are solved using the method of lines employing a 4th order Runge-Kutta time integrator.
Spatial components are computed on 3-dimensional, nested Cartesian meshes and (spatial) derivatives
are realized by centered or lop-sided finite difference (FD) stencils.
While the codes provide FD stencils up to sixth order, in practice we employ centered fourth order FD
stencils for regular derivatives (in the interior of the numerical domain) 
and fourth order lop-sided FD stencils to realize advection derivatives.

The \textsc{Lean--SR} code is based on the \textsc{Cactus} computational toolkit~\cite{Goodale:2002a,Cactuscode:web},
part of the \textsc{Einstein Toolkit}~\cite{Loffler:2011ay,EinsteinToolkit:web},
and Sperhake's \textsc{Lean} code~\cite{Sperhake:2006cy}.
Initial configurations are set up either by using analytic data or by solving the constraints
using the spectral \textsc{TwoPunctures} solver~\cite{Ansorg:2004ds}.
In order to compute the AH we employ the \textsc{AHFinderDirect}~\cite{Thornburg:2003sf,Thornburg:1995cp}.
The code furthermore incorporates adaptive mesh refinement (AMR) 
provided by the \textsc{Carpet} package~\cite{Schnetter:2003rb,CarpetCode:web}
and uses boxes moving across the numerical domain tracking the motion of the BHs.
Parallelization is implemented with MPI.

The algorithm of the \textsc{COSMOS} code is based on the \textsc{SACRA}
code written by T.~Yamamoto~{\it et al.}~\cite{Yamamoto:2008js} and developed to solve binary problems.
Initial data is given either analytically or constructed by solving the
constraints numerically using the Multi-Grid solver~\cite{Okawa:2013afa}.
In order to compute the AH we implement the AH
finder based on Refs.~\cite{Shibata:1997nc,Shibata:2000nw}.
The code employs fixed mesh refinement (FMR) and the BH is located at the center of the computational domain.
Parallelization is implemented with OpenMP.

We set up our numerical domain using (fixed or adaptive) mesh refinement. Then, the numerical grid contains
$N_{RL}$ refinement levels centered around the BH. We denote the egde length of the $n-th$ refinement
level as $2 x_{n}$. Its resolution is given by $h_{n}=2^{n-1} h_{1}$, where $h_{1}$ is the resolution of the 
innermost refinement level. During the presentation of the results we 
adopt the notation of Sec.~II~E in Ref.~\cite{Sperhake:2006cy}
and we summarize the grid setup as
\begin{align}
\label{eq:gridsetup}
& \{(x_{N_{RL}}, x_{N_{RL}-1},\ldots,x_{1}),h=h_{1}\}
\,.
\end{align}
In all tables summarizing the numerical setup we use units in which the BH bare mass $M=1$ (see Section~\ref{sec:initdata}, in particular Eqs.~\eqref{eq:LineElSchw} and \eqref{eq:KerrBL}). 
%%%%%%%%%%%%%%%%%%%%%%%%%%%%%%%%%%%%%%%%%%%%%%%%%%%%%%%%%%%%%%%%%%%%%%%%%%%%%%%%%%%%%%%%%%%%%%%%%
\section{Initial data construction for black holes immersed in scalar fields}\label{sec:initdata}
%%%%%%%%%%%%%%%%%%%%%%%%%%%%%%%%%%%%%%%%%%%%%%%%%%%%%%%%%%%%%%%%%%%%%%%%%%%%%%%%%%%%%%%%%%%%%%%%%
%
\begin{figure*}[htpb!]
\subfloat[Constraint satisfying initial data]{\includegraphics[width=0.33\textwidth,clip]{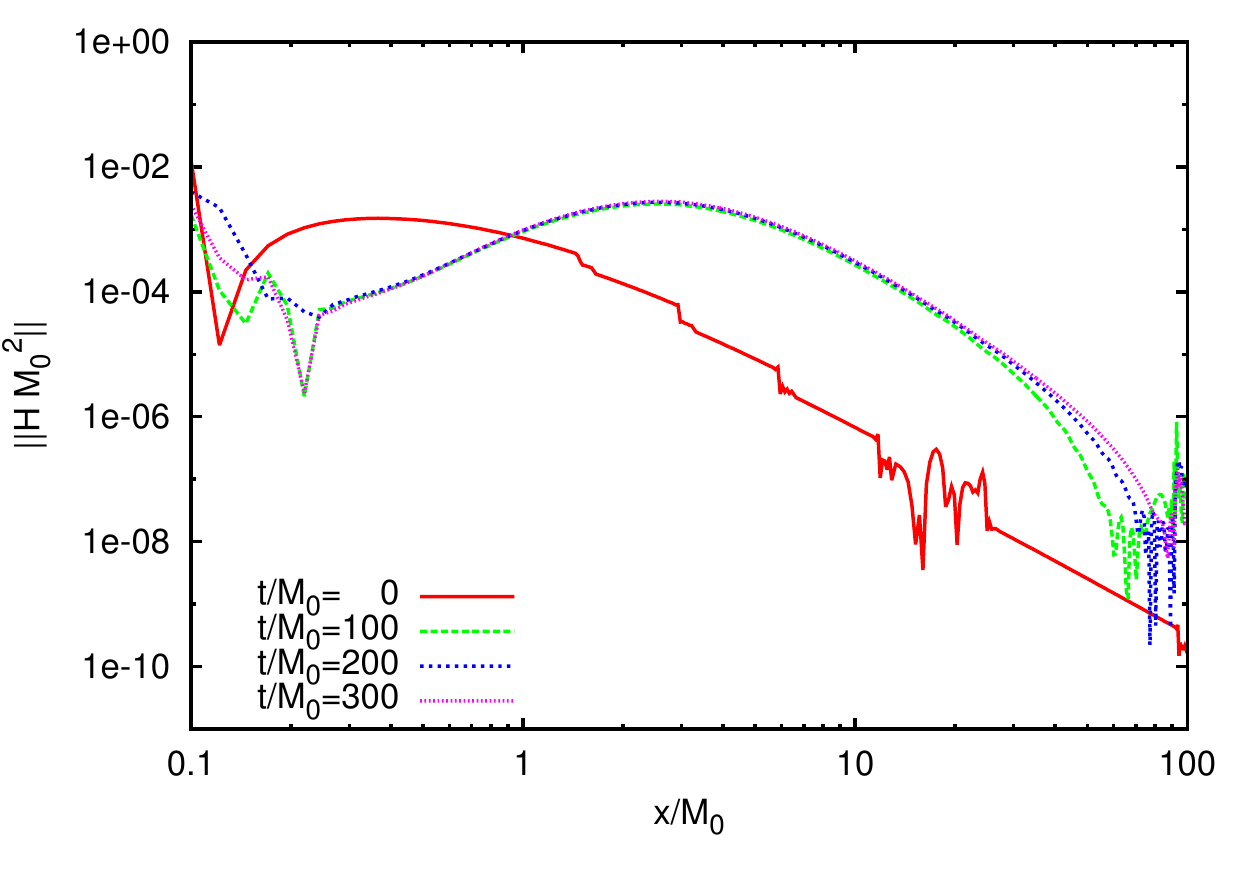}\label{fig:SchwarzschildMasslessHamilitonian}}
\subfloat[Constraint violating initial data]{\includegraphics[width=0.33\textwidth,clip]{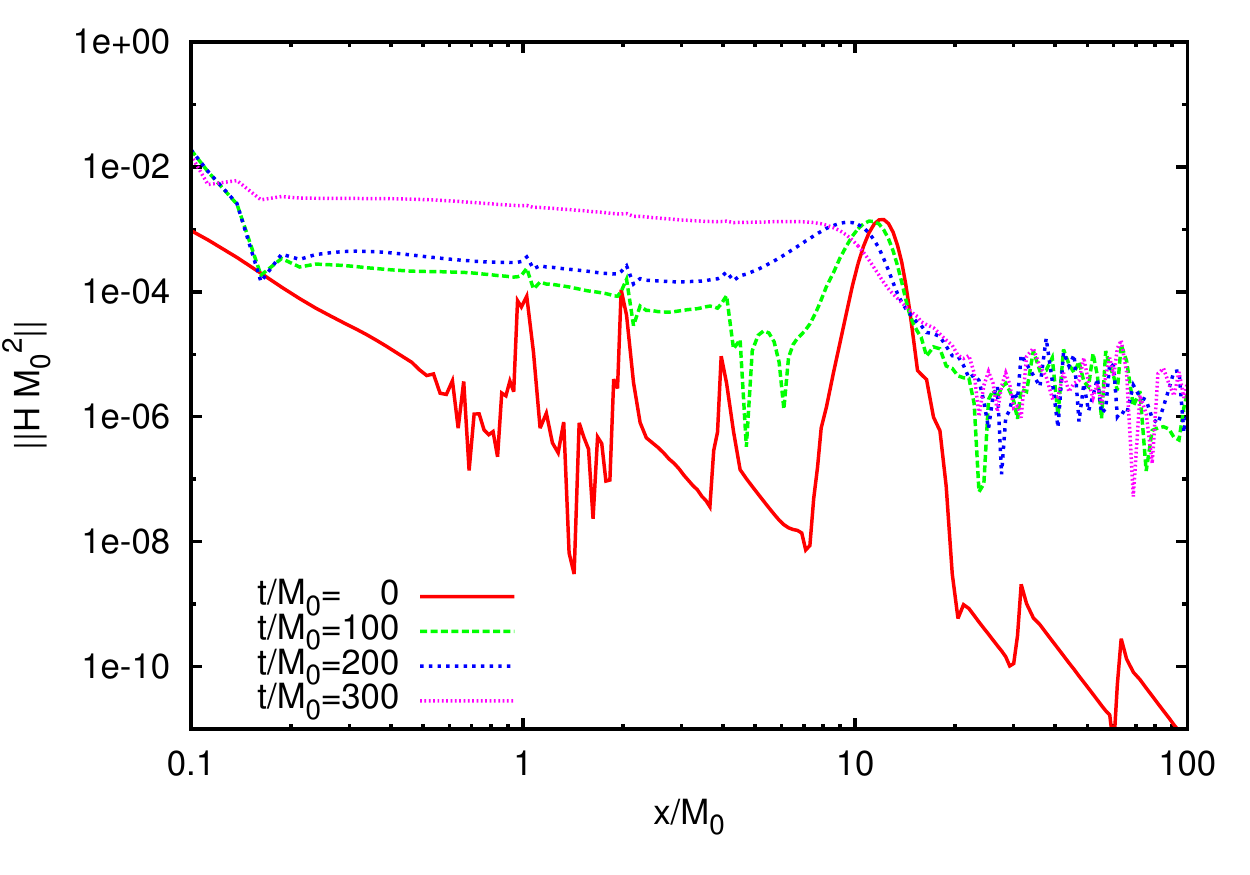}\label{fig:violation_H_time}}
\subfloat[Apparent horizon area]{\includegraphics[width=0.33\textwidth,clip]{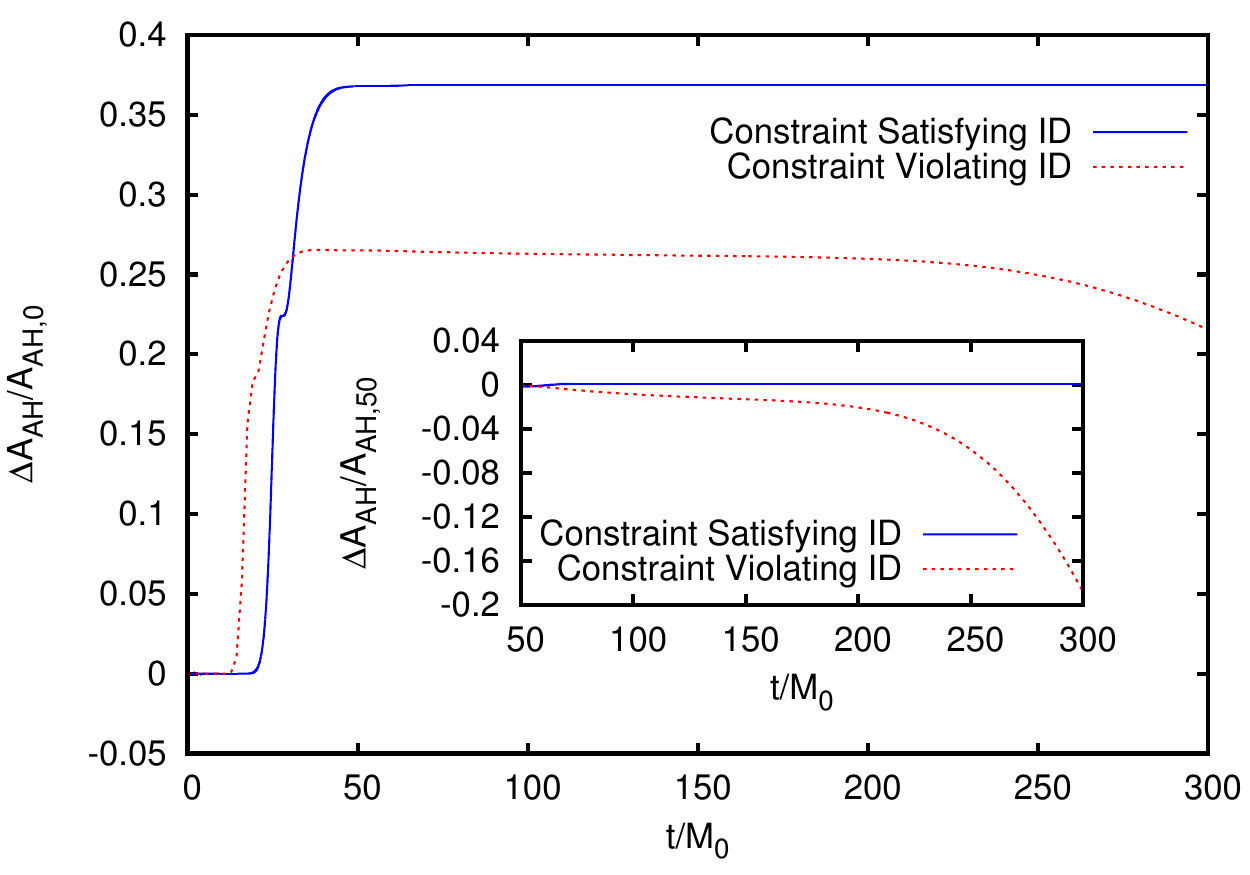}\label{fig:violation_AH}}
\caption{ \label{fig:constraint_violation} 
 Evolution of constraint satisfying (left panel, for further details on the construction see Table~\ref{tab:SetupSchwarzschildmassless} and section~\ref{ssec:InitDataSBH}) 
 and constraint violating (mid) initial data, 
 corresponding to run {\textit{S00\_m0\_e}} and {\textit{S00\_CV}} in Table ~\ref{tab:SetupSchwarzschildmassless}. 
 In both cases the spherically symmetric scalar shells are located at $r_0=12M$ with a width of $w=2M$ and amplitude $M A=0.075$.
 Figs.~\protect\subref{fig:SchwarzschildMasslessHamilitonian} and~\protect\subref{fig:violation_H_time} 
 show the Hamiltonian constraint along the $x$-axis at different times throughout the evolution.
 Constraints become violated with larger magnitude at late times for constraint-violating initial data, specially close to the horizon. 
 At late-times this impacts on physical quantities such as the AH area, shown in Fig.~\protect\subref{fig:violation_AH},
 where $A_{AH,t}$ denotes the AH area at time $t$.
 The violation of the constraints causes the AH area to \textit{decrease} (red dotted line) violating the area theorem while this 
 unphysical feature does not occur with constraint satisfying initial data (blue solid line). The inset normalizes the area variation to zero at $t=50$, in order to better gauge
 the changes induced by both types of initial data.
}
\end{figure*}

The approach to the time evolution problem outlined in the previous section has to be completed
by setting up initial states for the dynamical quantities $(\gamma_{ij},\Phi_{R,I},K_{ij},\Pi_{R,I})$.
Any initial data (describing GR systems) have to be a solution of Einstein's equations~\eqref{eq:EoM4D} and,
therefore, have to satisfy the constraints~\eqref{eq:constraintsADM}. 
In general, the construction of these solutions implies solving a set of four coupled, elliptic PDEs
which is difficult but nowadays well understood for vacuum spacetimes 
(see, e.g., Refs.~\cite{Cook:2000vr,Alcubierre:2008,Okawa:2013afa} and references therein).
However, more brainwork is needed to provide appropriate initial data for the more complicated case
that we are interested in and in which BH spacetimes are coupled to scalar fields.

In fact, some, if not all studies dealing with such a system have taken a crude approach and 
simply superposed a scalar field onto a BH spacetime~\cite{Healy:2011ef,Berti:2013gfa}, 
thus setting up \textit{constraint violating} initial data. 
However, the consequences of these procedures are not clear and, indeed, might contaminate the entire evolution of a system.
Bear in mind, that the constraint violation caused by the presence of the scalar field 
typically lies outside the BH, i.e., it is not hidden behind a horizon. Furthermore, 
the employed BSSN evolution scheme does not have any constraint damping mechanism
and, thus, their violation will remain present throughout the entire time evolution. 
The consequences of this argument will be clarified below, when we compare the evolution
of constraint-violating and constraint-satisfying initial data, summarized also in 
Fig.~\ref{fig:constraint_violation}.

For this reason it is mandatory to construct appropriate, {\textit{constraint satisfying}} initial data.
We have found novel, generic {\it analytic} or numerical solutions to this problem which may be very useful to study extensions
of GR to include self-interacting fields.
Our construction can also prove useful in completely different contexts such as the collapse of self-interacting fundamental fields in scalar-tensor theories or dynamical spacetimes in other modified theories of gravity involving couplings to dilaton or axionic fields. 
Solving the constraints (a set of four coupled, elliptic PDEs) is in general a demanding task.
In order to simplify the problem it is useful  to perform a conformal transformation of the ADM variables~\cite{Cook:2000vr}
\begin{subequations}
\label{eq:conformaltrafoID}
\begin{align}
\gamma_{ij} = & \psi^4\tgam_{ij} 
\,,\quad
\tgam={\rm det}\tgam_{ij} = 1 \,,\\
K_{ij} = & A_{ij} +\tfrac{1}{3}\gamma_{ij}K
\,,\quad
A_{ij} =   \psi^{-2}\tA_{ij}
\,.
\end{align}
\end{subequations}
Then, the constraints~\eqref{eq:constraintsADM} become 
\begin{subequations}
\label{eq:ConstraintsID}
\begin{align}
\label{eq:ham_const_t}
\mathcal{H} 
= & \tilde{\bigtriangleup}\psi -\tfrac{1}{8}\tR\psi -\tfrac{1}{12}K^2\psi^{5} 
        +\tfrac{1}{8}\tA^{ij}\tA_{ij}\psi^{-7}
\nonumber\\ &
+\pi\psi\left[ \tD^{i}\Phi^{\ast}\tD_{i}\Phi + \psi^{4}\left(\Pi^{\ast}\Pi + \mu_{S}^{2}\Phi^{\ast}\Phi\right) \right]
\,,\\
\label{eq:mom_const_t}
\mathcal{M}_i 
=& \tdif_j\tA^j_i -\tfrac{2}{3}\psi^{6}\tdif_i K -4\pi \psi^{6}\left(\Pi^{\ast}\tD_i\Phi+\Pi\tD_i\Phi^{\ast}\right)
\,,
\end{align}
\end{subequations}
in terms of the conformal variables, where $\tilde{\bigtriangleup}=\tgam^{ij}\tdif_i\tdif_j$
and $\tD$ and $\tR$ denote the conformal covariant derivative and Ricci scalar.
Note, that the constraints~\eqref{eq:ConstraintsID} fix four of the $16$ independent 
variables 
$(\psi, \tgam_{ij}, \tA_{ij}, K, \Phi_{R,I}, \Pi_{R,I})$ 
at $t=0$.
This leaves the freedom to specify the remaining quantities motivated by the physical scenario under consideration
and allowing for further simplifications of the PDE problem.

%%%%%%%%%%%%%%%%%%%%%%%%%%%%%%%%%%%%%%%%%%%%%%%%%%%%%%%%%%%%%%%%%%%%%%%%%%%%%%%%%%%%%%%%
\subsection{Non-rotating black holes: analytic initial data}
\label{ssec:InitDataSBH}
%%%%%%%%%%%%%%%%%%%%%%%%%%%%%%%%%%%%%%%%%%%%%%%%%%%%%%%%%%%%%%%%%%%%%%%%%%%%%%%%%%%%%%%%
We start by considering a single, non-rotating BH surrounded by a scalar field.
The line element of the Schwarzschild BH in isotropic coordinates is given by
\begin{align}
\label{eq:LineElSchw}
\dif s^2 = & -\left(\frac{1-\frac{M}{2r}}{1+\frac{M}{2r}}\right)^2\dif t^2
             +\left(1+\frac{M}{2r}\right)^4\eta_{ij}\dif x^i\dif x^j 
\nonumber\\ 
        = & -\alpha^2\dif t^2 +\psi_{S}^4\eta_{ij}\dif x^i\dif x^j
\,,
\end{align}
where $M$ is the BH bare mass parameter and $r$ is the isotropic, radial coordinate.
We impose conformal flatness, i.e., $\tgam_{ij}=\eta_{ij}$
and set $\tA_{ij}=0$,
because the BH has neither linear nor angular momentum.
Then, the constraints~\eqref{eq:ConstraintsID} become
\begin{subequations}
\begin{align}
\label{eq:ham_const_a}
0 = &\,
  \bigtriangleup_{\rm flat}\psi -\tfrac{1}{12}K^2\psi^{5} 
 +\pi\psi\eta^{ij}\left(\partial_i\Phi_R\partial_j\Phi_R +\partial_i\Phi_I\partial_j\Phi_I\right)
\nonumber\\ &
+\pi \psi^{5}\left[ \left(\Pi^{2}_{R} + \Pi^{2}_{I}\right) + \mu_{S}^{2} \left( \Phi^{2}_{R} + \Phi^{2}_{I} \right) \right]
\,,\\
\label{eq:mom_const_a}
0 = &\, \partial_i K +6\pi\left(\Pi_R\partial_i\Phi_R +\Pi_I\partial_i\Phi_I\right) 
\,,
\end{align}
\end{subequations}
where $\bigtriangleup_{\rm flat}$ is the flat space Laplacian.
Note, that we cannot specify $K=0$ and $\psi=\psi_{S}$ simultaneously
because then the Hamiltonian constraint~\eqref{eq:HamiltonianADM} would not be
satisfied for an arbitrary non-zero scalar field surrounding a Schwarzschild BH.
Here, we adopt the maximal slicing condition $K=0$ and solve for 
deformations of the conformal factor from the Schwarzschild case.
With these specifications the momentum constraint~\eqref{eq:mom_const_a}
implies that at $t=0$ either $\Pi$ should vanish or $\Phi$ should be a constant.
We consider the latter case and, specifically, choose $\Phi=0$.
Then, the momentum constraint~\eqref{eq:mom_const_a} is satisfied trivially
and the Hamiltonian constraint~\eqref{eq:ham_const_a} in spherical coordinates becomes
\begin{align}
\label{eq:ham_const_k0}
 \bigtriangleup_{\rm flat} \psi = &
\left[
        \frac{1}{r^2}\tfrac{\partial}{\partial r}r^2\tfrac{\partial}{\partial r}
       +\frac{1}{r^2\sin\theta}\tfrac{\partial}{\partial \theta}\sin\theta\tfrac{\p}{\p \theta}
       +\frac{1}{r^2\sin^2\theta}\tfrac{\partial^2}{\partial \Phi^2}%\psi
\right] \psi
\nonumber\\ 
= & -\pi\psi^{5} \left(\Pi_R^2+\Pi_I^2\right)
\,.
\end{align}
The ansatz
\begin{subequations}
\begin{align}
\label{eq:ansatz_Pi_SBH}
\Pi  =& \frac{\psi^{-5/2}}{\sqrt{r\pi}} F(r)Z(\theta,\phi)
\,,\\
\label{eq:ansatz_conf_SBH}
\psi = & \psi_{S} +\sum_{lm} \frac{u_{lm}(r)}{r} Y_{lm}(\theta,\phi)
\nonumber\\
     = & 1 +\frac{M}{2r} +\sum_{lm} \frac{u_{lm}(r)}{r} Y_{lm}(\theta,\phi)
\,,
\end{align}
\end{subequations}
reduces the Hamiltonian constraint to
\begin{align}
\label{eq:usol}
\sum_{lm} \left(u_{lm}''-\frac{l(l+1)}{r^2}u_{lm}\right) Y_{lm} = & -F(r)^2Z(\theta, \phi)^2
\,,
\end{align}
where we (furthermore) impose that $Z(\theta,\phi)$ and $\psi$ are real functions. 
Thus, we have been able to reduce a complex problem to finding the solution
of an inhomogenous second order ordinary differential equation (ODE). 
This equation has a number of interesting analytic solutions.
Here, we focus on two different classes of Gaussian-type initial conditions, one of which is spherically symmetric, 
while the second class contains a dipole configuration:

\noindent{{\bf Initial data I:}} Let us start with Gaussian-type, spherically symmetric initial data,
\begin{align}
\label{eq:ansatz_l0m0}
Z(\theta,\phi)=&\frac{1}{\sqrt{4\pi}}
\,,\quad
F(r)= A_{00}\times \sqrt{r}e^{-\tfrac{(r-r_0)^2}{w^2}}
\,,
\end{align}
where $A_{00}$ is the scalar field amplitude and $r_0$ and $w$ are the location of the centre of the Gaussian and its width.
By solving Eq.~\eqref{eq:usol}, we obtain the only non-vanishing component of $u_{lm}(r)$
\begin{align}
\label{eq:u00}
u_{00}= & A_{00}^2\frac{w(w^2-4r_0(r-r_0))}{16\sqrt{2}}\left(\erf{\left(\frac{\sqrt{2}(r-r_0)}{w}\right)}-1\right)
\nonumber\\ & 
-A_{00}^2\frac{r_0w^2}{8\sqrt{\pi}}e^{-2(r-r_0)^2/w^2}
\,,
\end{align}
where we have imposed that $u_{lm}\to 0$ at infinity. Other solutions can be obtained by adding a constant to \eqref{eq:u00}. 
\noindent{{\bf Initial data II:}} The second case of interest concerns initial data for a dipole 
scalar field~\footnote{Note, that the scalar field momentum $\Pi$ is {\it never} a pure dipole, 
because the term in Eq.~\eqref{eq:ansatz_Pi_SBH} that is proportional to the conformal factor $\psi$
mixes multipoles.
For all practical purposes, we will refer to this configuration as ``dipole'' bearing in mind this caveat.}.
Specifically, we consider a superposition of $l=1,m=\pm1$ spherical harmonics for its angular dependency, such
that $Z(\theta,\phi)\in\mathbb{R}$ and set 
\begin{subequations}
\label{eq:ansatz_l1m1}
\begin{align}
Z(\theta,\phi)=& 
          Y_{1-1} - Y_{11} 
        = \sqrt{\tfrac{3}{2\pi}}\sin\theta \cos\phi
\,,\\ 
F(r)= & A_{11}\times r e^{-\tfrac{(r-r_0)^2}{w^2}}
\,.
\end{align}
\end{subequations}
As before $r_0$ and $w$ are the parameters of the Gaussian, whereas $A_{11}$ denotes the amplitude of the dipole scalar field.
The only non-vanishing Clebsch-Gordon coefficients $C_Y^{lm}\equiv \int d\Omega Z(\theta,\phi)^2 Y_{lm}(\theta,\phi)$ are
\begin{align}
\label{eq:coef_Cy}
C_Y^{00} = & \frac{1}{\sqrt{\pi}}   \,,\quad
C_Y^{20} =  -\frac{1}{\sqrt{5\pi}} 
\,,\nonumber\\
C_Y^{22} = &  C_Y^{2-2}=\sqrt{\frac{3}{10\pi}} 
\,.
\end{align}
We thus find the following ODE for $u_{lm}$
\begin{align}
u_{lm}''-\frac{l(l+1)}{r^2}u_{lm} = & -C_Y^{lm} F(r)^2
\,,
\end{align}
with non-vanishing contributions 
\begin{widetext}
\begin{subequations}
\label{eq:u_l1m1}
\begin{align}
\label{eq:u22}
u_{22} = u_{2-2} = & 
-\tfrac{A_{11}^2w^2}{80r^2}\sqrt{\tfrac{3}{10\pi}}e^{-2\tfrac{(r-r_0)^2}{w^2}}\left(4(r^4+r^3r_0+r^2r_0^2+rr_0^3+r_0^4)+w^2(4r^2+7rr_0+9r_0^2)+2w^4\right)
\nonumber \\ &
 + A_{11}^2\sqrt{\tfrac{3}{5}}\tfrac{w\left(-16r^5+16r_0^5+40r_0^3w^2+15r_0w^4\right)}{320r^2}\left(\erf{\left(\tfrac{\sqrt{2}(r-r_0)}{w}\right)}-1\right)
\nonumber \\ &
 + A_{11}^2\sqrt{\tfrac{3}{5}}\tfrac{w r_0\left(16r_0^4+40r_0^2w^2+15w^4\right)}{320r^2}\left(\erf{\left(\tfrac{\sqrt{2}\,r_0}{w}\right)}+1\right)
\nonumber \\ &
 + A_{11}^2\sqrt{\tfrac{6}{5\pi}}\,e^{-2r_0^2/w^2}\tfrac{2w^2 \left(4r_0^4+9r_0^2w^2+2w^4\right)}{320r^2}
\,, \\
u_{00} = & \tfrac{A_{11}^2 w}{16}\left(-\tfrac{2w(2r_0^2+w^2)e^{-2(r-r_0)^2/w^2}}{\sqrt{\pi}}-\sqrt{2}\left(4(r-r_0)r_0^2+(r-3r_0)w^2\right)\left(\erf{\left(\tfrac{\sqrt{2}(r-r_0)}{w}\right)}-1\right)\right)\,,
\\
\label{eq:u20}
u_{20}=&-\sqrt{\tfrac{2}{3}}u_{22}\,.
\end{align}
\end{subequations}
\end{widetext}

{\noindent{\bf{Comparison of constraint-satisfying and constraint-violating initial data:}}}
To gauge the usefulness of this initial data, we have compared the evolution of both constraint violating and
constraint satisfying setups, runs {\textit{S00\_CV}} and {\textit{S00\_m0\_e}} in Table~\ref{tab:SetupSchwarzschildmassless}, respectively. 
In both cases we considered a spherically symmetric scalar field surrounding an initially non-rotating BH. 
The massless scalar shell is localized around $r_0=12M$ with a width of $w=2.0M$ and amplitude $M A=0.075$.
This constraint-violating initial data closely resembles the setup used in Refs.~\cite{Healy:2011ef,Berti:2013gfa}. 
Figs.~\ref{fig:SchwarzschildMasslessHamilitonian} and~\ref{fig:violation_H_time}
show snapshots of the Hamiltonian constraint along the x-axis.
We find that constraint violating initial data evolves to solutions which violate the constraints even more severely with time.
This is particularly relevant close to the horizon, and it leads to unphysical properties, such as a strong {\textit{decrease}} 
in the AH area as illustrated in Fig.~\ref{fig:violation_AH}.
We note that this violates the first law of BH thermodynamics, which states that the area of the (event) horizon should always increase.
On the other hand, the analytic, constraint-satisfying initial data yields much better constraint-preserving evolutions,
leading to physically sensible results, as is apparent in Fig.~\ref{fig:violation_AH}.
\begin{figure*}[htpb!]
 \subfloat[Initial data I ]{\includegraphics[width=8cm,clip]{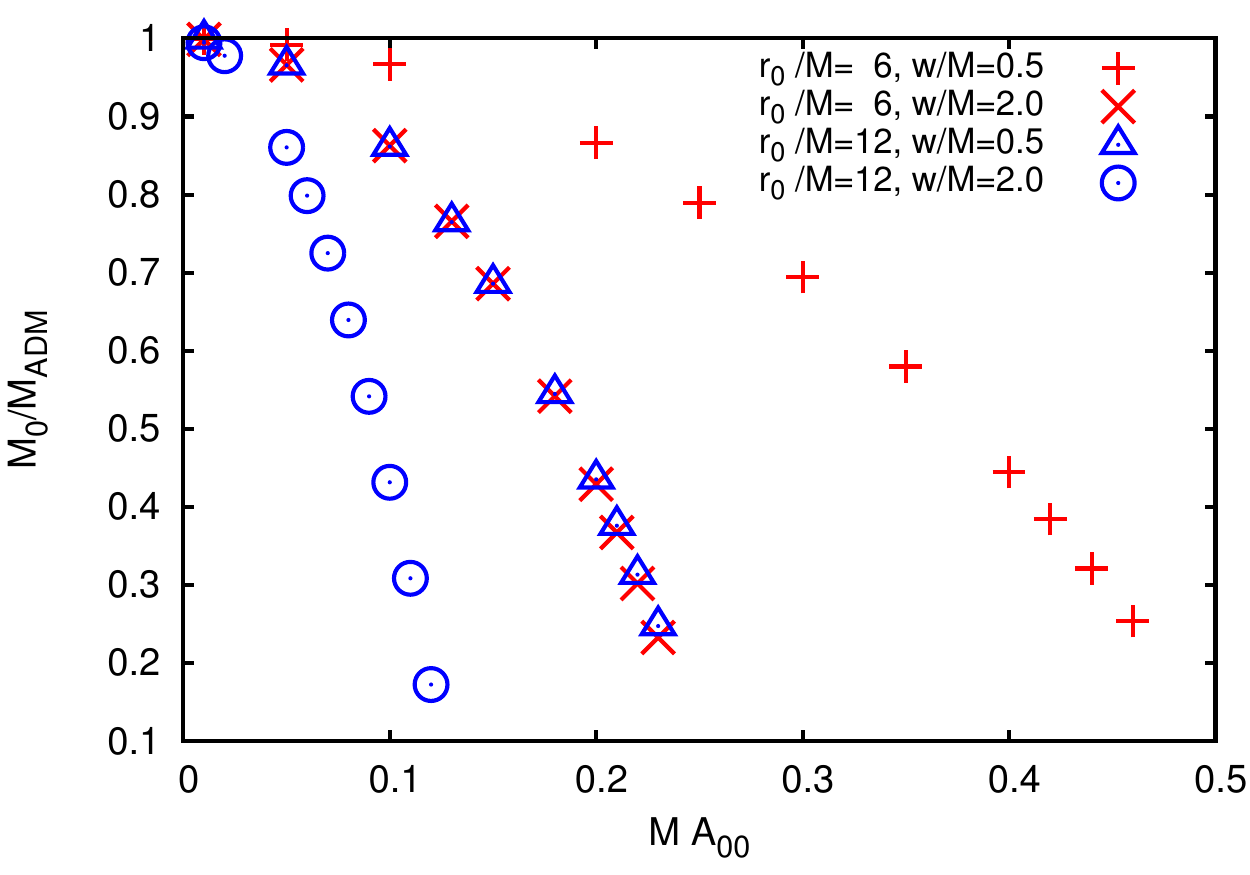}\label{fig:init_Y00}}
 \subfloat[Initial data II]{\includegraphics[width=8cm,clip]{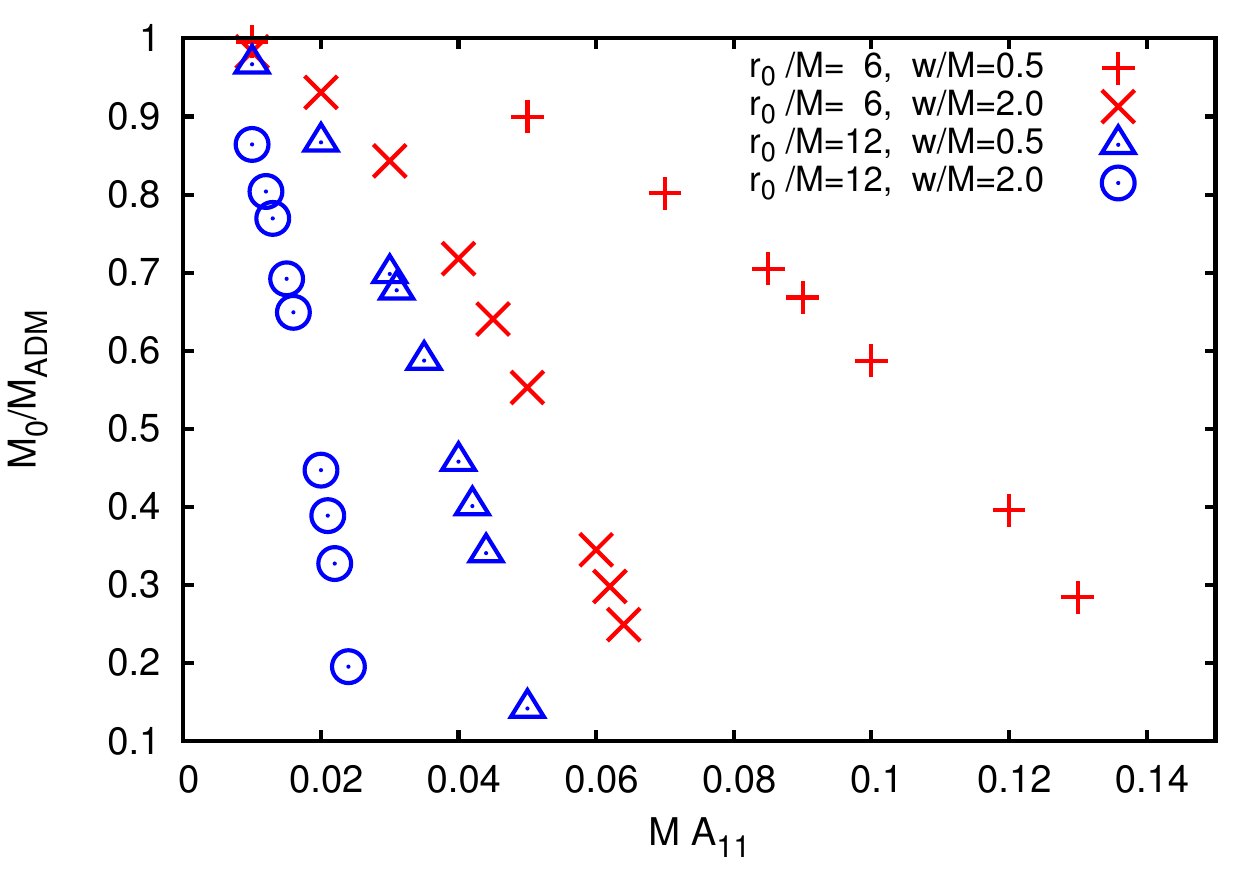}\label{fig:init_Y11}}
 \caption{\label{fig:SBH_initial_mass}
 We present the relation between the initial BH mass $M_{0}=M_{\rm{BH}}(0)$ normalized by the ADM mass.
 and the scalar field amplitude $A_{lm}$ for initial data I (left) and II (right)
 for various sets of parameters.
 Specifically, the Gaussian scalar shells with widths $w=0.5M$ or $w=2.0M$ 
 are centered around $r_0=6M$ or $r_0=12M$.
 The BH mass decreases as the amplitude increases because we have fixed the ADM mass.
 }
\end{figure*}
{\noindent{\bf{Validation of constraint satisfying initial data:}}}
Additionally, we have performed a series of simulations constructing both type-I and type-II initial data
for different sets $(r_0, w)$ of scalar field parameters and a wide range of amplitudes.
In Fig.~\ref{fig:SBH_initial_mass} we show the 
relation between the scalar field amplitude and the BH mass at $t=0$.
The total (ADM) mass $M_{\rm{ADM}}$ is fixed in these constructions because the term
$\sum_{lm} \tfrac{u_{lm}}{r}Y_{lm}$ in the conformal factor~\eqref{eq:ansatz_conf_SBH} falls off sharply at infinity 
and we have required $u_{lm}\rightarrow0$ asymptotically.
Therefore, the correction to the conformal factor, Eq.~\eqref{eq:ansatz_conf_SBH},
does not contribute to the mass of the system.
Both initial data sets have similar qualitative features, and specifically the BH Christodoulou mass decreases as the amplitude of the scalar field increases.

%%%%%%%%%%%%%%%%%%%%%%%%%%%%%%%%%%%%%%%%%%%%%%%%%%%%%%%%%%%%%%%%%%%%%%%%%%%%%%%%%%%%%%%%%%%%%%%%%%%%%%%%%%%%
\subsection{Extensions to generic conformally flat spacetimes and other scalar field profile}\label{ssec:InitDataCF}
%%%%%%%%%%%%%%%%%%%%%%%%%%%%%%%%%%%%%%%%%%%%%%%%%%%%%%%%%%%%%%%%%%%%%%%%%%%%%%%%%%%%%%%%%%%%%%%%%%%%%%%%%%%%

We note, that in the previous construction we considered scalar clouds around a spherically symmetric spacetime
with fixed ADM mass $M_{\rm{ADM}}$ by requiring that $u_{lm}$ vanishes asymptotically. 
However, this is only one possible choice. In particular, the framework is valid as long as 
we assume a conformally flat metric and can, therefore, easily be extended to more general cases.

The simplest example is the Minkowski spacetime with line element
\begin{align}
\dif s^2 =& -\dif t^2 +\eta_{ij} \dif x^i\dif x^j
\,,
\end{align}
for which $\tgam_{ij}=\eta_{ij}$ and $K=0$ by construction.
Again, we choose $\Phi=0$ and adopt Eq.~\eqref{eq:ansatz_Pi_SBH} for the scalar field momentum.
Because initially there is no BH present in this spacetime we take the ansatz
\begin{align}
\psi =& 1+\sum_{lm}\frac{u_{lm}}{r}Y_{lm}(\theta,\phi)
\,,
\end{align}
for the conformal factor, which follows directly from Eq.~\eqref{eq:ansatz_conf_SBH}
in the limit $M\rightarrow0$.
Then, the Hamiltonian constraint results in the ODE given in Eq.~\eqref{eq:usol}.
This implies that the solutions $u_{lm}$ of the initial data I and II
are also the analytic solutions for the Minkowski ``background'' spacetimes
with the ansatz~\eqref{eq:ansatz_l0m0} and \eqref{eq:ansatz_l1m1}.
We note that, in particular, the solution with the ansatz~\eqref{eq:ansatz_l0m0} is
spherically symmetric and the dynamical problem can be
reduced to a $1+1$ problem~\cite{Okawa:2013jba}.
In this case, the solution is written with $u_{00}$ given in Eq.~\eqref{eq:u00} by
\begin{subequations}
\label{eqinit2}
\begin{align}
 \psi = & 1+\frac{u_{00}(r)-u_{00}(0)}{2r\sqrt{\pi}}
 \,,\\ 
 \Pi = & \frac{A_{00}}{2\pi}e^{-\tfrac{(r-r_0)^2}{w^2}}\psi^{-\tfrac{5}{2}}
\,,
\end{align}
\end{subequations}
where we subtracted $u_{00}(0)$ in order to regularize the conformal
factor at the origin (because $u_{00}(0)$ is a nonzero constant, making the term $u_{00}(r)/r$ singular at the origin).
The system has ADM energy $M_{\rm ADM}=-u_{00}(0)/\sqrt{\pi}$ due to the scalar field.

%%%%%%%%%%%%%%%%%%%%%%%%%%%%%%%%%%%%%%%%%%%%%%%%%%%%%%%%%%%%%%%%%%%%%%%%%%%%%%%%%%%%%%%%%%%%%%%%%%%%%%%%%%%%
\subsection{Superposition of scalar field profiles and multiple black holes} \label{sec:InitDataSSF}
%%%%%%%%%%%%%%%%%%%%%%%%%%%%%%%%%%%%%%%%%%%%%%%%%%%%%%%%%%%%%%%%%%%%%%%%%%%%%%%%%%%%%%%%%%%%%%%%%%%%%%%%%%%%
%
Another interesting setup would be the superposition of various scalar fields
motivated by the possible presence of more than one fundamental, ultralight scalar field
or by the prospect of coupling fundamental fields to spacetimes in modified gravity, such as 
scalar--tensor theories.
Here, we will focus on scalar clouds in Minkowski or around Schwarzschild BHs 
which allows us to provide analytic solutions describing this setup.
In particular, we can keep the assumptions that we made in the last section, namely
\begin{align}
\tgam_{ij} = & \eta_{ij} \,,\quad
K=0 \,,\quad
\Phi=0 \,.
\end{align}
Instead, we consider non-trivial profiles of the scalar field momenta and will solve the Hamiltonian
constraint for the deformation of the conformal factor 
from its value $\psi_{\rm{CF}}$ in the absence of any additional field.
In particular, for Minkowski we have $\psi_{\rm{CF}}=1$. In a spacetime containing $N_{\rm{BH}}$ non-rotating BHs 
without any momenta the solution is given by Brill-Lindquist initial data~\cite{Brill:1963yv,Lindquist1963}
\begin{align}
\psi_{\rm{CF}}= & 1+\sum_{(a)=1}^{N_{\rm{BH}}}\frac{M_{(a)}}{2r_{(a)}}
\,,
\end{align}
where $M_{(a)}$ and $r_{(a)}$ are the bare mass parameter and position of the $(a)$-th BH.
For example, the calculation of the previous section can be generalized to two spherically symmetric profiles (with equal amplitude), 
\begin{align}
\Pi   =& \Pi_{1} +\Pi_{2}\,,\quad
\Pi_{i} =\frac{A_{00}}{2\pi}e^{-\tfrac{(r-r_{i})^2}{w^2}}\psi^{-\tfrac{5}{2}}\,,
\end{align}
where $r_{i}$ denotes the distance of the peak of the $i$-th spherical shell from the origin.
We take the ansatz for the conformal factor
\begin{align}
 \psi =& \psi_{\rm{CF}} + \Bigl[u_{00,1} + u_{00,2} + u_{00,\rm{CF}}\Bigr] \frac{Y_{00}(\theta,\phi)}{r}
\,.
\end{align}
Although the Hamiltonian constraint becomes more complicated with this ansatz
we already know the solution of the following equations by choosing
appropriate parameters in Eq.~\eqref{eq:u00},
\begin{align}
\bigtriangleup_{\rm{flat}} \frac{u_{00,i}(r; A_{00},r_{i})}{r}Y_{00}
= &  -\pi\psi^{5} \Pi^{2}_{i}
\,.
\end{align}
The solution of the remaining equation can be described by
the parameters $\bar{A}_{00}$ and $\bar{r}_{12}$,%$\bar{r}_0$,
\begin{align}
\bigtriangleup_{\rm flat} \frac{u_{00,\rm{CF}}(r;\bar{A}_{00},\bar{r}_{12})}{r}Y_{00} = &  
          - 2\pi\psi^{5} \Pi_1\Pi_2
\\ = &
          -\frac{\bar{A}_{00}^2}{4\pi}e^{-\tfrac{2\left(r-\bar{r}_{12}\right)^2}{w^2}}
\,,\nonumber
\end{align}
where
\begin{align}
\label{eq:def_cross_amp}
\bar{A}_{00}^2 = & 2 A_{00}^2 e^{-\tfrac{(r_1-r_2)^2}{2w^2}}
\,,\quad
\bar{r}_{12} = \frac{r_1+r_2}{2}
\,.
\end{align}
Thus, the initial data describing the superposition of two scalar fields is given by
\begin{subequations}
 \begin{align}
\psi =& \psi_{\rm{CF}}+\tfrac{1}{2r\sqrt{\pi}}\Bigl[u_{00,1}(r;A_{00},r_1) -u_{00,1}(0;A_{00},r_1)
\Bigr.\nonumber\\ & \Bigl.
  +u_{00,2}(r;A_{00},r_2) -u_{00,2}(0;A_{00},r_2)
\Bigr.\nonumber\\ & \Bigl.
  +u_{00,CF}(r;\bar{A}_{00},\bar{r}_{12})\Bigr]\,,\\
\Pi =&
  \frac{A_{00}}{2\pi}\psi^{-\tfrac{5}{2}}
  \Bigl[e^{-\tfrac{(r-r_1)^2}{w^2}} +e^{-\tfrac{(r-r_2)^2}{w^2}}\Bigr]
  \,.
 \end{align}
\end{subequations}
In addition, one can extend this construction to $N_{\rm{SF}}$ scalar field profiles
in a straight-forward calculation
\begin{subequations}
\begin{align}
\psi = & \psi_{\rm{CF}} 
        + \tfrac{1}{2r\sqrt{\pi}}\Bigl[\sum_{i=1}^{N_{\rm{SF}}}\left\{u_{00,i}(r;A_{00},r_i) -u_{00,i}(0;A_{00},r_i)\right\}
\Bigr.\nonumber\\ & \Bigl.
                +\sum_{i<j}u_{00,ij}(r;\bar{A}_{00,ij},\bar{r}_{ij})\Bigr]
\,,\\
\Pi = & \frac{A_{00}}{2\pi}\psi^{-\tfrac{5}{2}} \sum_{i=1}^{N_{\rm{SF}}}e^{-\tfrac{(r-r_i)^2}{w^2}}
\,,
\end{align}
\end{subequations}
with
\begin{align}
\bar{A}_{00,ij}^2 = & 2 A_{00}^2 e^{-\tfrac{(r_i-r_j)^2}{2w^2}}
\,,\quad
\bar{r}_{ij} = \frac{r_i+r_j}{2}
\,,
\end{align}
where the constant $u_{00,i}(r=0)$ was added to $u_{00,i}(r)$ in order to have a positive ADM mass.
In principle, it is possible to generalize the construction outlined in this section to include
type II initial data, modeling the superposition of $N_{\rm{SF}}$ dipole scalar fields,
or to allow for arbitrary amplitude $A_{i,00}\neq A_{j,00}$.

%
%%%%%%%%%%%%%%%%%%%%%%%%%%%%%%%%%%%%%%%%%%%%%%%%%%%%%%%%%%%%%%%%%%%%%%%%%%%%%%%%%%%%%%%
\subsection{A rotating black hole surrounded by a scalar field}\label{ssec:InitDataKBH}
%%%%%%%%%%%%%%%%%%%%%%%%%%%%%%%%%%%%%%%%%%%%%%%%%%%%%%%%%%%%%%%%%%%%%%%%%%%%%%%%%%%%%%%
A particularly interesting scenario that we have not yet captured with our analytic initial data
construction are scalar fields surrounding rotating BHs. This is an especially attractive
setup, because it opens up the possibility to explore superradiant effects and the
BH bomb mechanism \textit{non-linearly}.
For this purpose we consider the Kerr solution as underlying BH spacetime.
The line element of the Kerr BH in Boyer-Lindquist (BL) coordinates is given by
\begin{subequations}
\label{eq:KerrBL}
\begin{align}
\label{eq:KerrBLLE}
\dif s^2 = & -\left(1-\frac{2Mr_{\rm{BL}}}{\Sigma}\right)\dif t^2
        -\frac{4aMr_{\rm{BL}}\sin^2\theta}{\Sigma}\dif t\dif\phi
\nonumber\\ &
        +\frac{\Sigma}{\Delta}\dif r^2_{\rm{BL}} +\Sigma\dif \theta^2
        +\frac{\A}{\Sigma}\sin^2\theta\dif\phi^2
\,,\\
\label{eq:KerrAA_1}
\A = & \left(r_{\rm{BL}}^2+a^2\right)^2 -\Delta a^2\sin^2\theta
\,,\\ 
\label{eq:KerrSig_1}
\Sigma = & r_{\rm{BL}}^2 +a^2\cos^2\theta
\,,\\ 
\label{eq:KerrDelta_1}
\Delta = & r_{\rm{BL}}^2 -2Mr_{\rm{BL}}+a^2
\,,
\end{align}
\end{subequations}
where $M$ is again the BH bare mass parameter and $a/M$ is the dimensionless spin parameter.
However, BL coordinates are not favourable from a numerical viewpoint,
because there are coordinate singularities at the location of horizon $r_{\pm}=M\pm \sqrt{M^2-a^2}$.
Brandt \& Seidel \cite{Brandt:1994ee,Brandt:1996si} and Liu et al \cite{Liu:2009al}
proposed to employ quasi-isotropic coordinates in which
the spatial metric components remain regular.
Following Liu et al \cite{Liu:2009al}, we introduce the 
quasi-isotropic radial coordinate $R$ which is related to the BL radial coordinate $r_{\rm{BL}}$ via
\begin{align}
\label{eq:KerrRBL_V2}
r_{\rm{BL}} = & R\left( 1+\frac{r_{+}}{4R} \right)^2
\,.
\end{align}
In this coordinate the outer horizon is located at $R=\tfrac{r_{+}}{4}$ which
results in a \textit{finite} value $\lim_{a/M\rightarrow1}R = \tfrac{M}{4}$ in the extremal limit.
Thus, we are able to stably evolve highly spinning BHs -- with initial spin up to $a_{0}/M \sim 0.95$ --
represented by punctures with high accuracy. To be precise, the deviation from the expected spin parameter (in the pure Kerr case) 
is less than $1\%$ even for the highest spinning case, 
as we will discuss in Appendix~\ref{app:SFresultsIIIpureKerr} (see also Ref.~\cite{Liu:2009al}).
In these quasi-isotropic coordinates $(R,\theta,\phi)$ the spatial part of the Kerr solution~\eqref{eq:KerrBL}
writes
\begin{align}
\label{eq:KerrQI}
\dif l^2 = & \gamma_{ij}^{\rm{BG}} \dif x^{i} \dif x^{j}
\nonumber\\ = &
         \psi_{0}^4\left[
 \frac{\left(R+\tfrac{r_{+}}{4}\right)^2 }{R(r_{\rm{BL}} - r_{-})} \dif R^2
 + R^2 \dif\theta^2 + \frac{\A}{\Sigma^2} R^2\sin^2\theta \dif \phi^2
 \right]
\,,\nonumber\\
\tgam_{ij}^{\rm{BG}} =& \psi_{0}^{-4}\gamma_{ij}^{\rm{BG}}
\,,\quad
\psi^{4}_{0} = \tfrac{\Sigma}{R^2}
\nonumber\,,\\
 \alpha = & \sqrt{\tfrac{\Delta \Sigma}{\A}}
\,,\quad
\beta^{\phi} = -2 a M \tfrac{r_{\rm{BL}}}{\A}
\end{align}
where $\A$, $\Sigma$ and $\Delta$ are given by Eqs.~\eqref{eq:KerrBL}.
The extrinsic curvature becomes
\begin{subequations}
\label{eq:KerrQI_ext}
\begin{align}
K^{\rm{BG}}_{R\phi} = & \frac{a M \sin^2\theta}{\Sigma\sqrt{\A\Sigma}}
        \left( 2 r_{\rm{BL}}^2\left[r_{\rm{BL}}^2 + a^2\right] + \Sigma\left[r_{\rm{BL}}^2 - a^2\right] \right)
\nonumber\\ & \times
        \left( 1+\frac{r_{+}}{4R} \right) \frac{1}{\sqrt{R}\sqrt{r_{\rm{BL}}-r_{-}}} 
\,,\\
K^{\rm{BG}}_{\theta\phi} = & -2 a^3 M \frac{r_{\rm{BL}} \cos\theta \sin^3\theta}{\Sigma\sqrt{\A\Sigma}}
\nonumber\\ & \times
        \left(1 - \frac{r_{+}}{4R} \right) \sqrt{R}\sqrt{r_{\rm{BL}}-r_{-}}
\,.
\end{align}
\end{subequations}

Let us now consider the presence of a scalar field around a Kerr BH.
In order to prepare this type of configuration properly
we need to solve the constraint Eqs.~\eqref{eq:ConstraintsID}.
Specifically, we adopt the Kerr metric in quasi-isotropic coordinates, 
Eq.~\eqref{eq:KerrQI}, as conformal metric with the curvature given in Eq.~\eqref{eq:KerrQI_ext}.
We still have the freedom to impose the maximal slicing condition and, furthermore,
choose a vanishing scalar field initially, yielding the ansatz
\begin{align}
% K    = & 0 \,,\\ 
% \Phi = & 0 \,,\\
K = & 0 \,,\quad 
\Phi = 0 \,,\quad
\Pi  = \frac{A_{\rm{G}}}{\sqrt{\pi}} \psi^{-\tfrac{5}{2}} Z(\theta, \phi) e^{-\tfrac{(r-r_0)^2}{w^2}}
\end{align}
where the different angular profiles $Z (\theta,\phi)$ are given in Eqs.~\eqref{eq:ansatz_l0m0} or~\eqref{eq:ansatz_l1m1}.
The constraint Eqs.~\eqref{eq:ConstraintsID} become
\begin{subequations}
\label{eq:ConstraintsKerrCNF}
\begin{align}
\label{eq:ham_const_cnf0}
\mathcal{H} = & \tilde{\bigtriangleup}\psi -\tfrac{1}{8}\tR^{\rm{BG}}\psi
        +\tfrac{1}{8}\tA^{ij}_{\rm{BG}}\tA_{ij}^{\rm{BG}}\psi^{-7} +\pi\Pi^2\psi^{5} = 0
\,,\\
\label{eq:mom_const_cnf}
\mathcal{M}_i = & \tD_j \tA^{{\rm{BG}}j}{}_{i} = 0
\,,
\end{align}
\end{subequations}
where $\tR^{\rm{BG}}$ corresponds to $\tg^{\rm{BG}}_{ij}$ in Eq.~\eqref{eq:KerrQI}
and the traceless extrinsic curvature of the Kerr BH is
defined by using background quantities as $\tA^{\rm{BG}}_{ij} = \psi^{2}_{0} K^{\rm{BG}}_{ij}$.
As long as we use the background conformal metric and the traceless component of the background extrinsic
curvature, the momentum constraints~\eqref{eq:mom_const_t} are trivially satisfied by the maximal slicing condition.
Furthermore, we take the ansatz for the conformal factor
\begin{align}
\label{eq:conformal_cnf}
\psi \equiv & \psi_0 \left(1+u(r,\theta,\phi)\right)
\,,
\end{align}
where $\psi_0$ is given in Eq.~\eqref{eq:KerrQI}.
Then, Eq.~\eqref{eq:ham_const_cnf0} becomes
\begin{align}
\label{eq:ham_const_cnf}
\bigtriangleup_{\rm flat} u =& 
        \left[\eta^{ij}-\tgam^{ij}_{\rm{BG}}\right]\partial_i\partial_ju
        -\tgam^{ij}_{{\rm{BG}}\,,i}\partial_ju 
\nonumber\\ &
        -2\tgam^{ij}_{\rm{BG}}\psi_0^{-1}\partial_iu\partial_j\psi_0
        -A_{G}^2 \psi^{-1}_0 Z^2 (\theta,\phi) e^{-\tfrac{2(r-r_0)^2}{w^2}}
\nonumber\\ &
 +\tfrac{1}{8}\tA^{ij}_{\rm{BG}}\tA_{ij}^{\rm{BG}}\psi_0^{-8}u(2+u)(2+2u+u^2)
\nonumber\\ & \quad\times
(2+4u+6u^2+4u^3+u^4)(1+u)^{-7}
\,,
\end{align}
where we have used the vacuum Hamiltonian constraint to eliminate the Ricci scalar $\tilde{R}^{\rm{BG}}$.
As we have shown, this procedure yields a single elliptic PDE for the regular function $u$ 
which we will solve numerically using standard numerical methods under
the outer boundary condition $u\rightarrow r^{-1}$.

%%%%%%%%%%%%%%%%%%%%%%%%%%%%%%%%%%%%%%%%%%%%%%%%%%%%%%%%%%%%%%%%%%%%%%%%%%%%%%%%%%%%%%%%%%%%
\subsection{A black hole and a pseudo-bound state of a scalar field}\label{ssec:InitDataFBS}
%%%%%%%%%%%%%%%%%%%%%%%%%%%%%%%%%%%%%%%%%%%%%%%%%%%%%%%%%%%%%%%%%%%%%%%%%%%%%%%%%%%%%%%%%%%%
We have just explored different ways of providing initial data either in closed analytic form
or requiring numerical integrations assuming Gaussian scalar field profiles. 
However elegant and appealing, these constructions still describe very contrived physical situations.
Accordingly, the time evolution of these initial conditions might lead to arbitrary,
potentially large absorption and scattering of radiation to infinity.

However, linearized perturbations of massive scalar fields around Kerr BHs have revealed the existence 
of long-lived modes, or quasi-bound states, which have been investigated in the frequency domain in 
the past \cite{Cardoso:2005vk,Konoplya:2006br,Dolan:2007mj,Berti:2009kk}.
The quasi-bound states are useful because they prescribe a clean state
and, in fact, numerical simulations 
in the time domain have uncovered
interesting phenomena~\cite{Yoshino:2012kn,Witek:2012tr,Dolan:2012yt}.

We now turn to the question of constructing initial data that corresponds roughly to the scenario under consideration,
i.e., describing long-lived quasi-stationary states around BHs, for which the scalar field is almost monochromatic and in a very specific angular momentum state. We will call this initial data {\it pseudo-bound states}, as they will solve the full non-linear problem but only mimick
the {\it quasi-bound states} seen in linearized studies \cite{Cardoso:2004nk,Witek:2010qc,Dolan:2012yt}. The evolution of this data will lead to very small initial accretion or loss to infinity.

Initial data describing pseudo-bound states requires both a non-vanishing scalar field~$\Phi$ and conjugated 
momentum $\Pi$. In this case the momentum constraints~\eqref{eq:mom_const_t} become (in principle) non-trivial 
and are difficult to solve even numerically.
We show how to lighten this burden by taking the ansatz for 
$\Phi$~\footnote{for convenience we use the opposite sign in the time dependence as compared to 
standard conventions~\cite{Berti:2009kk}.}
\begin{align}
\label{eq:FBS_Phi}%\\
\Phi (t,r,\theta,\phi) = & \tfrac{A_P}{\sqrt{\pi}} \exp\left[i(\omega t+m\phi)-\tfrac{(r-r_0)^2}{w^2}\right] Z(\theta)
\,,
\end{align}
where $r_0$ and $w$ are the location and width of the Gaussian and $A_P$ is the scalar field amplitude.
We get, by definition, the relation,
\begin{align}
\label{eq:FBS_Pi}
\Pi (t,r,\theta,\phi) = & \tfrac{1}{\alpha}\left[\beta\partial_{\phi}\Phi-\partial_t\Phi\right]
        =\tfrac{i}{\alpha}\left(m\beta-\omega\right)\Phi
\,,
\end{align}
where we defined $\beta=\beta^{\phi}$. 
We now insert Eq.~\eqref{eq:FBS_Pi} into the momentum constraints~\eqref{eq:mom_const_t} and recall that 
$\tA^{\rm{BG}j}{}_{i}$ solves the vacuum momentum constraints. Then,
Eq.~\eqref{eq:mom_const_t} yields the relation
\begin{align}
\label{eq:mom_const_FBS}
\partial_i K = & -6\pi \Bigl[ \Pi\partial_i\Phi^{*}+\Pi^{*}\partial_i\Phi \Bigr]
\nonumber\\  = &
                 -i\frac{6\pi (m\beta-\omega)}{\alpha}\Bigl[ \Phi\partial_i\Phi^{*} -\Phi^{*}\partial_i\Phi\Bigr]
\nonumber\\  = &
                 -\frac{12\pi A_P^2}{\alpha} (m\beta-\omega) e^{-\tfrac{(r-r_0)^2}{w^2}} Z(\theta)^2 \delta^{\phi}_{i}
\,.
\end{align}
Here, we solve the (momentum) constraints~\eqref{eq:mom_const_t} by imposing the minimal slicing condition, i.e.,
$\p_{i} K = 0$. Then, the relation~\eqref{eq:mom_const_FBS} implies that $m\beta = \omega$ must hold.
Naively, one could fix the frequency $\omega=\omega_{B}=const$ to be the characteristic frequency of a quasi-bound state.
However, $\beta=const$ would not be consistent with demanding asymptotic flatness as well as 
the requirement that the shift vector decreases sufficiently fast at large distances.
Therefore, it is convenient to consider a position dependent shift and, consequently, a position dependent
frequency $\omega/m=\beta$, such that Eq.~\eqref{eq:mom_const_FBS} remains satisfied.
Guided by perturbative studies we relate the space-dependent frequency $\omega$ to the 
characteristic frequency $\omega_{B}$ of quasi-bound states 
(which is always of the order of the mass of the field), 
and specifically we set
\begin{align}
\label{eq:FBS_beta}
\beta (r,\theta,\phi)\Big|_{t=0}  = & \frac{\omega (r,\theta,\phi)}{m}
\nonumber\\ = &
        \frac{\omega_{B}}{m}\sqrt{\tfrac{3}{2\pi}}\exp\left[-\tfrac{(r-r_0)^2}{w^2}\right] \sin\theta\cos\phi
\,.
\end{align}

We now substitute the metric~\eqref{eq:KerrQI} and the extrinsic curvature~\eqref{eq:KerrQI_ext}
in the Hamiltonian constraint~(\ref{eq:ham_const_t}),
and obtain 
\begin{align}
 \label{eq:ham_const_FBS}
  \mathcal{H} = & \tilde{\bigtriangleup}\psi -\tfrac{1}{8}\tR^{\rm{BG}}\psi
  +\tfrac{1}{8}\tA^{ij}_{\rm{BG}}\tA^{\rm{BG}}_{ij}\psi^{-7}
\nonumber\\ & 
  +\pi\psi\tgam^{ij}_{\rm{BG}}\tD_i\Phi\tD_j\Phi^{*}
  +\pi\mu_S^2\psi^5\Phi\Phi^{*}=0
\,.
\end{align}
Note, that now the initial data depends on the mass parameter~$\mu_S$.
In addition, we split the conformal factor according to Eq.~\eqref{eq:conformal_cnf}
and eliminate the Ricci scalar term using the vacuum Hamiltonian. 
Under these conditions we end up with a single elliptic PDE for the regular function $u$
\begin{align}
\label{eq:pBSuPDE}
\bigtriangleup_{\rm flat} u =&
\left[\eta^{ij}-\tgam^{ij}_{\rm{BG}}\right]\partial_i\partial_ju
-\tgam^{ij}_{\rm{BG}\,,i}\partial_ju -2\tgam^{ij}_{\rm{BG}}\psi_0^{-1}\partial_iu\partial_j\psi_0
\nonumber\\ &
+\tfrac{1}{8}\tA^{ij}_{\rm{BG}}\tA_{ij}^{\rm{BG}}\psi_0^{-8}u(2+u)(2+2u+u^2)
\nonumber\\ & \quad\times 
        (2+4u+6u^2+4u^3+u^4)(1+u)^{-7}
\nonumber\\ & 
-\pi (1+u) \tgam^{ij}_{\rm{BG}}\left(\Phi^{R}_{,i}\Phi^{R}_{,j}+\Phi^{I}_{,i}\Phi^{I}_{,j}\right)
\nonumber\\ & 
-\pi \mu_S^2\psi^4_0(1+u)^5 \left(\Phi^{R}\Phi^{R}+\Phi^{I}\Phi^{I}\right)
\,.
\end{align}
We can now solve the PDE~\eqref{eq:pBSuPDE} for $u$ by employing standard numerical methods such as
a multigrid or spectral solver.

\begin{figure}[htpb!]
 \subfloat[Non-rotating BH Initial Data]{\includegraphics[width=8cm,clip]{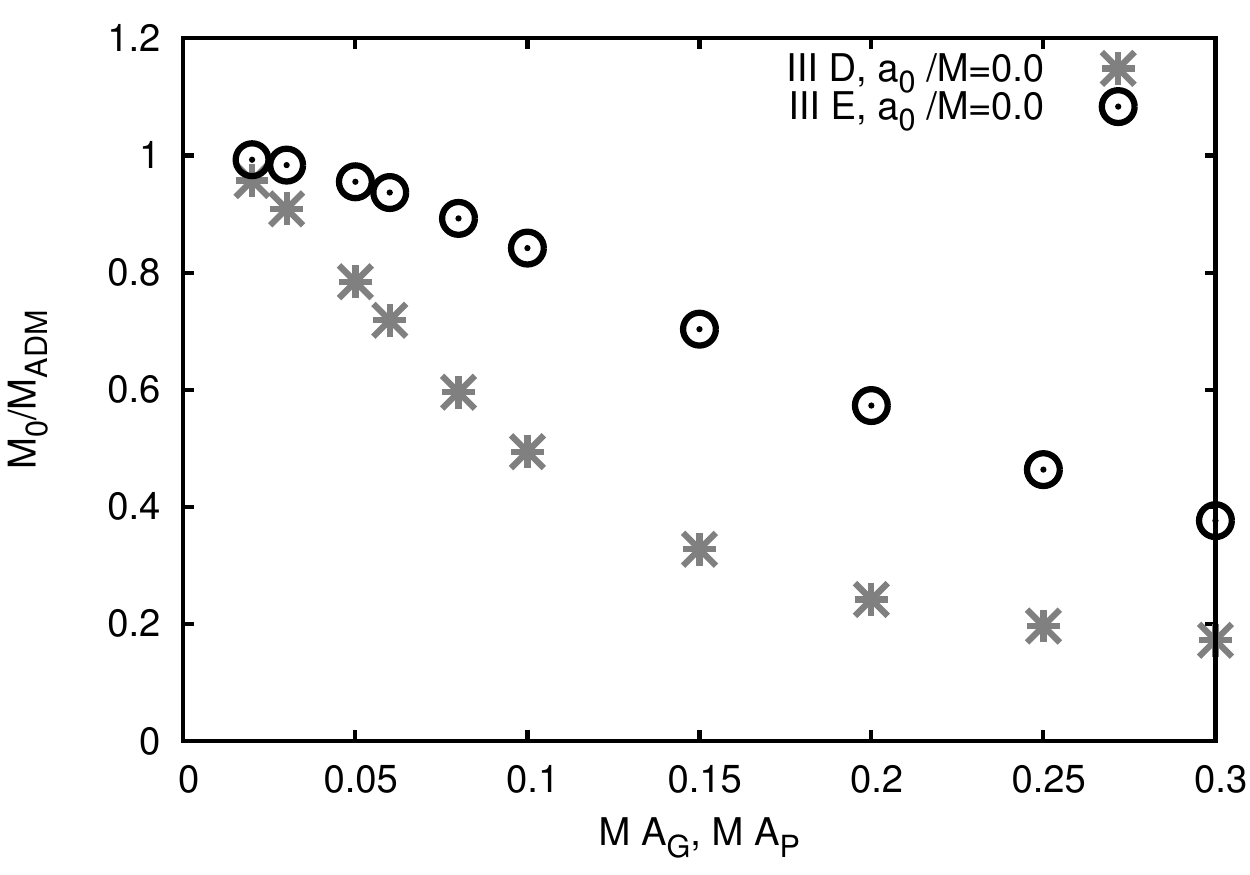}\label{fig:init_SBH_ID34}} \\
 \subfloat[Rotating BH Initial Data]{\includegraphics[width=8cm,clip]{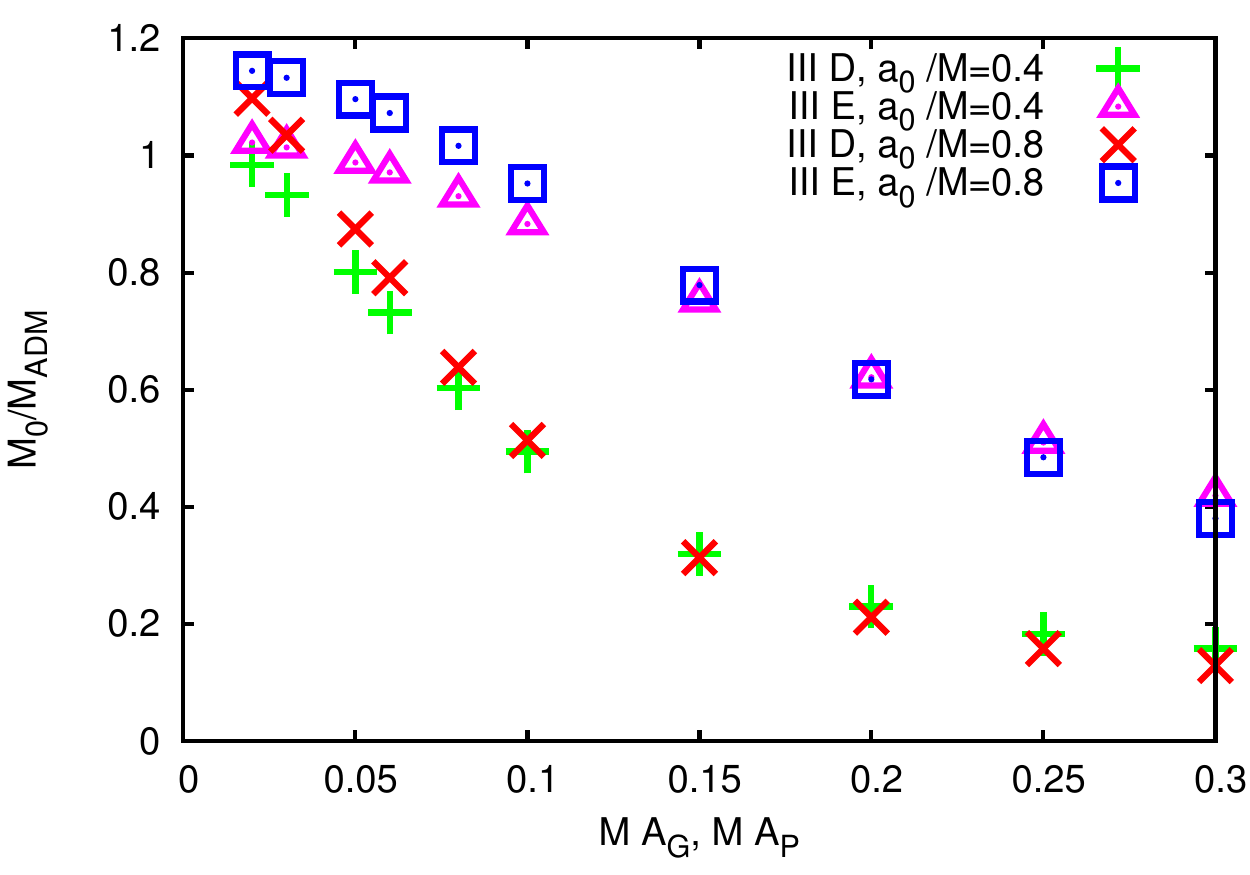}\label{fig:init_KBH_ID34}}
 \caption{\label{fig:KBH_initial_data}
 The relation between the amplitude of a scalar field and the initial BH mass 
 for a non-rotating (top) and rotating BH (bottom).
 The BH mass (in units of the ADM mass) decreases as the scalar field amplitude increases.
 Fig.~\protect\subref{fig:init_SBH_ID34} and Fig.~\protect\subref{fig:init_KBH_ID34}
 show a dipole Gaussian or pseudo-bound state scalar field coupled to a 
 non-rotating BH and a rotating BH with $a_0/M = 0.4$ and $0.8$. 
 The Gaussian scalar wave packet is localized at $r_0=12M$ and has width $w=2M$.
 }
\end{figure}

We have checked our initial data implementation by constructing Gaussian or pseudo-bound state scalar
clouds using the formalism presented in Sec.~\ref{ssec:InitDataKBH}
and~\ref{ssec:InitDataFBS}.
In particular we have computed the initial states of scalar shells with width $w=2M$ and localized at $r_0=12M$ 
for both non-rotating and rotating BH spacetimes.
We present the relation between the BH mass and the amplitude of the scalar cloud
in Fig.~\ref{fig:KBH_initial_data} for various values of the BH spin.
The present construction always yields a BH solution for any scalar field amplitude.

%%%%%%%%%%%%%%%%%%%%%%%%%%%%%%%%%%%%%%%%%%%%%%%%%%%%%%%%%%%%%%%%%%%%%%%%%%%%%%
%\clearpage
\section{Results I -- scalar clouds around non-rotating black holes}
\label{sec:SFresultsII}
%%%%%%%%%%%%%%%%%%%%%%%%%%%%%%%%%%%%%%%%%%%%%%%%%%%%%%%%%%%%%%%%%%%%%%%%%%%%%%
This section is devoted to the analysis of scalar clouds interacting with an initially non-rotating BH. 
We note that even though superradiance is absent in this case, long-lived modes
of massive fields still exist, thus involving a potentially rich phenomenology.

The numerical error is estimated to be at most $6\%$ for the scalar and gravitational waveforms at late times 
while the BH mass, area and spin have at most a $(0.19,8.5\cdot10^{-3},0.16)\%$ error. 
A discussion of the error and convergence analysis is done in Appendix~\ref{app:convergence}.

%%%%%%%%%%%%%%%%%%%%%%%%%%%%%%%%%%%%%%%%%%%%%%%%%%%%%%%%%%%%%%%%%%%%%%%%%%%%%%
\subsection{Massless scalars around non-rotating black holes}
\label{ssec:SFresultsIImassless}
%%%%%%%%%%%%%%%%%%%%%%%%%%%%%%%%%%%%%%%%%%%%%%%%%%%%%%%%%%%%%%%%%%%%%%%%%%%%%%
%
\begin{table*}[htpb!]
\begin{center}
\caption{\label{tab:SetupSchwarzschildmassless}
Setup and initial parameters for massless scalar fields around a non-rotating BH
with initial mass $M_{0}=M_{\rm{BH}}(0)$ and bare mass parameter $M=1$. 
We denote the dimensionless location $r_0/M$, width $w/M$ and amplitude $M A$
of the scalar shell with type I or II initial profile in Sec.~\ref{ssec:InitDataSBH}, 
refering, respectively, to $Z(\theta,\varphi)\sim Y_{00}$ or $Z(\theta,\varphi)\sim Y_{1-1}-Y_{11}$.
Note, that we have run one simulation using {\textit{constraint violating}} initial data, denoted as ``CV''.
We present the grid setup in the notation given by Eq.~\eqref{eq:gridsetup}, where the ``radii'' of the
refinement boxes are given in units of the bare mass $x_{i}/M$.
}
\begin{tabular}{l|ccccc|c}
\hline
Run        & type & $r_0/M$ & $w/M$ & $M\,A$  & $M_0$    & Grid setup 
\\ \hline
S00\_m0\_a & I    & $6.0$   & $2.0$ & $0.15$  & $0.6859$ & $\{(384,192,96,48,24,12,6,3,1.5),h=M/32\}$
\\
S00\_m0\_b & I    & $6.0$   & $2.0$ & $0.1$   & $0.8626$ & $\{(384,192,96,48,24,12,6,3,1.5),h=M/32\}$
\\
S00\_m0\_c & I    & $12.0$  & $0.5$ & $0.15$  & $0.6858$ & $\{(256,128,64,16,8,4,2,1),h=M/60\}$
\\
S00\_m0\_d & I    & $12.0$  & $0.5$ & $0.015$ & $0.9969$ & $\{(256,128,64,16,8,4,2,1),h=M/60\}$
\\
S00\_m0\_e & I    & $12.0$  & $2.0$ & $0.075$ & $0.6837$ & $\{(256,128,64,16,8,4,2,1),h=M/60\}$
\\
S00\_CV    & CV   & $12.0$  & $2.0$ & $0.075$ &  $1.0$ & $\{(256,128,64,32,16,8,4,2,1),h=M/40\}$
\\
\hline
S11\_m0\_a & II   & $6.0$   & $2.0$ & $0.04$  & $0.7178$ & $\{(384,192,96,48,24,12,6,3,1.5),h=M/32\}$
\\
S11\_m0\_b & II   & $6.0$   & $2.0$ & $0.03$  & $0.8430$ & $\{(384,192,96,48,24,12,6,3,1.5),h=M/40\}$
\\
S11\_m0\_c & II   & $12.0$  & $0.5$ & $0.03$  & $0.6983$ & $\{(192,96,48,24,12,6,3,1.5),h=M/60\}$
\\
S11\_m0\_d & II   & $12.0$  & $0.5$ & $0.003$ & $0.9970$ & $\{(192,96,48,24,12,6,3,1.5),h=M/60\}$
\\
\hline
\end{tabular}
\end{center}
\end{table*}

We start by investigating massless scalars coupled to a Schwarzschild BH.
Therefore, we set up initial data describing a spherically symmetric or dipole
scalar field provided, respectively, by {\textit{Initial data I}} and {\textit{Initial data II}} in Sec.~\ref{ssec:InitDataSBH}, 
Eqs.~\eqref{eq:u00} and~\eqref{eq:u_l1m1}.
We have performed a series of simulations with varying location $r_0/M$, width $w/M$ and amplitude 
$M A$
of the Gaussian wave packet, where $A$ refers to the amplitude $A_{00}$ or $A_{11}$ for type I or II initial data in
Sec.~\ref{ssec:InitDataSBH}.
The specific type and set of parameters of the initial scalar, the mass $M_{0}=M_{\rm{BH}}(0)$ 
of the initial BH as well as the grid setup of our numerical domain are summarized in Table~\ref{tab:SetupSchwarzschildmassless}. 

%-------------- Results -----------------------------------------------
%
\begin{figure*}[htpb!]
\begin{center}
\subfloat[BH properties]{\label{fig:SchMasslessBHprop}
\includegraphics[width=0.45\textwidth]{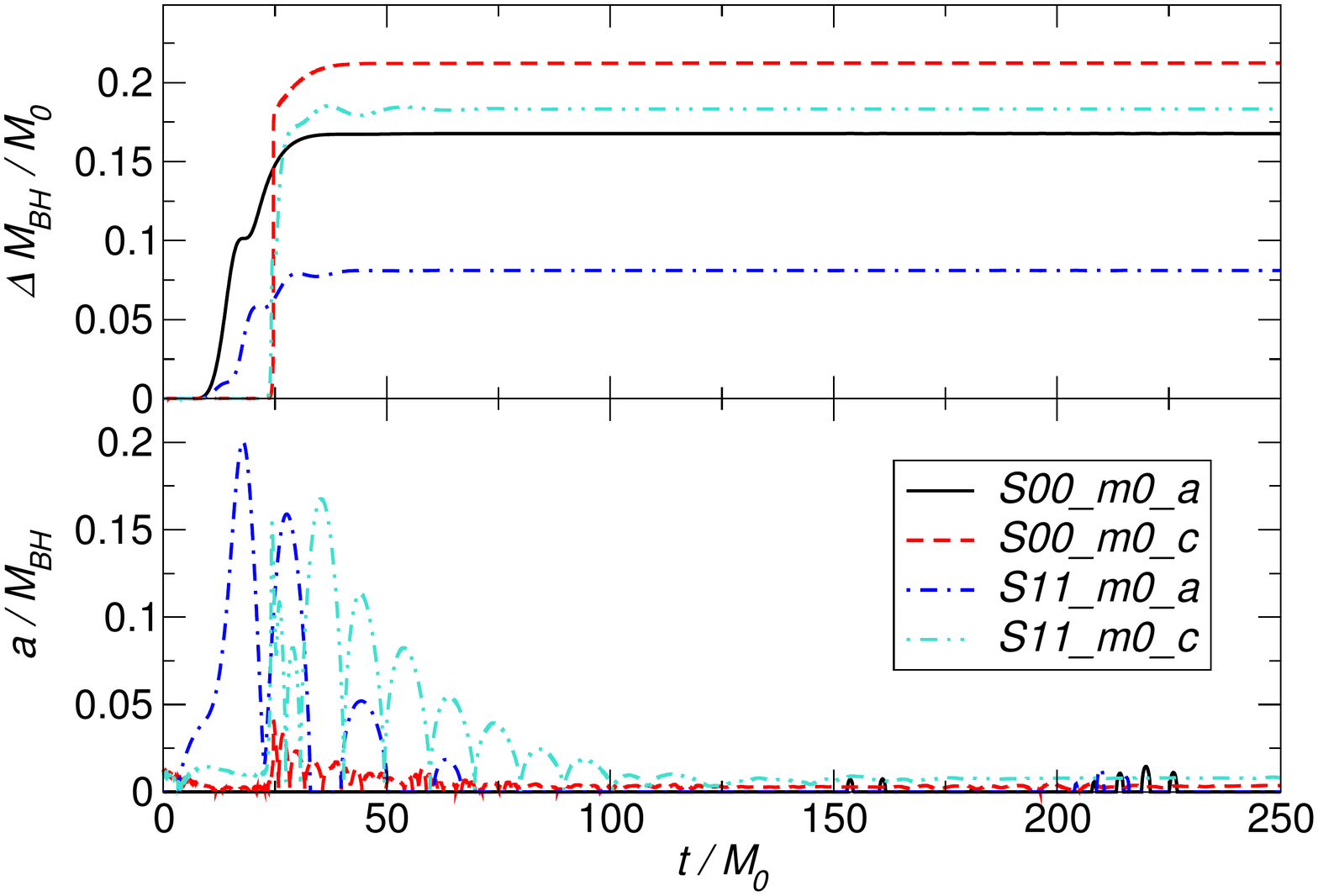}} 
\subfloat[Waveforms]{\label{fig:SchMasslessWaveforms}
\includegraphics[width=0.45\textwidth]{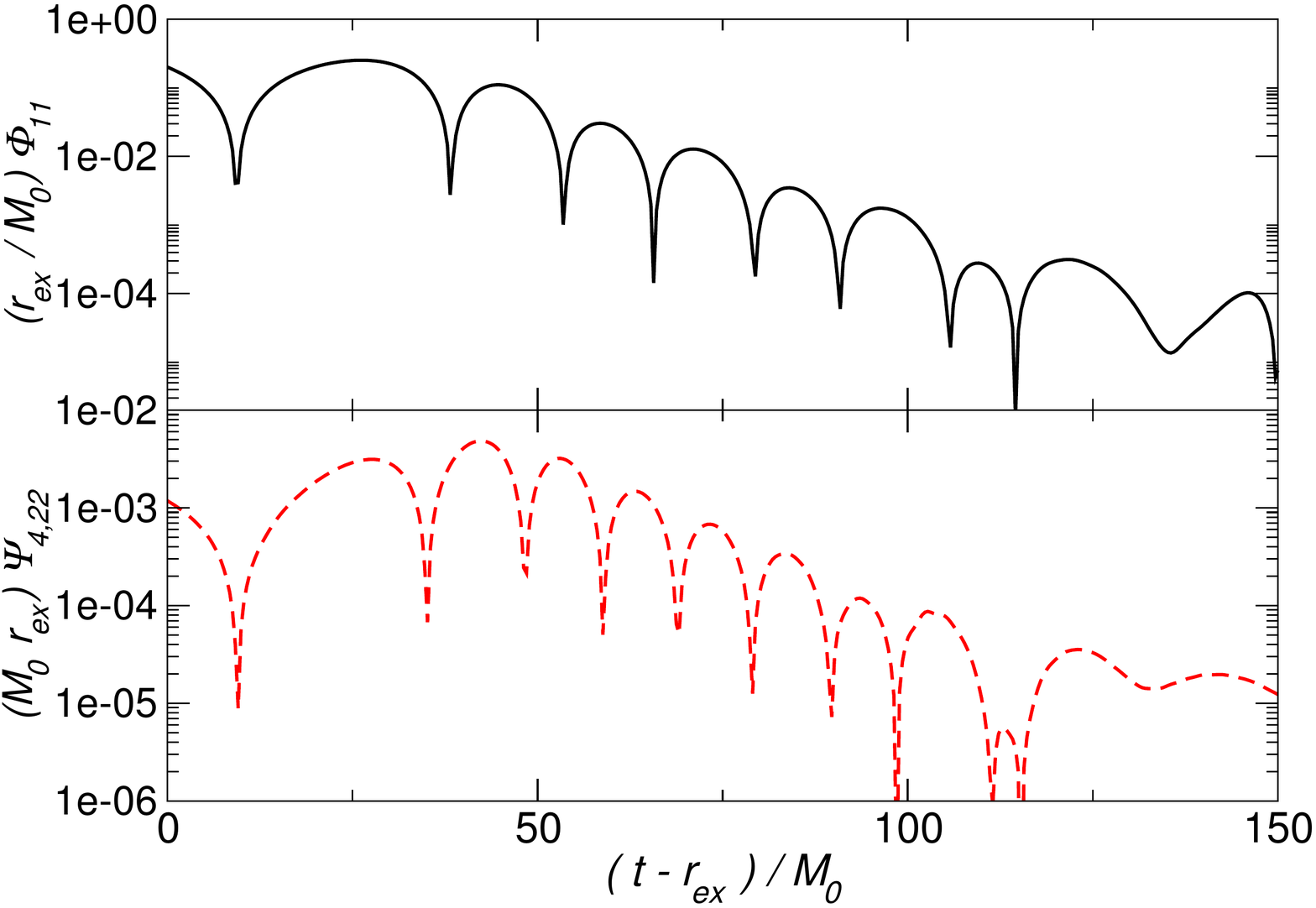}}
 \caption{ 
 Results for a massless scalar field around a non-rotating BH. 
 Fig.~\protect\subref{fig:SchMasslessBHprop} depicts the change in the BH mass (top) 
 and the (dimensionless) angular momentum $a/M_{\rm{BH}}$ of the BH (bottom) as a function of time. 
 The field has been set up as a shell with spherically symmetric or dipole angular configuration.
 Fig.~\protect\subref{fig:SchMasslessWaveforms} illustrates the $l=m=1$ waveform (top) of the massless scalar field with 
 an initial dipole configuration and the dominant $l=m=2$ gravitational waveform (bottom) emitted after the
 accretion. Exemplarily, we present results produced by model S11\_m0\_c in 
 Table~\ref{tab:SetupSchwarzschildmassless}.
 Both waveforms, shifted in time by the extraction radius $r_{\rm{ex}}=40M$, 
 show a clear quasi-normal ringdown signal.
}
\end{center}
\end{figure*}
We have evolved these configurations in time, following 
the infall of the scalar shell into the non-rotating BH and their interaction
and monitoring the BH's response. Our simulations allow us to explore effects of backreaction onto the spacetime,
one effect of which is BH mass increase via scalar field accretion. Depending on the initial energy content of the scalar field 
we find that the BH mass increases by up to $21\%$ (for our setups)
as is illustrated in the top panel of Fig.~\ref{fig:SchMasslessBHprop}.

Accretion is generically accompanied by BH ringdown -- both in the scalar and the GW channel -- which 
we find is in good agreement with linearized BH perturbation
calculations~\cite{Berti:2009kk}. For spherically symmetric profiles there is no gravitational signal due to
the symmetry and the scalar waveform exhibits a short quasi-normal ringdown followed by a power-law fall-off $t^{-p}$ with $p=3.2$ 
which is in good agreement with linearized calculations of the power-law tail \cite{Price:1971fb,Ching:1995tj,Gundlach:1993tn}. 
Instead, the response to the dipole scalar field consists of scalar as well as gravitational quasi-normal 
ringdown  as is shown in Fig.~\ref{fig:SchMasslessWaveforms} where we present the respective
dominant multipoles $\Phi_{11}$ and $\Psi_{4,22}$. 
As illustrated in Fig.~\ref{fig:SchMasslessBHprop} even this apparently simple configuration 
exposes new features of the BH response:
the BH is distorted by the infalling scalar wave and experiences a small spin-up especially from the dipole scalar
as can been seen in the bottom panel of Fig.~\ref{fig:SchMasslessBHprop}.
The excited BH rings down, shedding off most of the  angular momentum and settles down to a Kerr BH 
with a small spin and larger final mass.

%%%%%%%%%%%%%%%%%%%%%%%%%%%%%%%%%%%%%%%%%%%%%%%%%%%%%%%%%%%%%%%%%%%%%%%%%%%%%%
\subsection{Massive scalars around non-rotating black holes}
\label{ssec:SFresultsIImassive}
%%%%%%%%%%%%%%%%%%%%%%%%%%%%%%%%%%%%%%%%%%%%%%%%%%%%%%%%%%%%%%%%%%%%%%%%%%%%%%
%
\begin{table*}[htpb!]
\begin{center}
\caption{\label{tab:SetupSchwarzschildmassive}
Setup and initial parameters for massive scalar fields 
around a non-rotating BH with initial mass $M_{0}=M_{\rm{BH}}(0)$ and bare mass $M=1$.
We denote the initial mass coupling $M_0\mu_S$, the dimensionless 
location $r_0/M$, width $w/M$ and amplitude $M A$ of the scalar shell
with type I or II initial profile in Sec.~\ref{ssec:InitDataSBH}, Eqs.\eqref{eq:u00} and \eqref{eq:u_l1m1}.
We present the grid setup in the notation given by Eq.~\eqref{eq:gridsetup} with the ``radii'' of the 
refinement levels given in units of the bare mass $M$.
}
\begin{tabular}{l|cccccc|c}
\hline
Run          & type  & $M_0\mu_S$ & $r_0/M$ & $w/M$ & $M\,A$  & $M_0$    & Grid setup 
\\ \hline
S00\_m42\_a  & I     & $0.29$     & $6.0$   & $2.0$ & $0.15$  & $0.6859$ & $\{(1536,768,384,192,96,48,24,12,6,3,1.5),h=M/32\}$
\\
S00\_m42\_b  & I     & $0.36$     & $6.0$   & $2.0$ & $0.1$   & $0.8626$ & $\{(1536,768,384,192,96,48,24,12,6,3,1.5),h=3\,M/80\}$
\\
S00\_m42\_c  & I     & $0.29$     & $12.0$  & $0.5$ & $0.15$  & $0.6858$ & $\{(1024,256,128,64,16,8,4,2,1),h=M/60\}$
\\
S00\_m42\_d  & I     & $0.42$     & $12.0$  & $0.5$ & $0.015$ & $0.9969$ & $\{(1024,256,128,64,16,8,4,2,1),h=M/60\}$
\\
\hline
S11\_m42\_a  & II    & $0.30$     & $6.0$   & $2.0$ & $0.04$  & $0.7178$ & $\{(3072,1536,768,384,192,96,48,24,12,6,3,1.5),h=M/32\}$
\\
S11\_m42\_b  & II    & $0.35$     & $6.0$   & $2.0$ & $0.03$  & $0.8430$ & $\{(1536,768,384,192,96,48,24,12,6,3,1.5),h=M/32\}$
\\
S11\_m42\_c1 & II    & $0.29$     & $12.0$  & $0.5$ & $0.03$  & $0.6983$ & $\{(1536,384,192,32,16,8,4,2,1),h=M/52\}$
\\
S11\_m42\_c2 & II    & $0.29$     & $12.0$  & $0.5$ & $0.03$  & $0.6983$ & $\{(1536,384,192,32,16,8,4,2,1),h=M/56\}$
\\
S11\_m42\_c3 & II    & $0.29$     & $12.0$  & $0.5$ & $0.03$  & $0.6983$ & $\{(1536,384,192,32,16,8,4,2,1),h=M/60\}$
\\
S11\_m42\_d  & II    & $0.41$     & $12.0$  & $0.5$ & $0.01$  & $0.9669$ & $\{(1536,384,192,32,16,8,4,2,1),h=M/60\}$
\\
S11\_m42\_e  & II    & $0.42$     & $12.0$  & $0.5$ & $0.003$ & $0.9970$ & $\{(1024,256,128,64,16,8,4,2,1),h=M/60\}$
\\
S11\_m77     & II    & $0.54$     & $12.0$  & $0.5$ & $0.03$  & $0.6983$ & $\{(1536,384,192,32,16,8,4,2,1),h=M/60\}$
\\
\hline
\end{tabular}
\end{center}
\end{table*}
\begin{figure*}[htpb!]
\begin{center}
\subfloat[BH properties]{\label{fig:SchMassiveBHpropl0}
\includegraphics[width=0.45\textwidth]{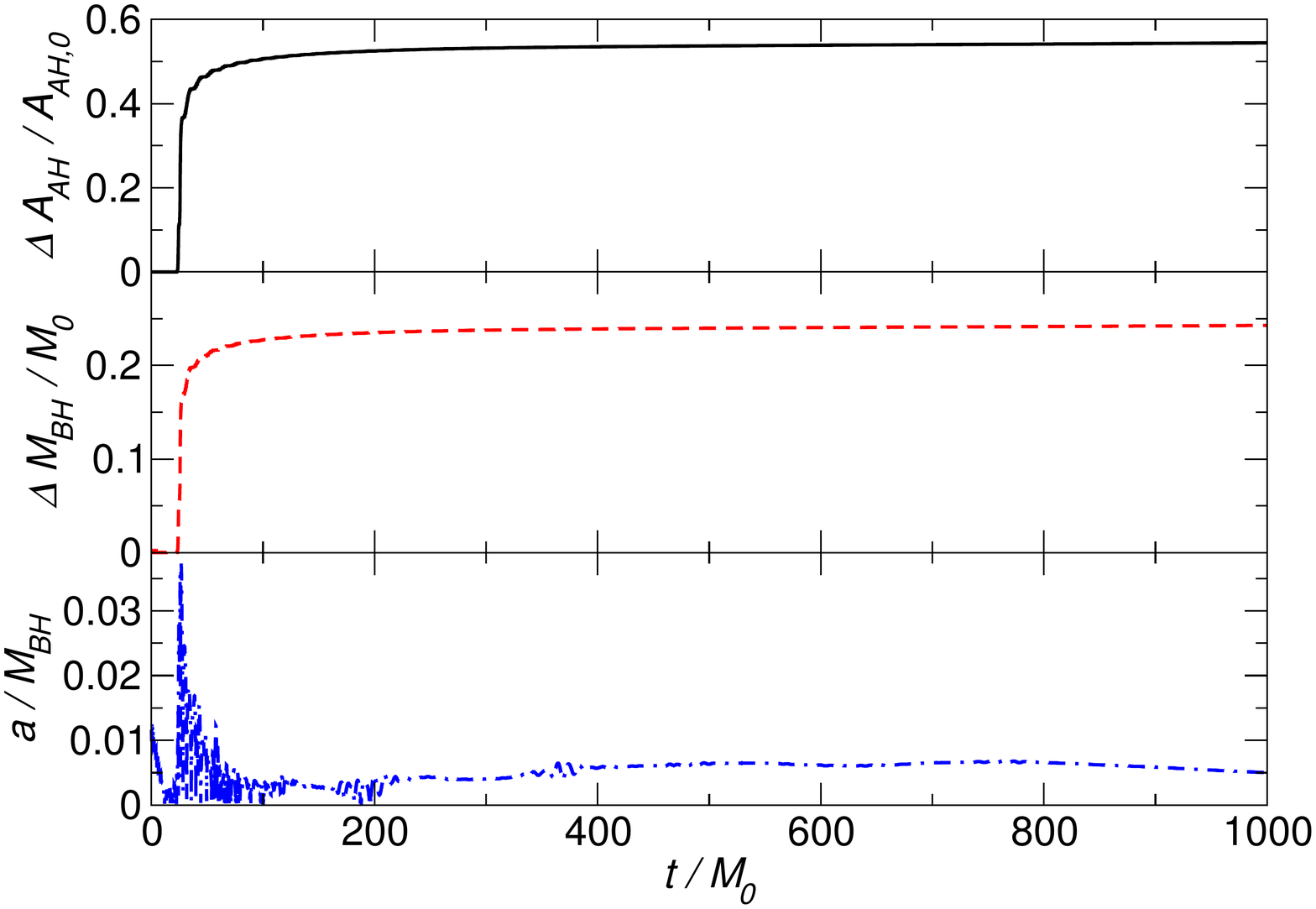}}
\subfloat[waveforms and tail]{\label{fig:SchMassiveWaveformsl0}
\includegraphics[width=0.45\textwidth]{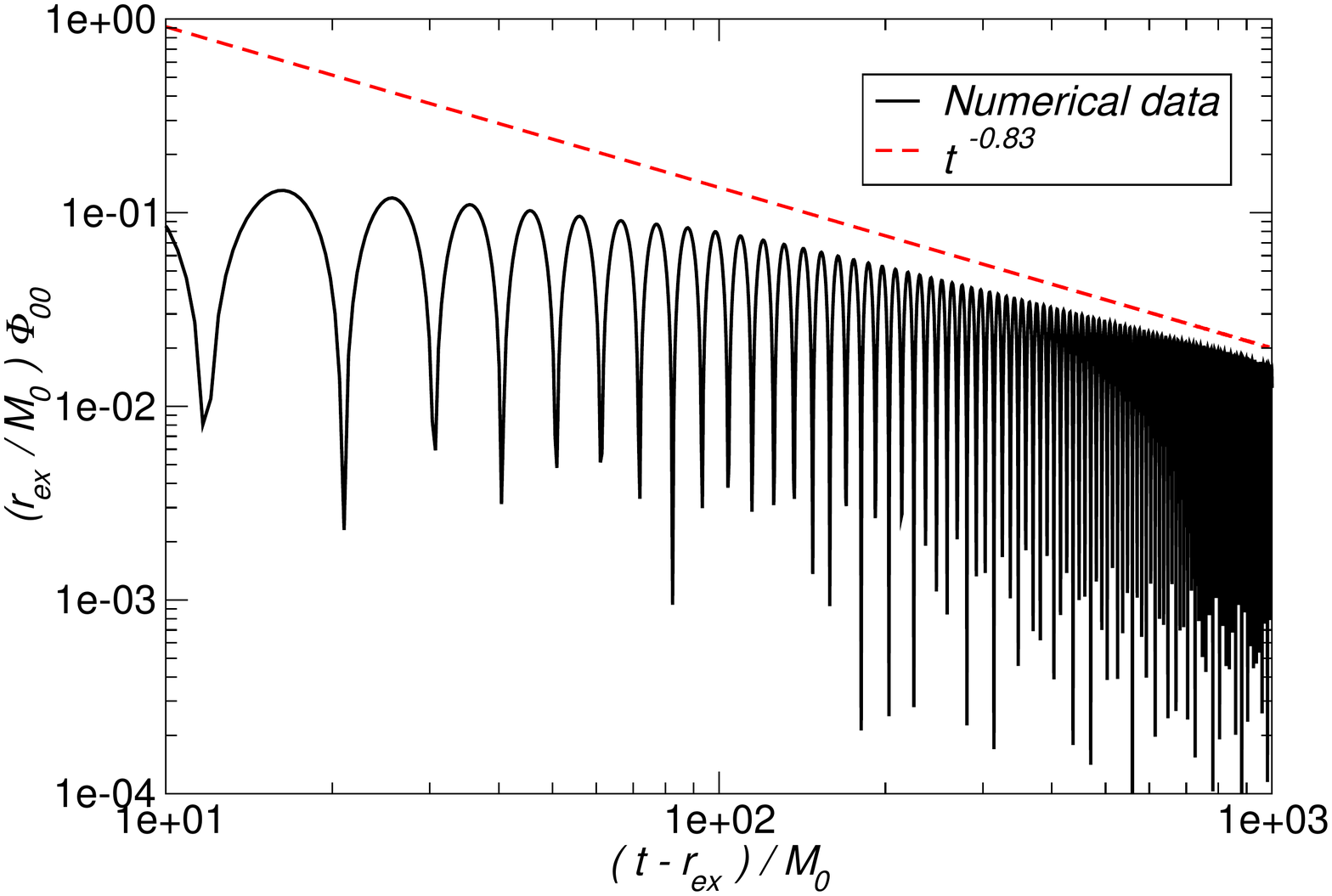}}
 \caption{ 
 Evolution of a spherically symmetric massive scalar field with $M_{0}\mu_S=0.29$ around a non-rotating BH.
 Fig.~\protect\subref{fig:SchMassiveBHpropl0} depicts the relative area of the AH (top),
 the relative BH mass (middle) as compared to its value at $t=0$ and the dimensionless
 spin parameter $a/M_{\rm{BH}}$ (bottom) as functions of time.
 Fig.~\protect\subref{fig:SchMassiveWaveformsl0} presents the $l=m=0$ waveform of the scalar field
 measured at $r_{\rm{ex}}=40M$. In addition to the numerical data (black solid curve) we show a fit to the late-time tail 
 (red dashed curve) with $t^{-0.83}$ in excellent agreement with linearized analysis. 
}
\end{center}
\end{figure*}

We next focus on the time evolution of massive scalar fields around non-rotating BHs.
Although this system is not subject to superradiant amplification, it does give rise to long-lived scalar clouds,
as illustrated in an animation available at~\cite{DyBHo:web} and Fig.~\ref{fig:SnapshotsSchwarzschild} 
in Appendix~\ref{app:snapshots}, 
which continuously feed the BH and trigger long-lived GW emission.
Thus, even this simple configuration allows for interesting new phenomenology concerning smoking gun effects for ``gravitational atoms.''
%
%-------- Setup ---------------------
We take as initial data the construction outlined in Sec.~\ref{ssec:InitDataSBH}, Eqs.\eqref{eq:u00} and \eqref{eq:u_l1m1}.
Table~\ref{tab:SetupSchwarzschildmassive} summarizes the initial parameters.
Throughout this section, we will mainly present results obtained for the
scalar field with the largest energy content resulting in a mass coupling $M_{0}\mu_S=0.29$.

%------- Results l=0 -----------------
The BH response to a spherically symmetric configuration is summarized in Fig.~\ref{fig:SchMassiveBHpropl0}. A fraction of the field
is accreted by the BH, leading to an increase in BH mass of $24\%$. 
\begin{figure}[htpb!]
\begin{center}
\includegraphics[width=0.45\textwidth]{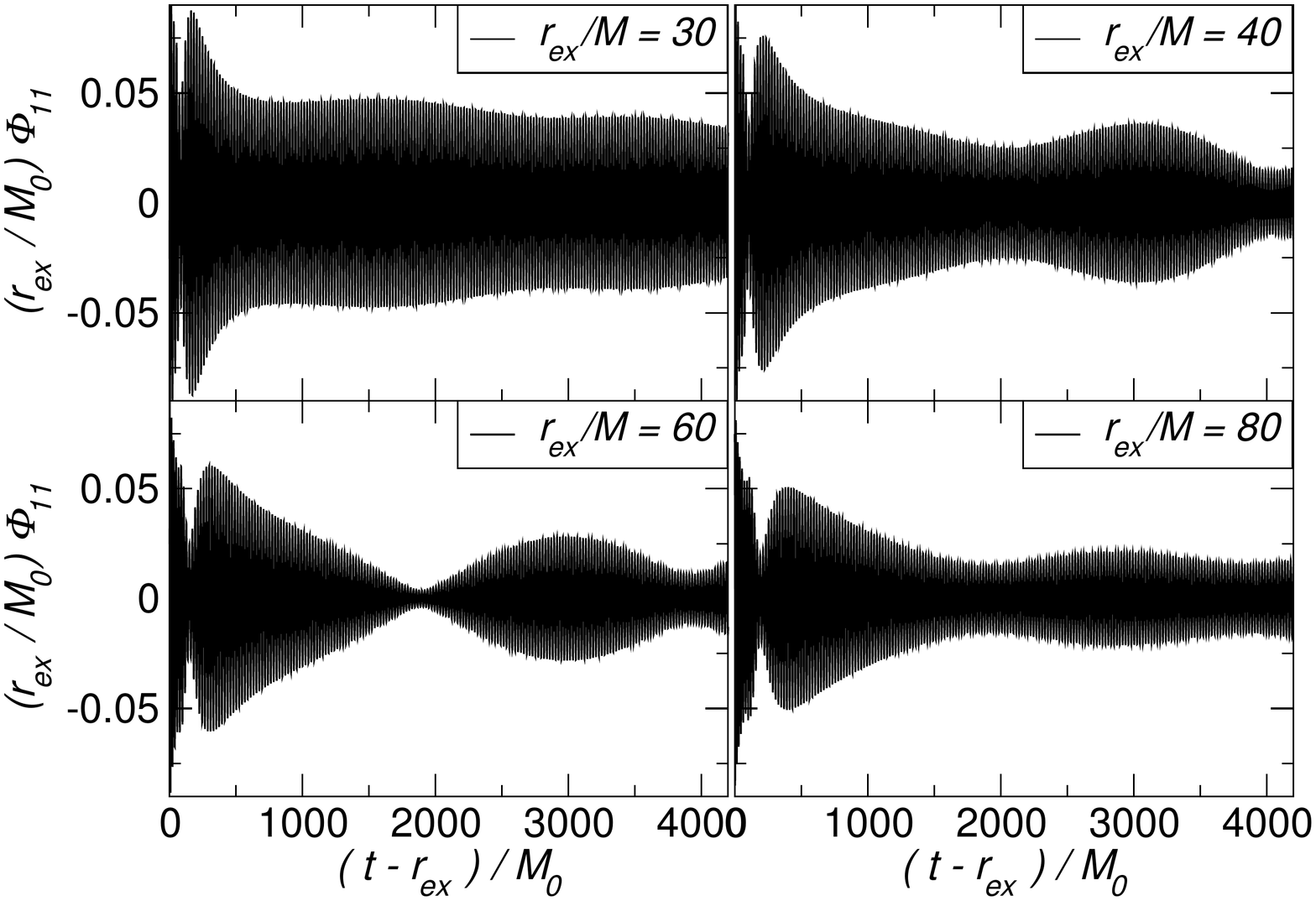}
\caption{\label{fig:SchMassiveSFPsil1}
 $l=m=1$ scalar field multipole for a massive, dipole scalar field with $M_0\mu_S=0.29$ around a non-rotating BH.
 The waveforms, extracted at different radii $r_{\rm{ex}}$ exhibit pronounced beating patterns.
 }
\end{center}
\end{figure}
\begin{figure}[htpb!]
\begin{center}
\includegraphics[width=0.45\textwidth]{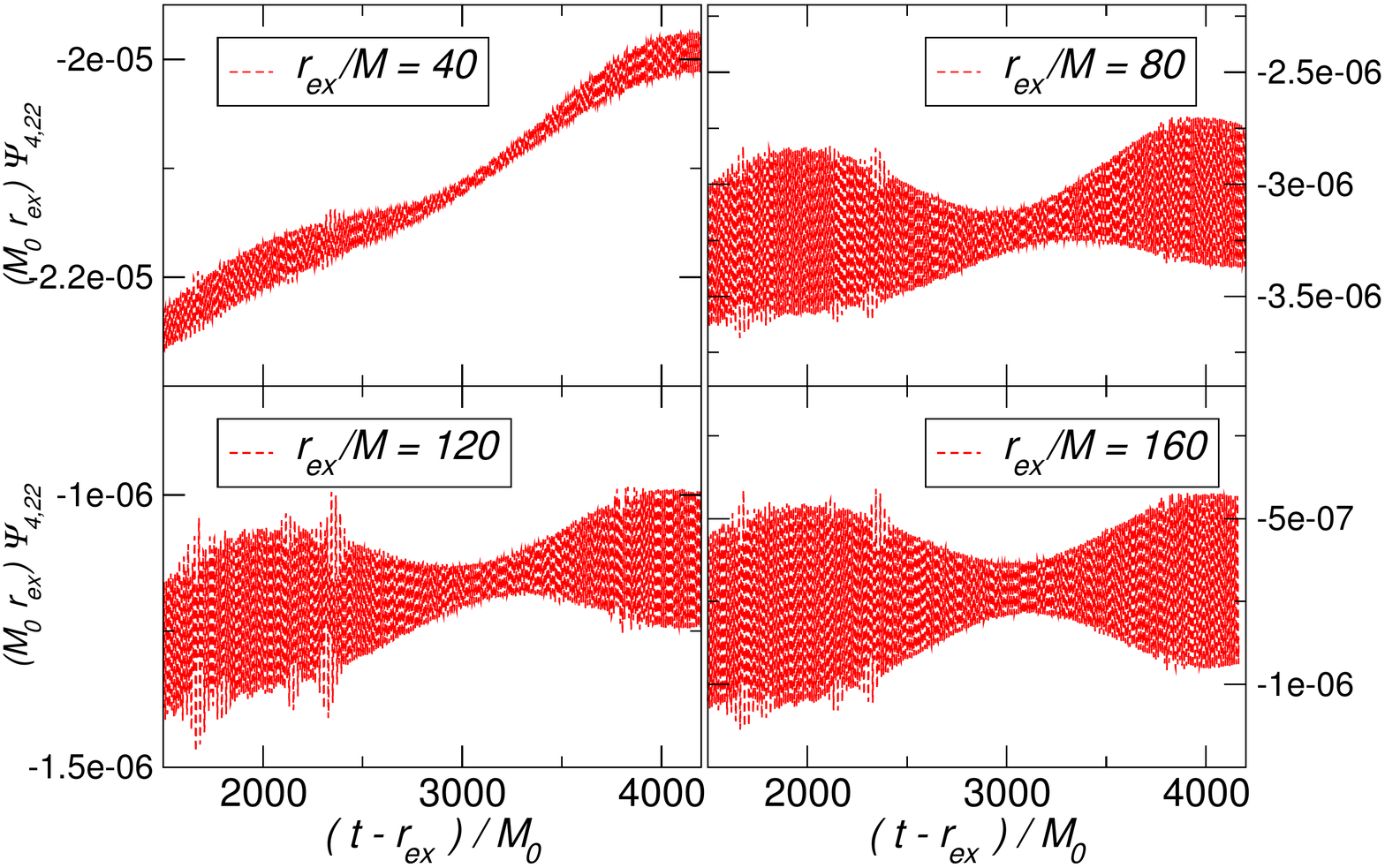}
\caption{\label{fig:SchMassivePsi4l2}
 $l=m=2$ gravitational waveform for the same setup as Fig.~\ref{fig:SchMassiveSFPsil1}.
 }
\end{center}
\end{figure}
The infalling scalar excites monopole BH ringdown, shown in Fig.~\ref{fig:SchMassiveWaveformsl0}.
The short quasinormal ringdown of the $l=m=0$ mode is succeeded by a 
late-time power law tail of the form $t^{-p}\sin(\omega t)$ with $p=0.83$ as illustrated by the red dashed line 
in Fig.~\ref{fig:SchMassiveWaveformsl0}. This numerical value is in excellent 
agreement with predictions of $p=\tfrac{5}{6}$ from the linearized 
analysis of massive scalar fields in BH spacetimes~\cite{Hod:1998ra,Koyama:2001ee,Koyama:2001qw,Burko:2004jn,Witek:2012tr}.
It is remarkable, and to the best of our knowledge the first time, that this late-time massive tail is recovered with such high accuracy in a fully non-linear simulation.
%

\begin{figure}[htpb!]
\begin{center}
\includegraphics[width=0.45\textwidth]{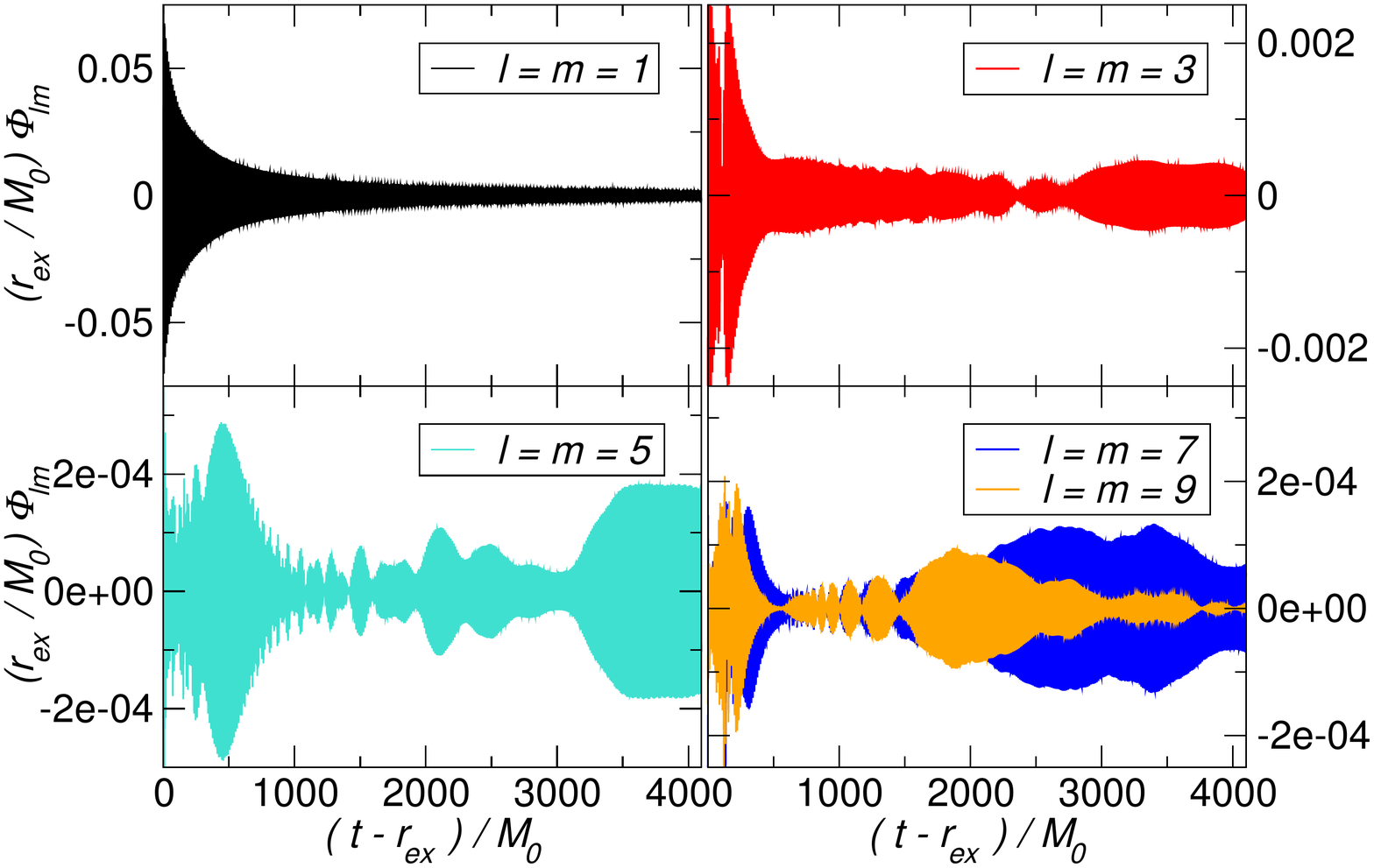}
\caption{\label{fig:multipoles} 
We display different $l=m$ scalar multipoles, measured at $r_{\rm{ex}}=40M$, induced by a massive scalar field
with initial coupling $M_0\mu_S=0.537$.
We observe, that higher multipoles are excited to a significant amount and they exhibit a colorful
beating pattern.
}
\end{center}
\end{figure}

%------- Results l=1 -----------------
Fields with non-trivial angular profile can be confined between the centrifugal barrier induced by the angular-momentum
and the massive barrier at large distances. They can thus exhibit more colorful effects
and we display the most striking one in Figs.~\ref{fig:SchMassiveSFPsil1},~\ref{fig:SchMassivePsi4l2}
and~\ref{fig:multipoles}: the formation of a long-lived scalar cloud, resembling a ``gravitational atom'',
also illustrated in the animation available at~\cite{DyBHo:web} and the snapshots %are displayed 
in Fig.~\ref{fig:SnapshotsSchwarzschild}.

At early times, the infalling scalar field excites the BH causing an outburst of 
gravitational and scalar radiation. The first pulse resembles the quasinormal ringdown 
of a Schwarzschild BH; 
in fact, we find good agreement (to within $3\%$) between the oscillation frequencies of this early time response and those
of Schwarzschild QNMs~\cite{Berti:2009kk}. This early quasinormal ringdown stage gives way
to a long-lived gravitational and scalar wave phase, as is apparent in Figs.~\ref{fig:SchMassiveSFPsil1}
and~\ref{fig:SchMassivePsi4l2}.

As reported recently, this long-lived state is an excitation of several closely-spaced overtones,
and therefore exhibits a modulation and space-dependent excitation compatible with a beating pattern which had so far only been observed in 
time evolutions of scalar fields in {\textit{fixed}} backgrounds~\cite{Witek:2012tr,Dolan:2012yt}.
The long-lived scalar cloud in turn non-linearly triggers the excitation of GWs which slowly leak to infinity, explaining the
long-lived states observed also in the $l=m=2$ gravitational sector
as illustrated in Fig.~\ref{fig:SchMassivePsi4l2}.
Our results indicate that the gravitational signal
is at late-times a long-lived exponentially decaying sinusoid, induced
by the (long-lived) quasi-stationary scalar field state.

So far, we have focused only on the dominant, i.e., dipole mode. Additionally, let us inspect higher multipoles
presented in Fig.~\ref{fig:multipoles}.
The waveforms exhibit beating between different modes, but also a seemingly non-linear excitation of higher multipoles. 
A linearized analysis shows that the beating period increases as $2l+1$ and is of the order of our simulation timescales already for $l=3$.
Thus, it seems that the timescales probed here are not sufficient to discriminate clearly between linear and non-linear effects in the complex pattern
observed in higher multipoles. The possibility that these are truly non-linear effects akin to recently reported turbulence effects which shift the radiation
to shorter scales and that might therefore lead to collapse of the field~\cite{Bizon:2011gg,
Maliborski:2012gx,Buchel:2013uba, Adams:2013vsa}
is certainly worth exploring more in the future. It is plausible that quasi-bound (``confined'') states are also prone to such effects, but now in asymptotically flat spacetimes~\cite{Okawa:2013jba,Yang:2014tla}.

%%%%%%%%%%%%%%%%%%%%%%%%%%%%%%%%%%%%%%%%%%%%%%%%%%%%%%%%%%%%%%%%%%%%%%%%%%%%%%
\section{Results II -- scalar clouds around rotating BHs}
\label{sec:SFresultsIII}
%%%%%%%%%%%%%%%%%%%%%%%%%%%%%%%%%%%%%%%%%%%%%%%%%%%%%%%%%%%%%%%%%%%%%%%%%%%%%%
This section concerns the evolution of scalar fields around spinning BHs.
Because this is mostly uncharted territory, we have compared our results to previous reports in the literature, 
by monitoring the time evolution of an isolated, highly rotating BH set up in quasi-isotropic coordinates.
This analysis is done in Appendix \ref{app:SFresultsIIIpureKerr}.
Instead, the main body of the present work is devoted to the evolution of massive scalars coupled
to a Kerr BH.

%%%%%%%%%%%%%%%%%%%%%%%%%%%%%%%%%%%%%%%%%%%%%%%%%%%%%%%%%%%%%%%%%%%%%%%%%%%%%%
\subsection{Massive scalars around spinning black holes}
\label{ssec:SFresultsIIImassiveLong}
%%%%%%%%%%%%%%%%%%%%%%%%%%%%%%%%%%%%%%%%%%%%%%%%%%%%%%%%%%%%%%%%%%%%%%%%%%%%%%
%
\begin{table*}[htpb!]
\begin{center}
\caption{\label{tab:SetupKerrmassiveLong}
Setup and initial parameters for massive scalar fields 
around a BH with initial spin parameter $a_0/M=0.95$ and initial BH mass $M_{0}=1$.
We set up pseudo-bound state (``BS'') or Gaussian (``Ga'') type initial data
described, respectively, in Sec.~\ref{ssec:InitDataFBS} and Sec.~\ref{ssec:InitDataKBH}.
The scalar shell with width $w=2.0M$ is located around $r_0=12.0M$
and has an amplitude $M A$. 
The (initial) mass coupling is given by $M_0\mu_S$.
In case of pseudo-bound state initial data we employ its characteristic frequency $M \omega_{B}=0.4929$.
We present the grid setup in the notation given by Eq.~\eqref{eq:gridsetup}, where the ``radii'' of the
refinement levels are given in units of the bare mass parameter $M=1$.
}
\begin{tabular}{l|ccc|c}
\hline
Run          & type  & $M_0\mu_S$ & $M\,A$  & Grid setup 
\\ \hline
AnimKerr     & BS    & $0.35$     & $0.075$ & $\{(192,64,32,16,8,4,2),h=M/48\}$
\\
KBl\_m35\_a  & BS    & $0.35$     & $0.05$  & $\{(384,192,64,32,16,8,4,2),h=M/60\}$ 
\\
KBl\_m35\_b  & BS    & $0.35$     & $0.075$ & $\{(1536,384,192,64,32,16,8,4,2),h=M/60\}$
\\
KGl\_m30\_a1 & Ga    & $0.30$     & $0.025$ & $\{(384,192,96,48,24,12,4,2),h=M/52\}$
\\
KGl\_m30\_a2 & Ga    & $0.30$     & $0.025$ & $\{(384,192,96,48,24,12,4,2),h=M/56\}$
\\
KGl\_m30\_a3 & Ga    & $0.30$     & $0.025$ & $\{(384,192,96,48,24,12,4,2),h=M/60\}$
\\
KGl\_m30\_b  & Ga    & $0.30$     & $0.075$ & $\{(384,192,96,48,24,12,4,2),h=M/60\}$
\\
\hline
\end{tabular}
\end{center}
\end{table*}
\begin{figure}[htpb!]
\begin{center}
\subfloat[Scalar waveforms]{\label{fig:KerrMassivePhi11Long}
\includegraphics[width=0.45\textwidth]{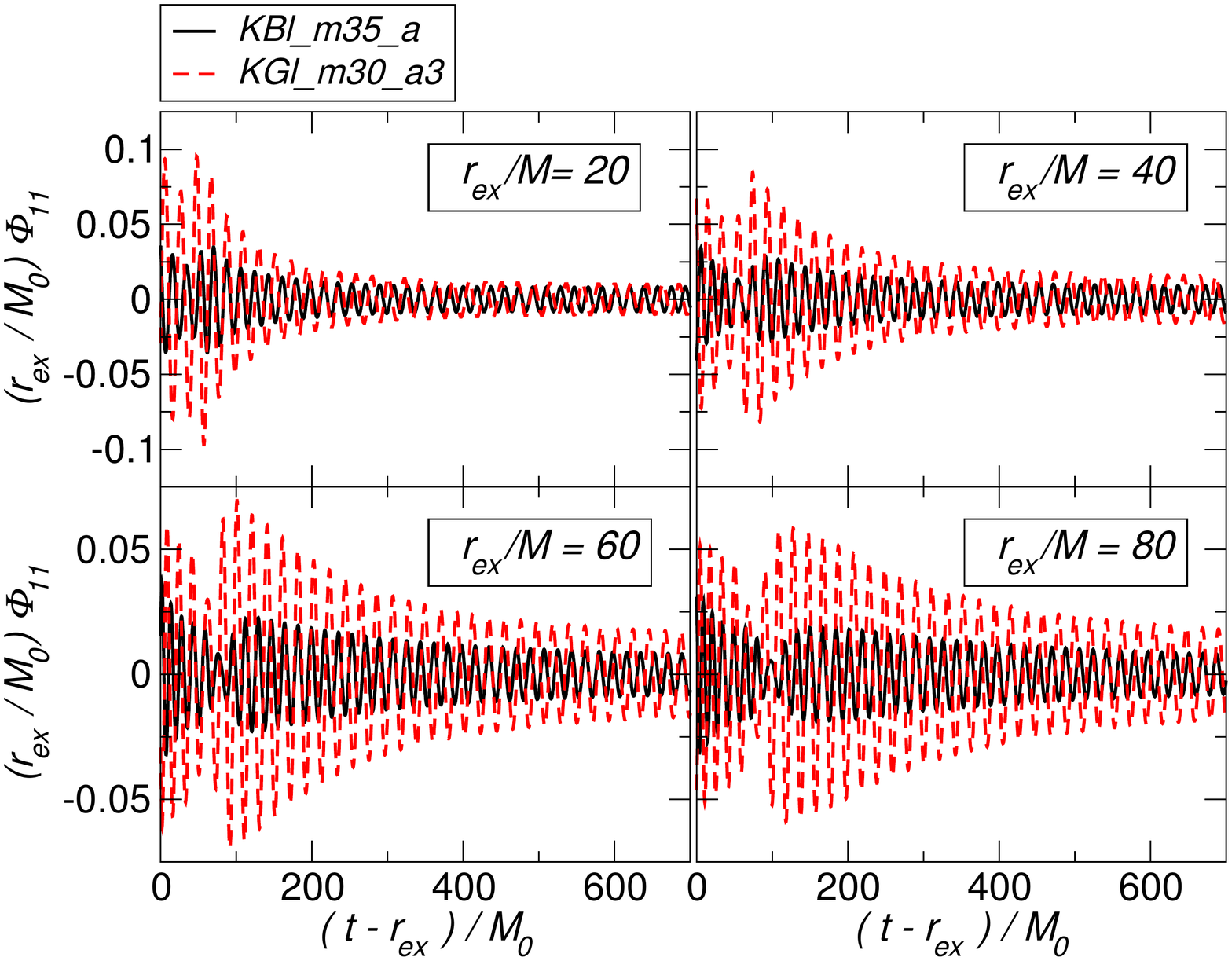}} \\
\subfloat[Gravitational waveforms]{\label{fig:KerrMassivePsi22LongLog}
\includegraphics[width=0.45\textwidth]{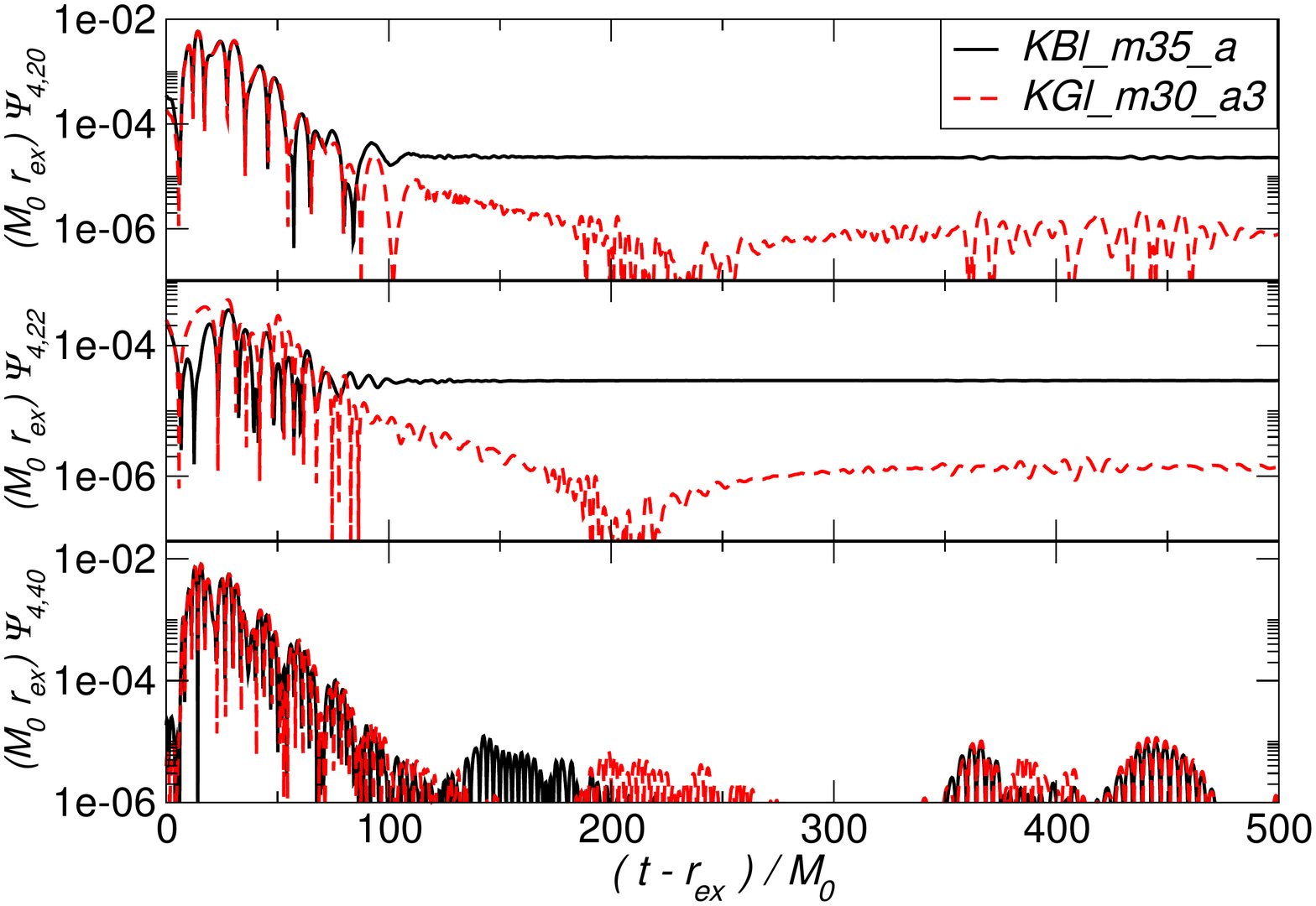}}
 \caption{\label{fig:KerrMassiveWaveformsLong}
 Waveforms for the massive scalar-cloud around a Kerr BH with $a_0/M=0.95$. The 
 scalar shell has been set up as pseudo-bound state (black solid curves) or Gaussian wavepacket (red dashed curves).
 Fig.~\protect\subref{fig:KerrMassivePhi11Long} presents the $l=m=1$ mode of the scalar field $\Phi_{lm}$, 
 rescaled by and presented at various extraction radii $r_{\rm{ex}}$.
 Fig.~\protect\subref{fig:KerrMassivePsi22LongLog} 
 shows the $l=2,m=0$ (top), $l=m=2$ (middle) and $l=4,m=0$ (bottom)
 modes of the gravitational waveform $\Psi_{4,lm}$ extracted at $r_{\rm{ex}} = 40 M$.
 }
\end{center}
\end{figure}
The evolution of massive scalar fields in rotating BH spacetimes is summarized in Fig.~\ref{fig:KerrMassiveWaveformsLong},
where we focus on a highly spinning BH with (initial) spin $a_0/M = 0.95$ and initial BH mass $M_0=1.0$.
We set up the scalar shell with a dipole angular configuration 
either as generic Gaussian wavepacket or pseudo-bound state 
according to Sec.~\ref{ssec:InitDataKBH} and Sec.~\ref{ssec:InitDataFBS}, respectively.
Unless denoted otherwise the shell has a width of $w=2.0M$ and its maximum
is centered around $r_0 = 12.0M$. 
In case of pseudo-bound state initial data we furthermore specify the eigenfrequency of the system
which is $M\omega_{B}=0.4929$ for our choice of parameters~\cite{Cardoso:2005vk,Dolan:2007mj}.
Further specific parameters, such as the (initial) mass coupling $M_0\mu_S$, 
the amplitude $A$ refering to $A_{\rm{G}}$ or $A_{\rm{P}}$ 
for Gaussian or pseudo-bound state initial data,
and the setup of the numerical domain are summarized
in Table~\ref{tab:SetupKerrmassiveLong}. An animation illustrating the process is available online~\cite{DyBHo:web}.

The time evolution has many characteristics in common with the non-rotating case discussed previously.
The scalar shell is initially accreted onto the BH, and it excites the system by initiating 
a large burst of scalar and gravitational radiation leading to ringdown.
Following this outburst of radiation a scalar cloud forms around the BH and we observe two interesting effects:

\noindent (i) the scalar cloud is dragged along with the BH, an effect which seems to be clearly related to frame-dragging;

\noindent (ii) the scalar cloud lightens up and dims periodically over time, like a light-house on the seashore.
This is due both to the azimuthal dependence of the field as well as to the beating effects already discussed 
(see also Refs.~\cite{Witek:2012tr,Dolan:2012yt} for a thorough linear analysis).

One interesting aspect of these simulations is that they embody the properties of the initial data discussed previously:
pseudo-bound state initial data shows less accretion than Gaussian initial data
for comparable initial energy densities, because it was built to behave as a quasi-bound state. 
This property is clearly seen in Fig.~\ref{fig:KerrMassiveWaveformsLong}: the GW response is similar in magnitude and has similar temporal behavior. However, the scalar field differs substantially, with much larger variations (and accretion) happening for Gaussian initial data.
This property is accordingly imprinted on the gravitational-wave signal: Gaussian-type initial data induces ringdown which is followed by a power-law decay and a transition to a small, almost constant GW emission.
Instead, those simulations starting with a pseudo-bound state configuration show an immediate transition
to an almost constant GW signal in the $l=2$ modes as shown in Fig.~\ref{fig:KerrMassiveWaveformsLong}.
This (almost) constant signal is powered by the continuing influx of the scalar field and thus induced 
stimulation of the BH. The different late-time response accommodates the fact that the generic Gaussian scalar field 
undergoes a transition to a pseudo-bound state, in which the scalar cloud localizes in the vicinity
of the BH.

%%%%%%%%%%%%%%%%%%%%%%%%%%%%%%%%%%%%%%%%%%%%%%%%%%%%%%%%%%%%%%%%%%%%%%%%%%%%%%
\subsection{Hunting for superradiance}
\label{ssec:SFresultsIIImassiveShort}
%%%%%%%%%%%%%%%%%%%%%%%%%%%%%%%%%%%%%%%%%%%%%%%%%%%%%%%%%%%%%%%%%%%%%%%%%%%%%%
%
\begin{figure*}[htpb!]
\begin{center}
\subfloat[Variation of (initial) BH spin parameter]{\label{fig:KerrMassiveBHshortSpin1}
\includegraphics[width=0.45\textwidth,clip]{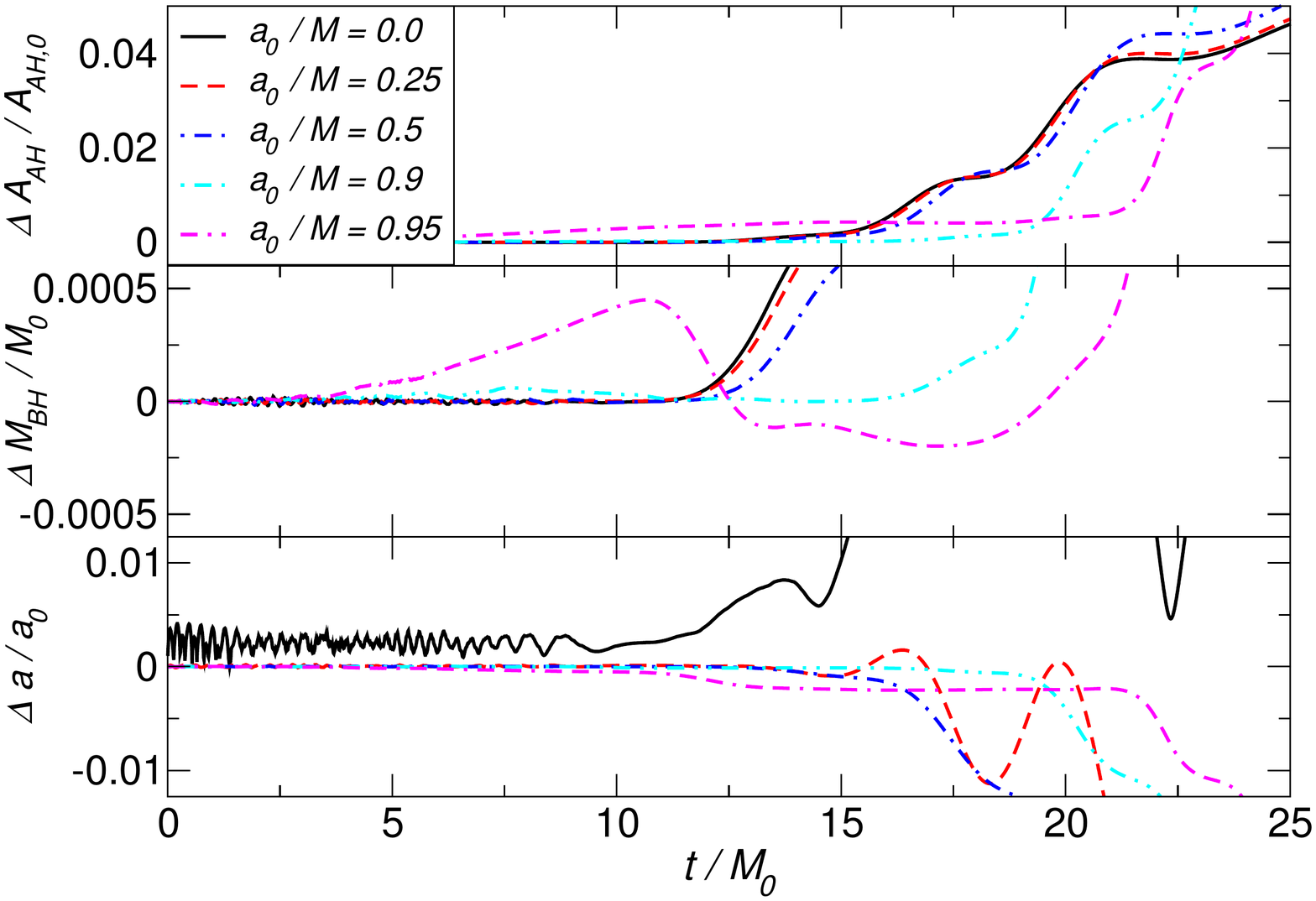}} 
\subfloat[Variation of (initial) scalar field amplitude]{\label{fig:KerrMassiveBHshortAmp1}
\includegraphics[width=0.45\textwidth,clip]{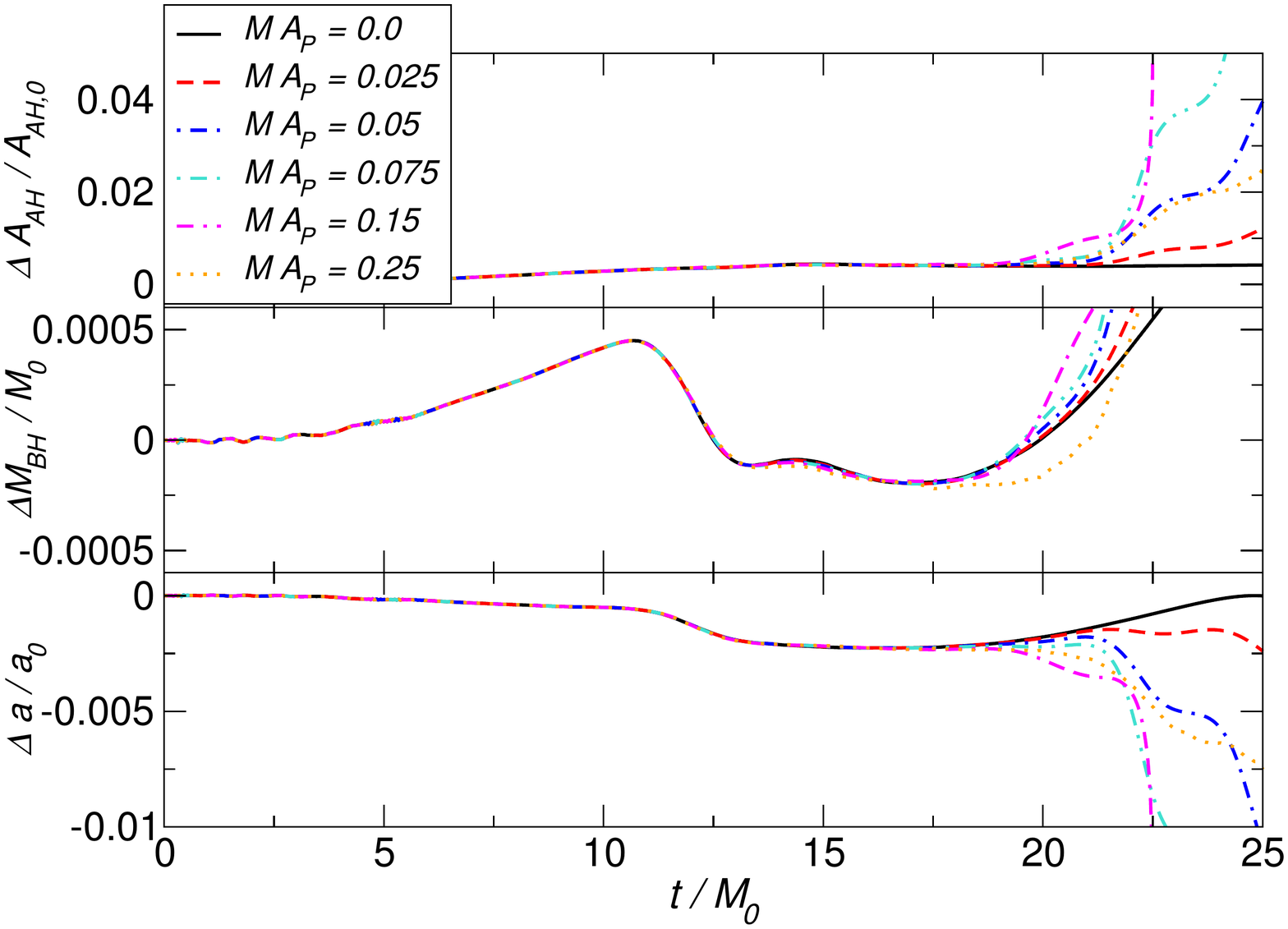}}
 \caption{\label{fig:KerrMassiveBHshortSpin}
We present the change in AH area (top), BH mass (middle) and BH spin (bottom) as compared to
their initial values for various initial spin parameters (left) and scalar field amplitudes (right). 
For $a_0/M=0.95$ we observe a decrease in both the BH mass and spin for any scalar field amplitude.
}
\end{center}
\end{figure*}
\begin{table*}[htpb!]
\begin{center}
\caption{\label{tab:SetupKerrmassiveShort}
Setup and initial parameters for massive scalar fields 
around a BH with initial spin parameter $a_0/M$ and initial BH mass $M_{0}=1$.
We give the corresponding critical frequency $ \omega_C= m\Omega_H$ for superradiance, Eq.~\eqref{eq:SRcond}.
We set up pseudo-bound states described in Sec.~\ref{ssec:InitDataFBS}
and specify its eigenfrequency $M \omega_{B}$ guided by linearized computations. 
The scalar shell with width $w=2.0M$ is located around $r_0=12.0M$
and has an amplitude $M\,A_{P}$. The (initial) mass coupling is $M_0\mu_S=0.35$.
We present the grid setup in the notation given by Eq.~\eqref{eq:gridsetup}, where the refinement box 
``radii'' are given in units of the bare mass $M$.
}
\begin{tabular}{l|cccc|c}
\hline
Run          & $a_0/M$ & $M \omega_C$ & $M\,A_{P}$ & $M \omega_{B}$ & Grid setup 
\\ \hline
KBs\_m35\_a  & $0.00$  & $0.0000$     & $0.075$    & $0.3929$       & $\{(96,48,24,12,4,2),h=M/60\}$
\\
KBs\_m35\_b  & $0.25$  & $0.2540$     & $0.075$    & $0.3929$       & $\{(96,48,24,12,4,2),h=M/60\}$
\\
KBs\_m35\_c  & $0.50$  & $0.5359$     & $0.075$    & $0.3929$       & $\{(96,48,24,12,4,2),h=M/60\}$
\\
KBs\_m35\_d  & $0.90$  & $1.2536$     & $0.075$    & $0.5878$       & $\{(96,48,24,12,4,2),h=M/60\}$
\\
KBs\_m35\_e1 & $0.95$  & $1.4479$     & $0.075$    & $0.4929$       & $\{(96,48,24,12,4,2),h=M/52\}$
\\
KBs\_m35\_e2 & $0.95$  & $1.4479$     & $0.075$    & $0.4929$       & $\{(96,48,24,12,4,2),h=M/56\}$
\\
KBs\_m35\_e3 & $0.95$  & $1.4479$     & $0.075$    & $0.4929$       & $\{(96,48,24,12,4,2),h=M/60\}$
\\
\hline
KBs\_m35\_f  & $0.95$  & $1.4479$     & $0.000$    & $0.4929$       & $\{(96,48,24,12,4,2),h=M/60\}$
\\
KBs\_m35\_g  & $0.95$  & $1.4479$     & $0.025$    & $0.4929$       & $\{(96,48,24,12,4,2),h=M/60\}$
\\
KBs\_m35\_h  & $0.95$  & $1.4479$     & $0.050$    & $0.4929$       & $\{(96,48,24,12,4,2),h=M/60\}$
\\
KBs\_m35\_i  & $0.95$  & $1.4479$     & $0.150$    & $0.4929$       & $\{(96,48,24,12,4,2),h=M/60\}$
\\
KBs\_m35\_j  & $0.95$  & $1.4479$     & $0.250$    & $0.4929$       & $\{(96,48,24,12,4,2),h=M/60\}$
\\
\hline
\end{tabular}
\end{center}
\end{table*}
As we mentioned previously it is extremely challenging to observe scalar-field superradiance
at the full non-linear level. This is due to the (relatively) short timescales over which BH systems can be evolved accurately 
but also due to the fact that scalar fields have a very small superradiant amplification factor 
of $0.04\%$~\cite{Press:1972zz}. 
However, we do find signs of 
%{\it gravitational-induced}
induced {\it{gravitational}} superradiance. 
This claim is supported by Fig.~\ref{fig:KerrMassiveBHshortSpin}, which shows the evolution
of BH mass, area and spin for several initial BH spins and scalar field amplitude.
These results refer to a series of simulations in which we consider a pseudo-bound state
initial scalar shell centered around $r_0=12.0M$ with a dipole angular dependence, a width of $w=2.0M$ 
and mass coupling $M_0\mu_S=0.35$. 
We estimate the numerical error in the AH area, BH mass and spin to be, respectively,
$\Delta A_{\rm{AH}}/A_{\rm{AH}}\leq 0.091\%$, $\Delta M_{\rm{BH}}/M_{\rm{BH}}\leq 0.0076\%$ and $\Delta a/a\leq 0.055\%$.

In the first set of runs we have fixed the scalar field amplitude to $A_{P}=0.075$
and varied the initial spin of the BH in the range $a_0/M=0,\ldots,0.95$.
In the second set we have fixed the BH spin $a_0/M=0.95$ and varied
the scalar field amplitude $A_{P}=0,\ldots,0.25$ and, thus, the energy content in
the initial scalar cloud. The specific setups are summarized in Table~\ref{tab:SetupKerrmassiveShort},
where we also give the critical frequency for superradiance
(for the initial setup)
\begin{align}
\label{eq:CritOmega}
\omega_C = & m \Omega_H = \tfrac{m}{2R_{+}}\left(\tfrac{a}{M}\right)
= 2m\left(\tfrac{a}{M}\right)\tfrac{1}{M+\sqrt{M^2-a^2}}
\,,
\end{align}
in terms of the quasi-isotropic radial coordinate used in our simulations 
(see Eq.~\eqref{eq:KerrRBL_V2}).

A close inspection of the BH parameters reveals 
possibly superradiant behaviour in various time intervals at early stages of the evolution.
Here, we focus on the interval $10\leq t_1/M_0\leq20$.

It is apparent from Fig.~\ref{fig:KerrMassiveBHshortSpin} that for large enough initial spins 
both the BH mass and spin {\it decrease}, while the horizon area keeps increasing during this time interval.
On the other hand, it is also clear that these changes are {\it not} dependent on the scalar field amplitude,
and thus its energy density.
We are thus inclined to interpret this as graviton superradiance of spurious gravitational radiation present in the initial slice. 
This interpretation is consistent with the superradiant amplification factors 
expected for spin-2 particles, whereas scalar fields have too low an amplification to explain the observed behavior~\cite{Bardeen:1972fi,Press:1972zz,Teukolsky:1974yv,Berti:2009kk}.

One of the main obstacles against observing scalar-field superradiance are the extremely
small amplification factors for these fields. Gravitational fields on the other hand, can have
amplification factors orders of magnitude larger. A recent study has shown clearly for the first time {\it gravitational} superradiance
at the nonlinear level, by scattering of gravitational wavepackets off a spinning BH~\cite{East:2013mfa}.

%%%%%%%%%%%%%%%%%%%%%%%%%%%%%%%%%%%%%%%%%%%%%%%%%%%%%%%%%%%%%%%%%%%%%%%%%%%%%%
%\clearpage
%\newpage
\section{Conclusions and Outlook}
\label{sec:conclusions}
%%%%%%%%%%%%%%%%%%%%%%%%%%%%%%%%%%%%%%%%%%%%%%%%%%%%%%%%%%%%%%%%%%%%%%%%%%%%%%
The physics and phenomenology of fundamental fields is extremely rich and fascinating.
Current models for the evolution of the universe, 
dark matter and string theory 
all advocate the existence of light scalar degrees of freedom.
Some of these ultra-light fields might have a dramatic impact on the evolution of BH systems,
thus making BHs perfect laboraties to search for physics beyond the standard model.
Therefore, investigating BH physics in the presence of
such kind of matter is more than an academic exercise.

In the present paper we have started to explore the rich phenomenology of 
BHs encompassed by a scalar field cloud in the fully dynamical, i.e., non-linear regime of gravity.
Technically, this requires to numerically evolve the coupled GR--Klein-Gordon system.
One fundamental ingredient for succesful numerical simulations is the construction of
appropriate, {\textit{constraint-satisfying}} initial configurations.
We have found novel ways to prescribe initial data describing BHs surrounded by scalar clouds, either in analytic
form or semi-analytically. These data provide solutions for both rotating or non-rotating BHs 
and (almost) monochromatic, pseudo-bound state or generic multi-frequency (Gaussian)
field configurations. 

These technical improvements find many interesting applications beyond the case studied in this paper.
In particular, they are of utmost importance for investigations of extensions of GR 
which are typically motivated by string theory compactifications in the low-energy limit and
involve a dilatonic or axion-like coupling. 
Probably the most straight-forward generalization of GR are scalar--tensor theories for which
our methods apply directly.

Our non-linear evolutions confirm the existence of long-lived states around BHs 
which slowly extract rotational energy from the BH. In previous, linearized studies
this phenomenology has been used to impose stringent bounds on dark matter candidates
or on the photon mass if considered as hidden U(1) vector field~\cite{Pani:2012bp}.
The fully dynamical evolutions show that the interaction of the 
scalar field with the central BH results in both scalar and gravitational radiation.
In particular, the accretion of the scalar field triggers quasi-normal ringdown in both 
excitation channels. Following this first burst of radiation
we witness the formation of a long-lived scalar cloud surrounding the BH which is 
illustrated in animations available at~\cite{DyBHo:web}.
These (long-lived) scalar modes induce gravitational radiation with approximately twice the frequency.

We note, that the typical frequencies emitted both in the scalar and gravitational wave channel
are $f\sim\mathcal{O}(10)kHz (M/M_{\odot})^{-1})$
which would potentially be observable with advLIGO or eLISA if the central BH is, respectively a solar-mass or 
intermediate to supermassive BH.

Furthermore, the beating pattern that we find in the scalar waveforms
due to the presence of several overtone modes~\cite{Witek:2012tr,Dolan:2012yt}
stimulates a similar behaviour of mode modulation and space dependend excitation in the 
gravitational channel.
Additionally, in the case of a Kerr BH we directly see frame-dragging effects due to the rotation.
Another exciting observation concerns the shift from an initially almost pure dipole mode 
towards higher multipoles. 
This shift hints at an energy cascade towards smaller scales
which, in turn, leaves room for the exciting possibility of (gravitational) turbulent
effects,
similar to those found recently in asymptotically AdS spacetimes~\cite{Bizon:2011gg,
Maliborski:2012gx,Adams:2013vsa}.

The present study is just a starting point to explore the rich phenomenology of BH -- scalar field configurations
and raise many interesting and important questions:
can BHs with scalar clouds be formed during a collapse of these 
fundamental fields or are there other formation mechanisms at play?
How would the presence of a fundamental (massive) field change the dynamics of BH binary systems?
What will very long-term evolutions of these scalar clouds around BHs yield, or in other words,
can we make predictions about the end-state and non-linear stability of the system?
Do different multipoles interact non-linearly and eventually cascade to smaller scales,
eventually collapsing and producing smaller BHs?

Other especially attractive models for future investigations -- because linearized computations have shown that 
the superradiant instability can be tuned to be orders of magnitude larger --
include mass-varying scalars and vectors, induced by coupling to matter, 
which arise in either scalar-tensor theories or even in the 
standard model~\cite{Witek:2012tr,Cardoso:2013opa,Pani:2013hpa,Cardoso:2013fwa}.

%%%%%%%%%%%%%%%%%%%%%%%%%%%%%%%%%%%%%%%%%%%%%%%%%%%%%%%%%%%%%%%%%%%%
\section{Acknowledgements}
%%%%%%%%%%%%%%%%%%%%%%%%%%%%%%%%%%%%%%%%%%%%%%%%%%%%%%%%%%%%%%%%%%%%
We warmly thank Joan Camps, Sam Dolan, Pau Figueras and Harvey Reall as well as all the participants of 
the ``Gravity - New perspectives from strings and higher dimensions'' Benasque workshop for useful discussions and feedback.
We wish to thank the anonymous referee for useful suggestions to improve the manuscript.
H.~O. and V.~C. acknowledge financial support provided under the European Union's FP7 ERC Starting Grant ``The dynamics of black holes:
testing the limits of Einstein's theory'' grant agreement no. DyBHo--256667.
H.~W. acknowledges financial support provided under
the {\it ERC-2011-StG 279363--HiDGR} ERC Starting Grant and the STFC GR Roller grant ST/I002006/1.
This research was supported in part by Perimeter Institute for Theoretical Physics. 
Research at Perimeter Institute is supported by the Government of Canada through 
Industry Canada and by the Province of Ontario through the Ministry of Economic Development 
$\&$ Innovation.
This work was supported by the NRHEP 295189 FP7-PEOPLE-2011-IRSES Grant, and by FCT-Portugal through projects
PTDC/FIS/116625/2010, CERN/FP/116341/2010 and CERN/FP/123593/2011.
Computations were performed on the ``Baltasar Sete-Sois'' cluster at IST,
on ``venus'' cluster at YITP, on the Altamira supercomputer in Cantabria
through BSC grant AECT-2012-3-0012 
and at the COSMOS supercomputer, part of the DiRAC HPC Facility which is funded by STFC and BIS.

\appendix
%%%%%%%%%%%%%%%%%%%%%%%%%%%%%%%%%%%%%%%%%%%%%%%%%%%%%%%%%%%%%%%%%%%%%%%%%%%
%\section{Convergence analysis and error estimates \label{app:convergence}}
\section{Convergence analysis}\label{app:convergence}
%%%%%%%%%%%%%%%%%%%%%%%%%%%%%%%%%%%%%%%%%%%%%%%%%%%%%%%%%%%%%%%%%%%%%%%%%%%
%
Here we briefly discuss the numerical accuracy of our simulations.
For this purpose we have evolved the dipole, massive scalar field system
denoted as S11\_m42\_c in Table~\ref{tab:SetupSchwarzschildmassive}
and the initial data with angular momentum, model KGl\_m30\_a in Table~\ref{tab:SetupKerrmassiveLong}
at three different resolutions $h_{c}=M/52$, $h_{m}=M/56$ and~$h_{h}=M/60$,
hereafter denoted as coarse, medium and high resolution runs.

\begin{figure}[htpb!]
\begin{center}
\subfloat[Scalar field waveform]{\label{fig:SchMassiveConvSFPsi}
\includegraphics[width=0.45\textwidth,clip]{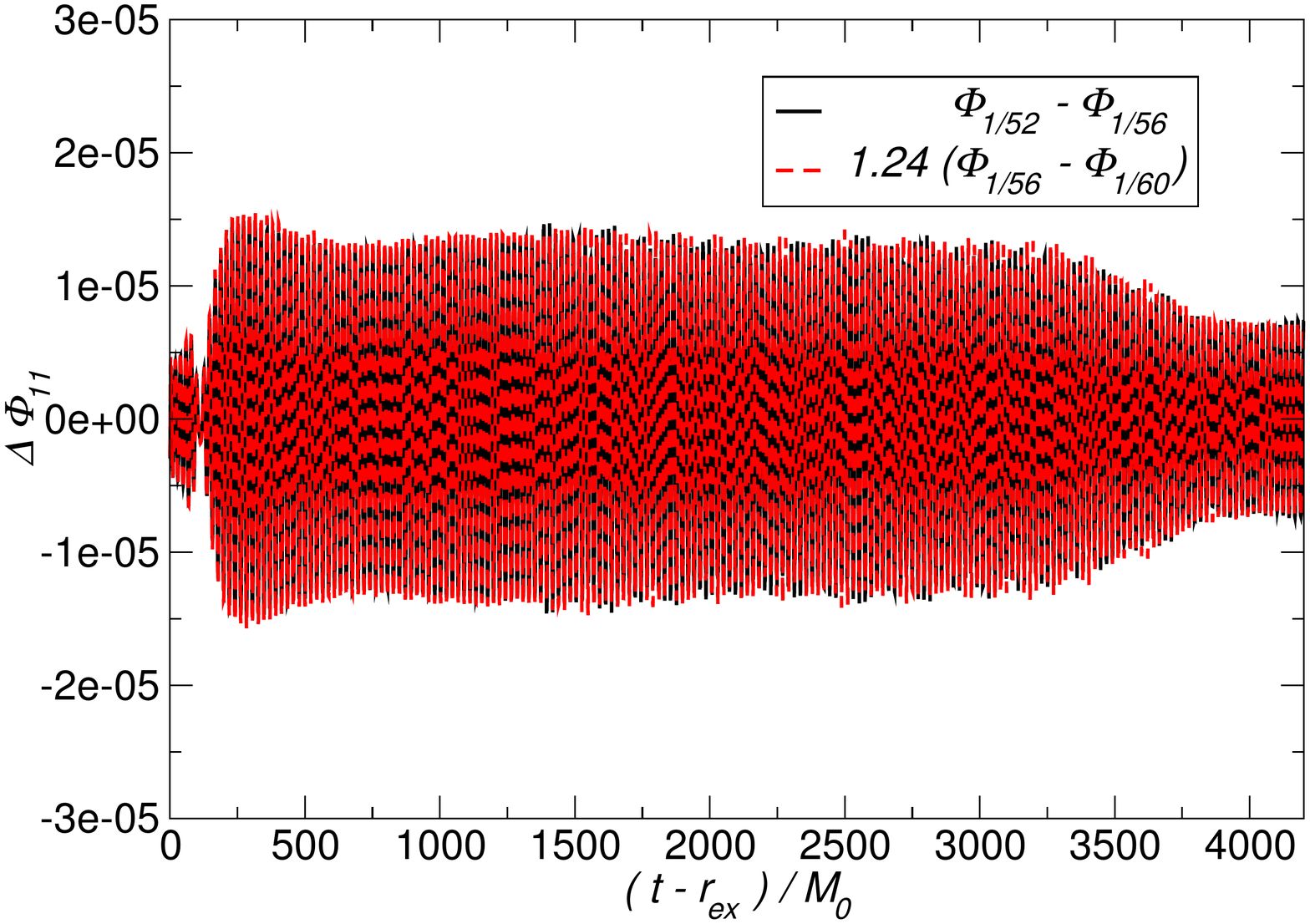}} \\
\subfloat[Gravitational waveform]{\label{fig:SchMassiveConvPsi4}
\includegraphics[width=0.45\textwidth,clip]{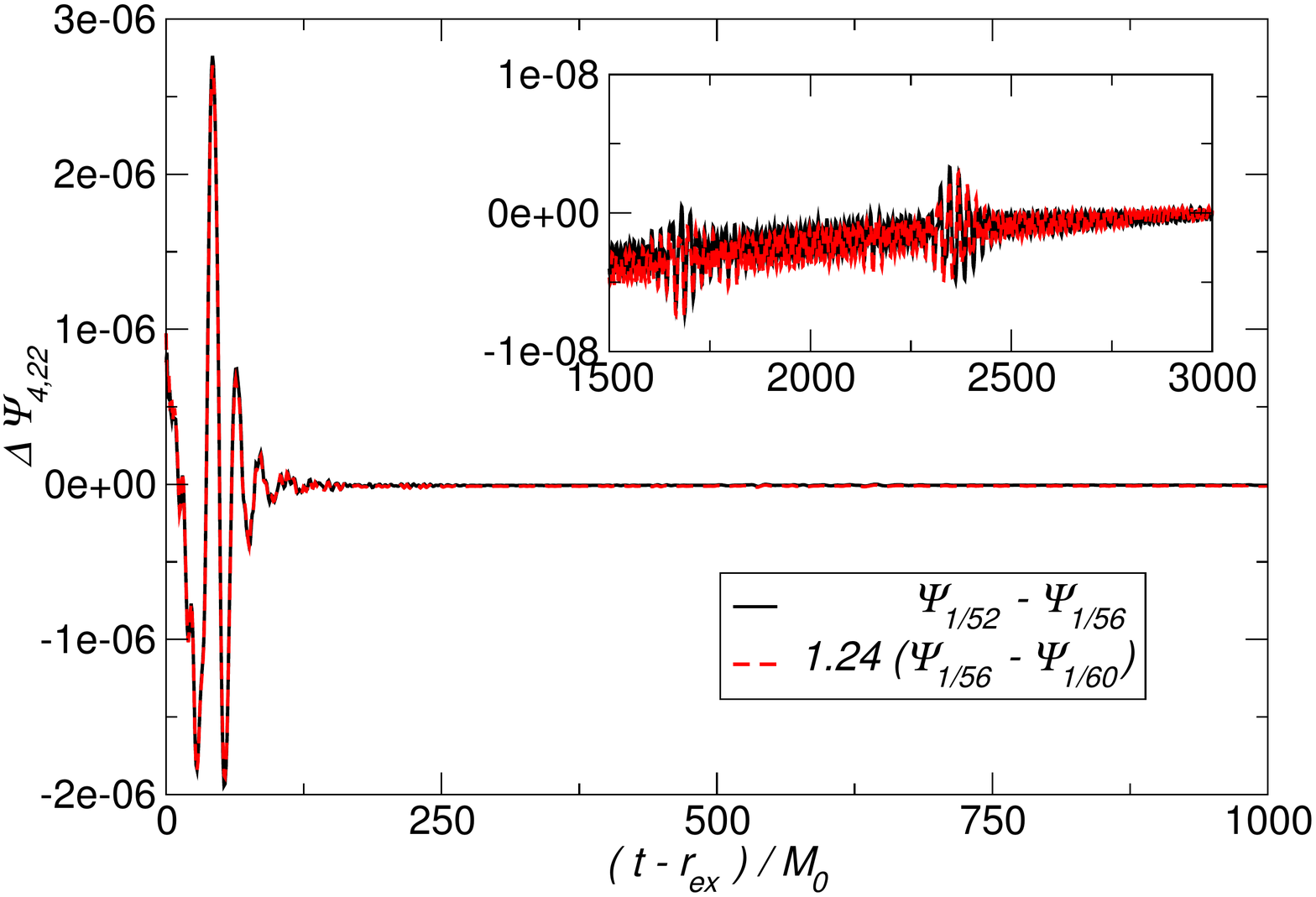}}
 \caption{\label{fig:SchMassiveConv}
 Convergence plots  
 for a massive scalar field with type II initial data and $M_0\mu_S=0.29$ around a non-spinning BH.
 We present the differences in the coarse--medium and medium--high resolutions runs 
 of the $l=m=1$ mode of the scalar field $\Phi$ in Fig.~\protect\subref{fig:SchMassiveConvSFPsi}
 and the $l=m=2$ mode of the gravitational waveform $\Psi_{4}$ in Fig.~\protect\subref{fig:SchMassiveConvPsi4},
 extracted at $r_{\rm{ex}}=40M$. 
 The latter difference has been rescaled by $Q_{2}=1.24$ indicating second order convergence.
 Both waveforms carry a numerical error of $<6\%$.
 }
\end{center}
\end{figure}

\begin{figure}[htpb!]
\begin{center}
\subfloat[Scalar field waveform]{\label{fig:KerrMassiveConvSFPsi}
\includegraphics[width=0.45\textwidth]{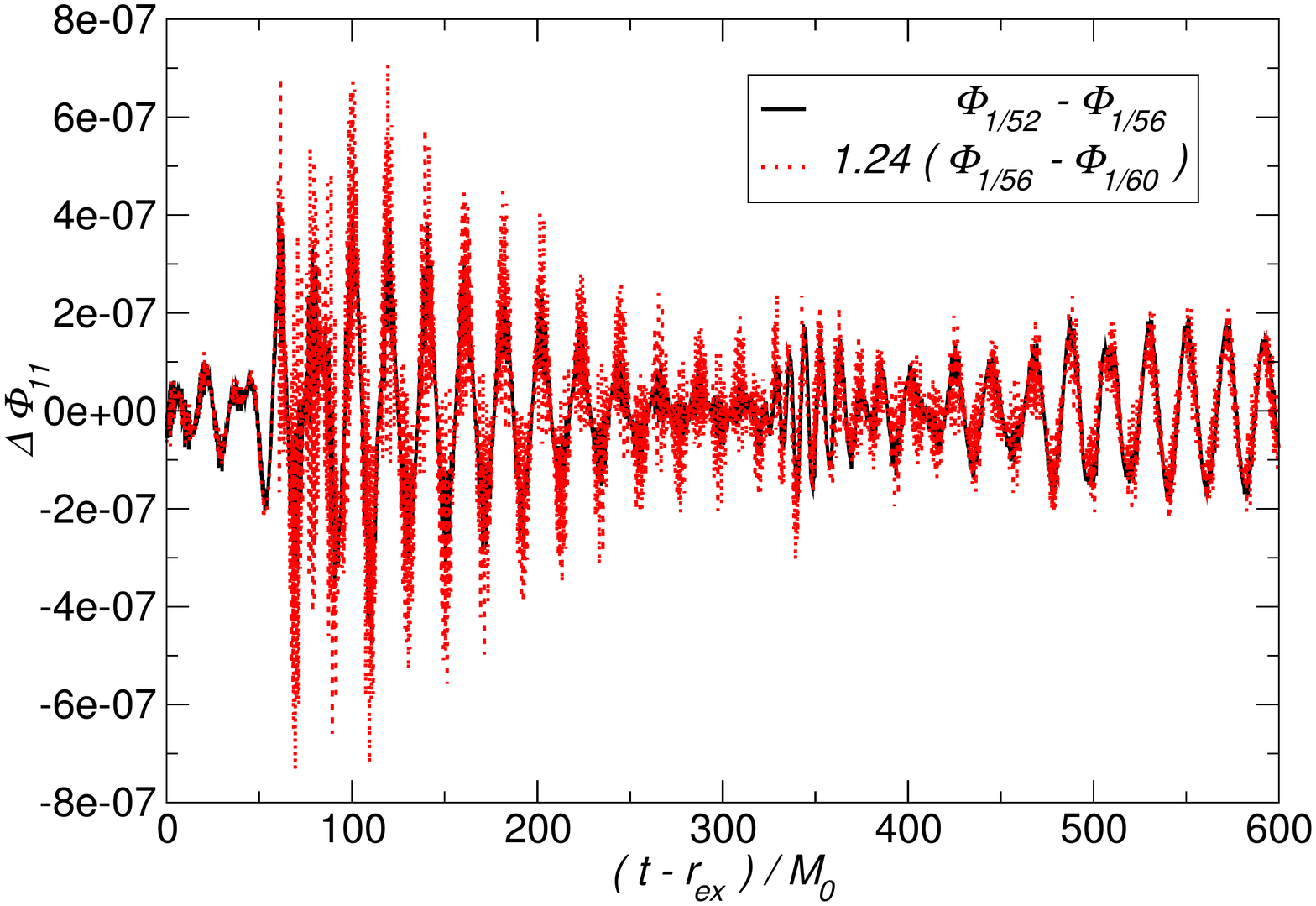}}\\
\subfloat[Gravitational waveform]{\label{fig:KerrMassiveConvPsi4}
\includegraphics[width=0.45\textwidth]{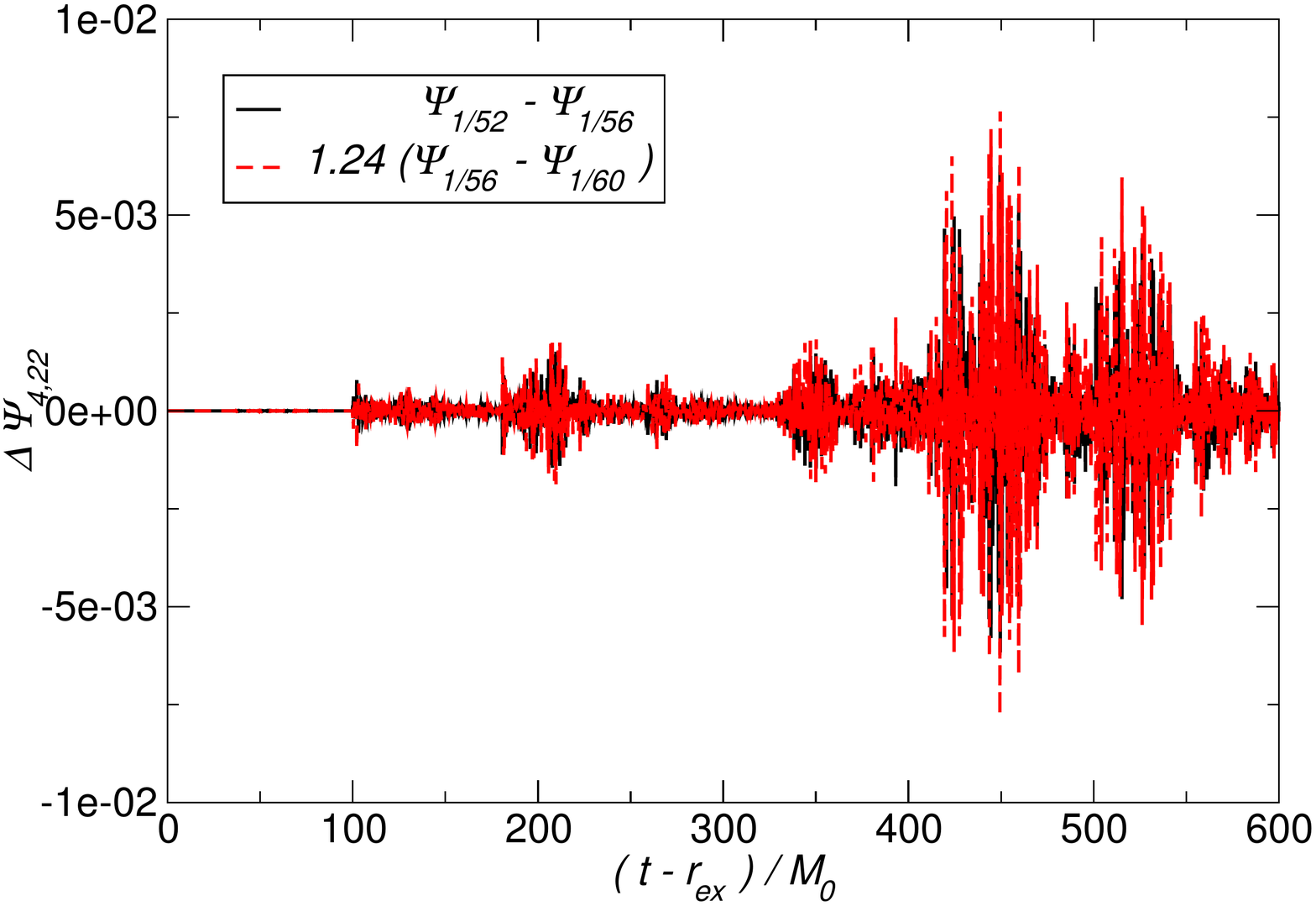}}
 \caption{\label{fig:KerrMassiveConvergence}
 Convergence plots for runs {\textit{KGl\_m30\_a}} in Table~\ref{tab:SetupKerrmassiveLong}, i.e.,
 for a massive scalar field with $M_0\mu_S=0.30$ around a Kerr BH with spin parameter $a_0/M=0.95$.
 We present the differences in the coarse--medium and medium--high resolutions runs 
 of the $l=m=1$ mode of the scalar field $\Phi$ in Fig.~\protect\subref{fig:KerrMassiveConvSFPsi}
 and the $l=m=2$ mode of the gravitational waveform $\Psi_{4}$ in Fig.~\protect\subref{fig:KerrMassiveConvPsi4},
 extracted at $r_{\rm{ex}}=40M$. 
 The latter difference has been rescaled by $Q_{2}=1.24$ indicating second order convergence.
 Both waveforms carry a numerical error of $<2\%$.
 }
\end{center}
\end{figure}

The corresponding convergence plots for the scalar and gravitational radiation 
are presented in Figs.~\ref{fig:SchMassiveConv} and \ref{fig:KerrMassiveConvergence}
refering, respectively, to the Schwarzschild or Kerr case.
Specifically, we consider the differences between the coarse and medium and
medium and high resolutions runs, where the latter difference is rescaled by the 
appropriate convergence factor. We present these tests exemplarily for 
the $l=m=1$ mode of the scalar field $\Phi$ in Figs.~\ref{fig:SchMassiveConvSFPsi} and~\ref{fig:KerrMassiveConvSFPsi}
and the $l=m=2$ mode of the gravitational waveform $\Psi_{4}$, extracted at $r_{\rm{ex}}=40M$, 
in Figs.\ref{fig:SchMassiveConvPsi4} and~\ref{fig:KerrMassiveConvPsi4}.
In both cases we find second order convergence as indicated by the factor $Q_{2}=1.24$.
The convergence order is not only determined by the fourth order FD stencils but also by 
interpolations schemes employed, e.g. at refinement boundaries. In the present simulations
it appears to dominate the numerical accuracy and, because the interpolation in time is only second order,
results in an overall convergence of second order.

In the main body of this paper we have investigated both the long-term behaviour of the scalar-field -- gravity system
as well as the short-term behaviour which gave us insight into the potentially superradiant regime.
Therefore, we estimate the numerical error for both regimes.

Specifically, the error at early times has been measured at $t\sim20M_{0}$.
In this case we have focused solely on the properties of the BH and estimate
the numerical error in the AH area, BH mass and spin to be, respectively,
$\Delta A_{\rm{AH}}/A_{\rm{AH}}\leq 0.091\%$, $\Delta M_{\rm{BH}}/M_{\rm{BH}}\leq 0.0076\%$ and $\Delta a/a\leq 0.055\%$.

Instead, the late time numerical error which is relevant for the long-term simulations has been measured
at $t\sim3000M_{0}$ in the Schwarzschild and $t\sim500M_{0}$ in the Kerr case.
In both cases we find a numerical error of about 
$(8.5\cdot10^{-3},0.19,0.16)\%$ in the AH area, BH mass and spin, respectively.
The dominant scalar and gravitational waveforms exhibit a numerical error of 
about $6\%$ in the Schwarzschild case and of about $2\%$ in the Kerr case.

Additionally, we present the violation of the Hamiltonian constraint along 
the x-axis in Fig.~\ref{fig:SchwarzschildMasslessHamilitonian}
exemplarily for run {\textit{S00\_m0\_e}} in Table~\ref{tab:SetupSchwarzschildmassless}
plotted at different instances during the evolution. 
We observe that initially the constraints are satisfied within less than $0.1\%$ close to the BH horizon
and better than $10^{-7}\%$ towards the outer boundary. 
It is important to note that the constraint violation remains small during the entire evolution;
specifically they are satisfied within less than $0.1\%$
near the BH and within less than $10^{-5}\%$ in the outer regions.
The spikes that can be seen in Fig.~\ref{fig:SchwarzschildMasslessHamilitonian}
correspond to the location of the refinement boundaries.

%

%%%%%%%%%%%%%%%%%%%%%%%%%%%%%%%%%%%%%%%%%%%%%%%%%%%%%%%%%%%%%%%%%%%%%%%%%%%%%%
\section{Benchmarking tests for a spinning BH}
\label{app:SFresultsIIIpureKerr}
%%%%%%%%%%%%%%%%%%%%%%%%%%%%%%%%%%%%%%%%%%%%%%%%%%%%%%%%%%%%%%%%%%%%%%%%%%%%%%
%
\begin{figure*}[htpb!]
\begin{center}
\subfloat[BH spin]{\label{fig:KerrTest1}
\includegraphics[width=0.33\textwidth]{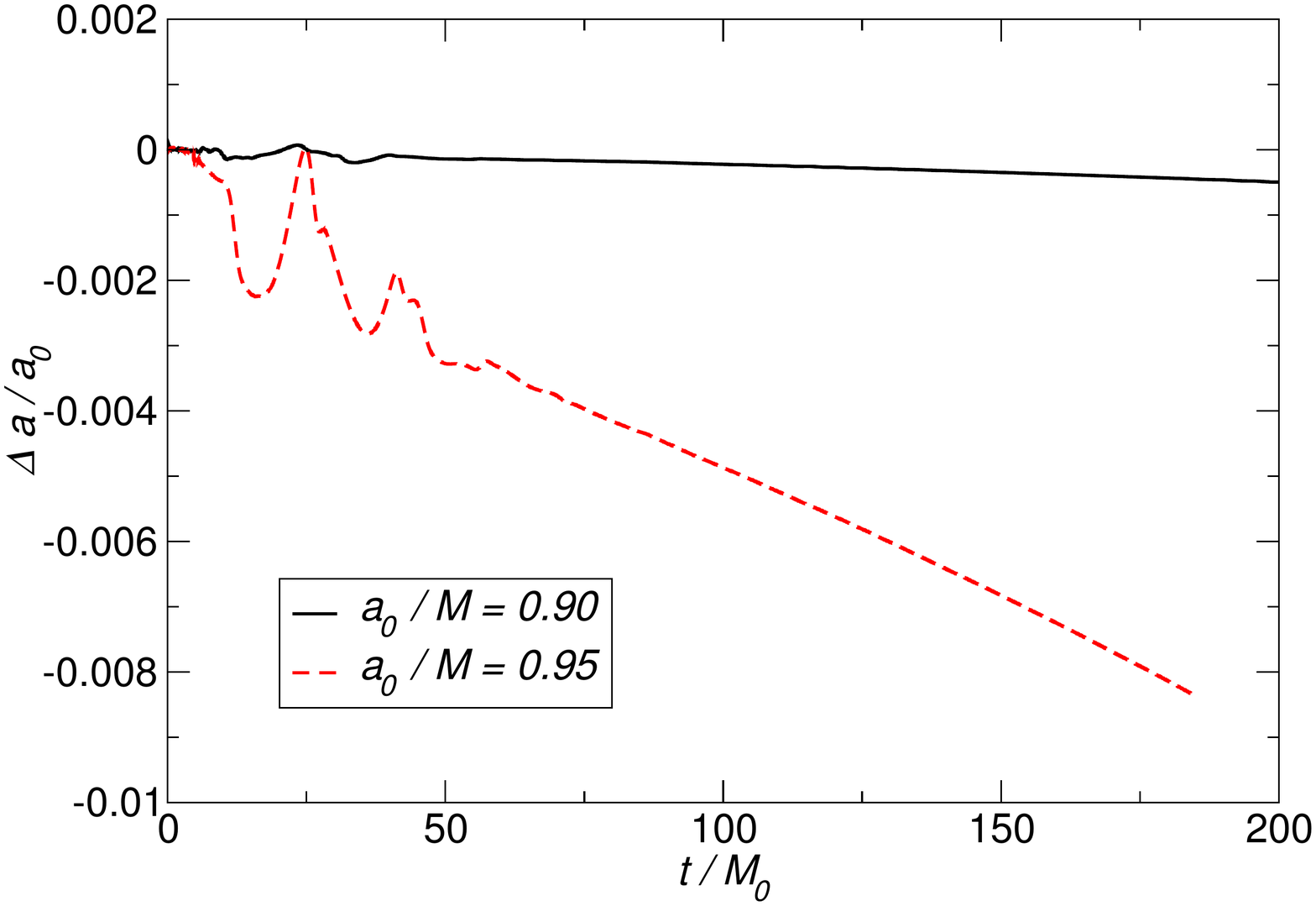}} %\\
\subfloat[AH area and BH mass]{\label{fig:KerrTest2}
\includegraphics[width=0.33\textwidth]{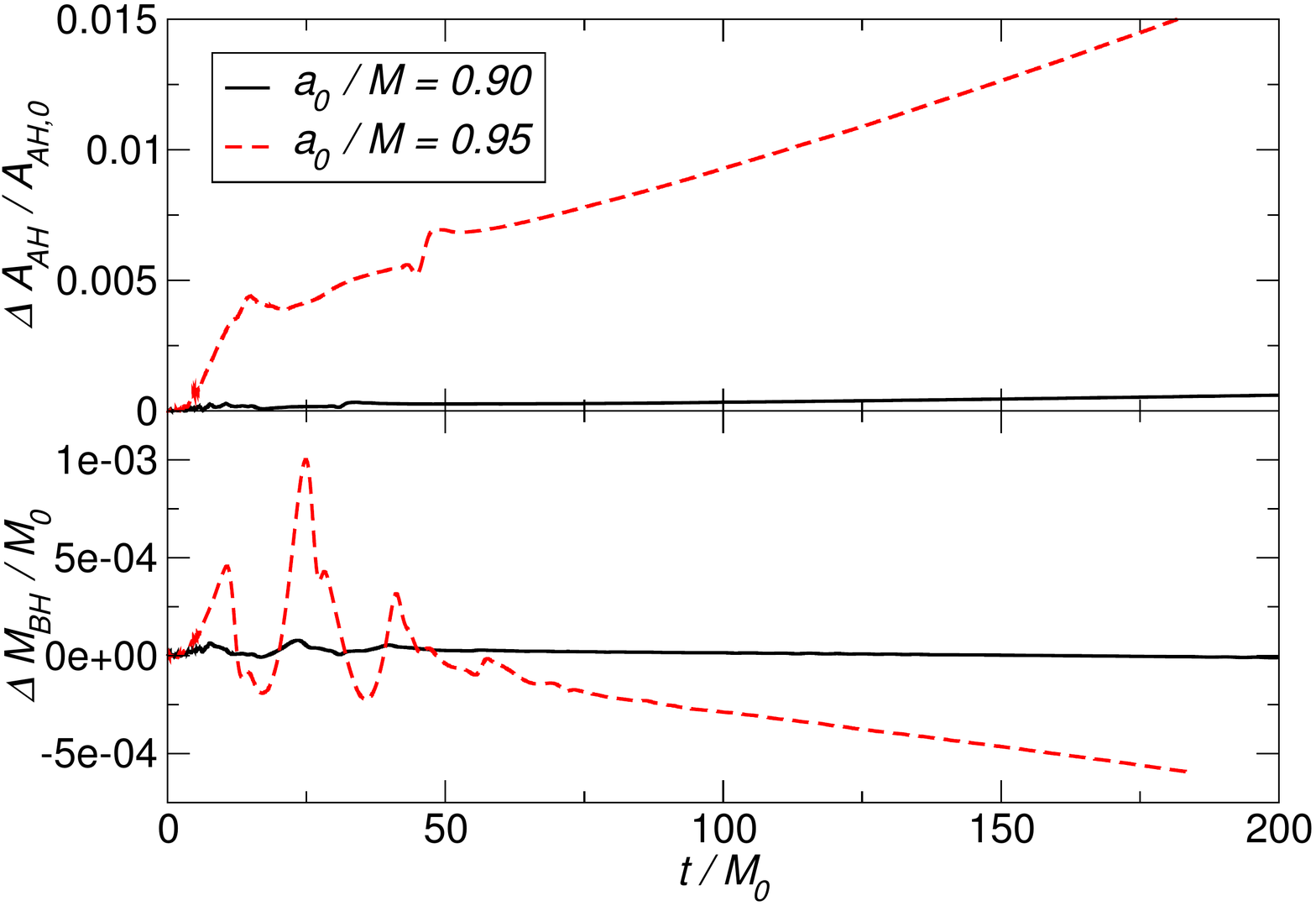}} %\\
\subfloat[Gravitational radiation]{\label{fig:KerrTest3}
\includegraphics[width=0.33\textwidth]{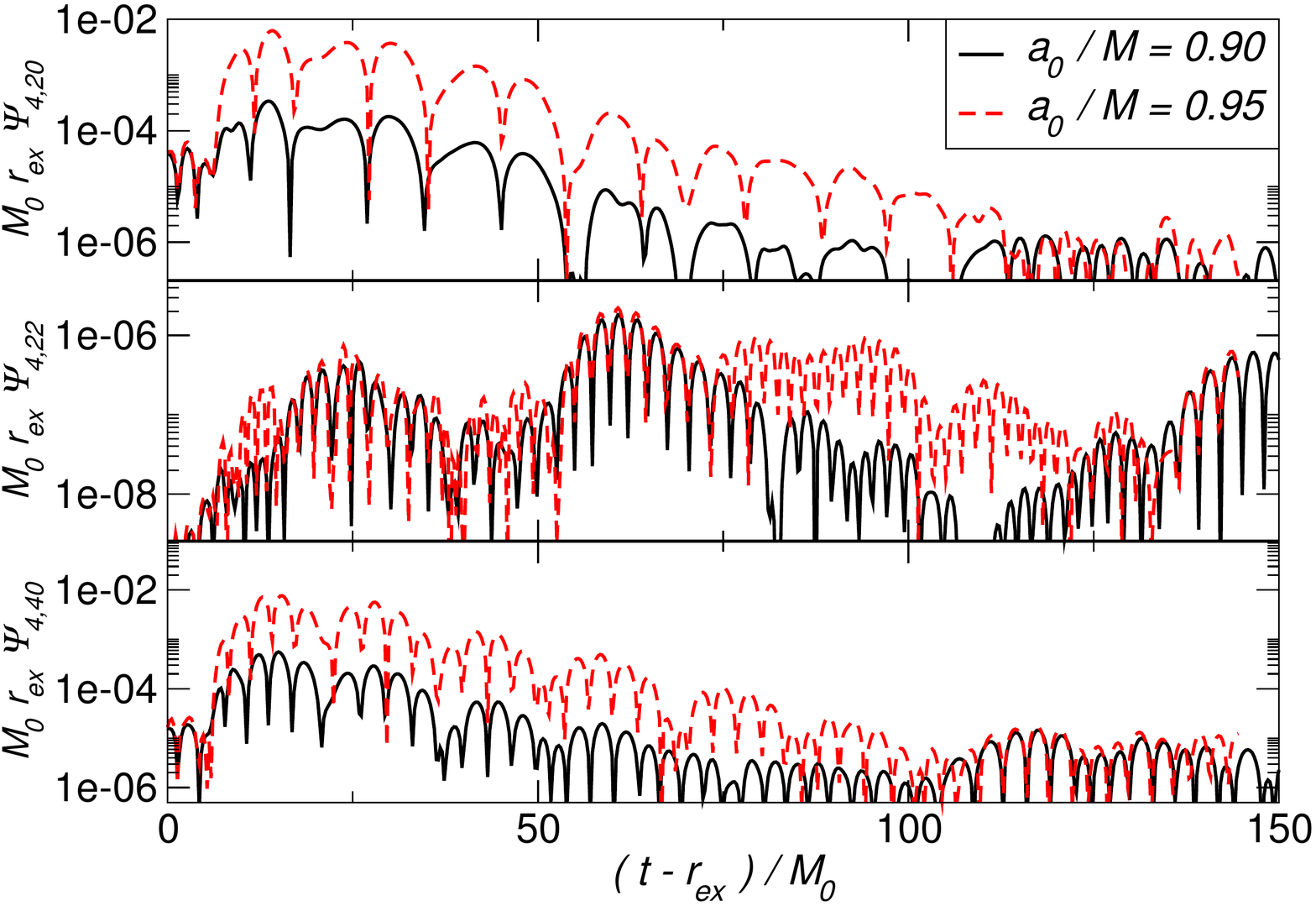}}
\caption{\label{fig:KerrTest}
Time evolution of the characteristic parameters spin (top), 
AH area and BH mass (middle) of a BH with initial spin $a_0/M=0.90, 0.95$.
Because the setup does not provide a pure Kerr BH gravitational radiation is present in the spacetime
as illustrated by the multipoles of $\Psi_{4,lm}$ (bottom).
}
\end{center}
\end{figure*}
In order to check the performance of our code and in particular our construction
of Kerr-like initial data as presented in Sec.~\ref{ssec:InitDataKBH},
we have evolved a single Kerr BH without a scalar cloud.
For this purpose we use a numerical domain given by
$\{(192,96,48,24,16,8,4,2),h=M/60\}$ in the notation of Ref.~\cite{Sperhake:2006cy}.
We focus on the most demanding cases 
that we have been able to evolve with good accuracy, namely Kerr BHs with initial spin parameters 
$a_0/M=0.90$ and $a_0/M=0.95$, where the bare mass parameter $M=1$.

Fig.~\ref{fig:KerrTest1} shows how much the BH spin changes in time, as the system evolves freely.  
Until $t\sim200M_0$ we find a decrease of $\Delta a/a_0\lesssim0.06\%$ (for $a_0/M=0.90$) 
and $\Delta a/a_0\lesssim0.9\%$ (for $a_0/M=0.95$).
The bottom panel of Fig.~\ref{fig:KerrTest2} depicts the change in the BH mass
which is almost constant (within $\lesssim 0.1\%$).
These error estimates are comparable with the results 
found by Liu et al~\cite{Liu:2009al} (see their Fig.~1).
We have furthermore verified that the AH area is indeed increasing as it should
and show the time evolution of the relative AH area in the top panel of Fig.~\ref{fig:KerrTest2}.

As we have seen, the construction of conformally Kerr-like puncture initial data allows to evolve
rotating BHs with spin parameters much closer to extremality than the widely used conformally flat approach which
is restricted by the Bowen-York limit of $a/M\sim0.93$ due to the presence of spurious radiation. 
However, even the improved, conformally Kerr-like numerical solutions contain spurious gravitational 
waves as we illustrate in Fig.~\ref{fig:KerrTest3}.
Thus the numerical solution does not represent an isolated Kerr BH, but rather
a rotating BH spacetime  which also contains spurious (gravitational) radiation.

%%%%%%%%%%%%%%%%%%%%%%%%%%%%%%%%%%%%%%%%%%%%%%%%%%%%%%%%%%%%%%%%%%%%%%%%%%%
\section{Snapshots of scalar clouds}\label{app:snapshots}
\begin{figure*}[htpb!]
 \begin{center}
 \subfloat[$t=14.32M_0$]{%\label{fig:SchMasslessBHprop}
 \includegraphics[width=0.25\textwidth,clip]{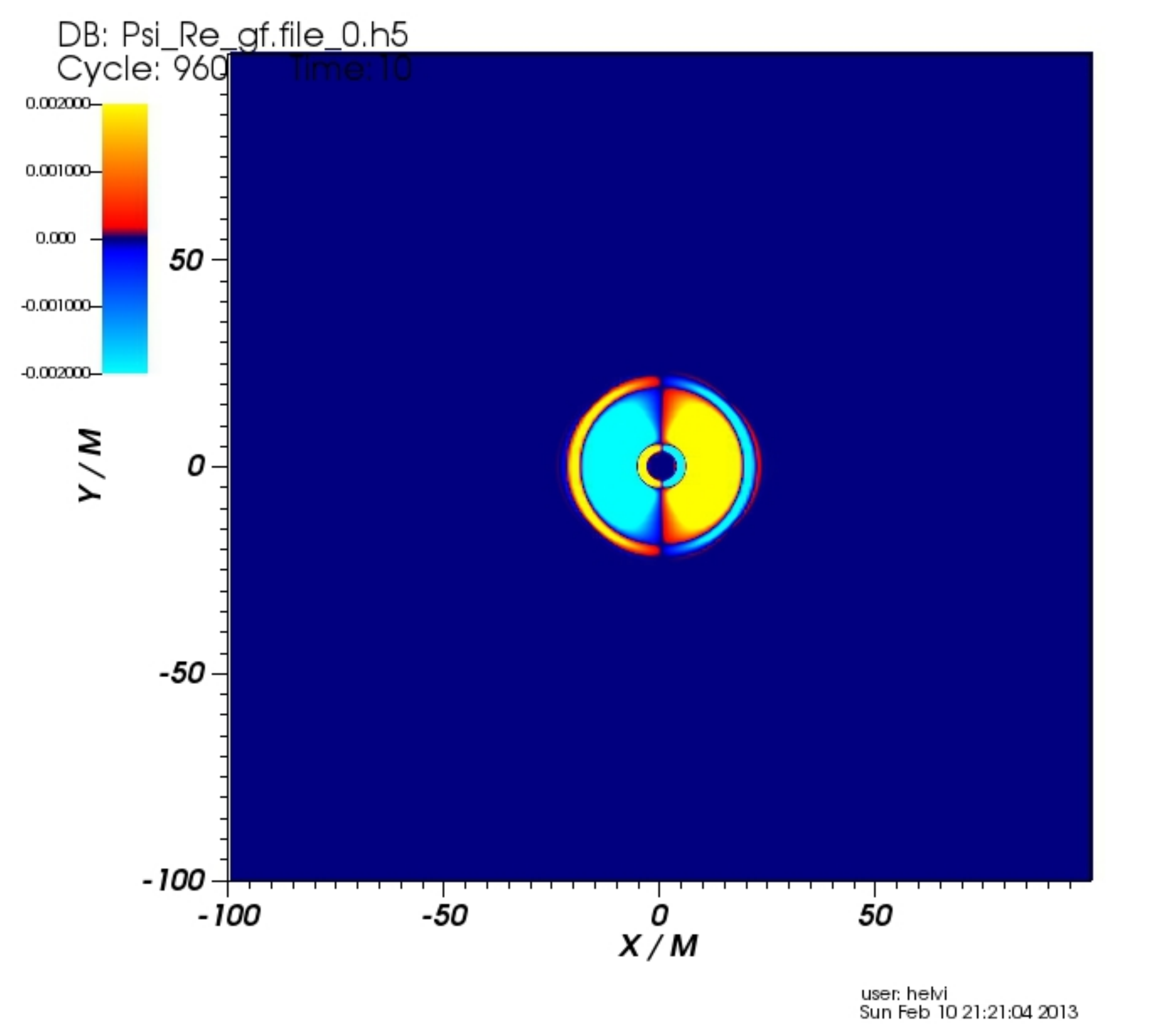}}
 \subfloat[$t=171.84M_0$]{%\label{fig:SchMasslessWaveforms}
 \includegraphics[width=0.25\textwidth,clip]{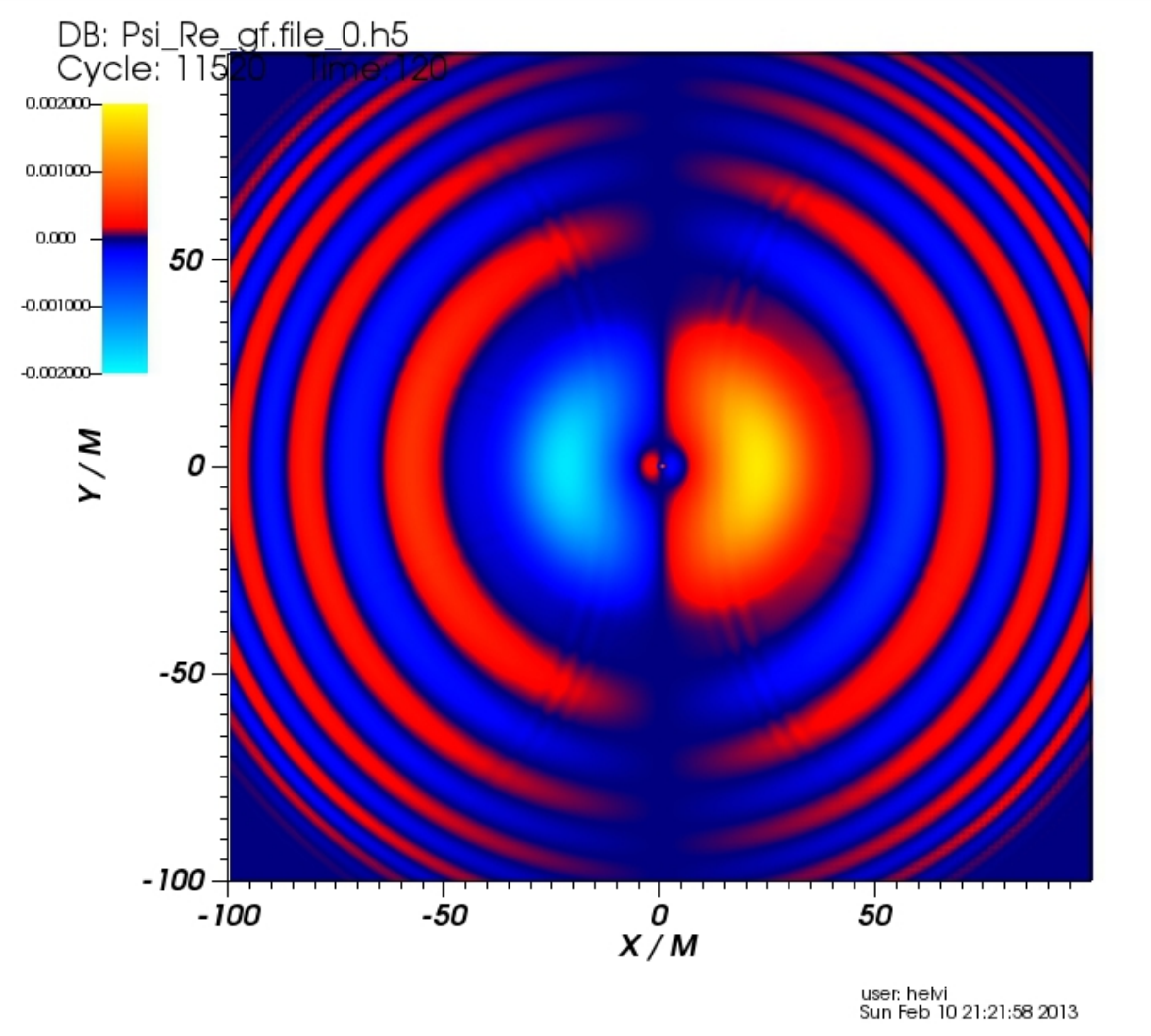}}
 \subfloat[$t=286.41M_0$]{%\label{fig:SchMasslessWaveforms}
 \includegraphics[width=0.25\textwidth,clip]{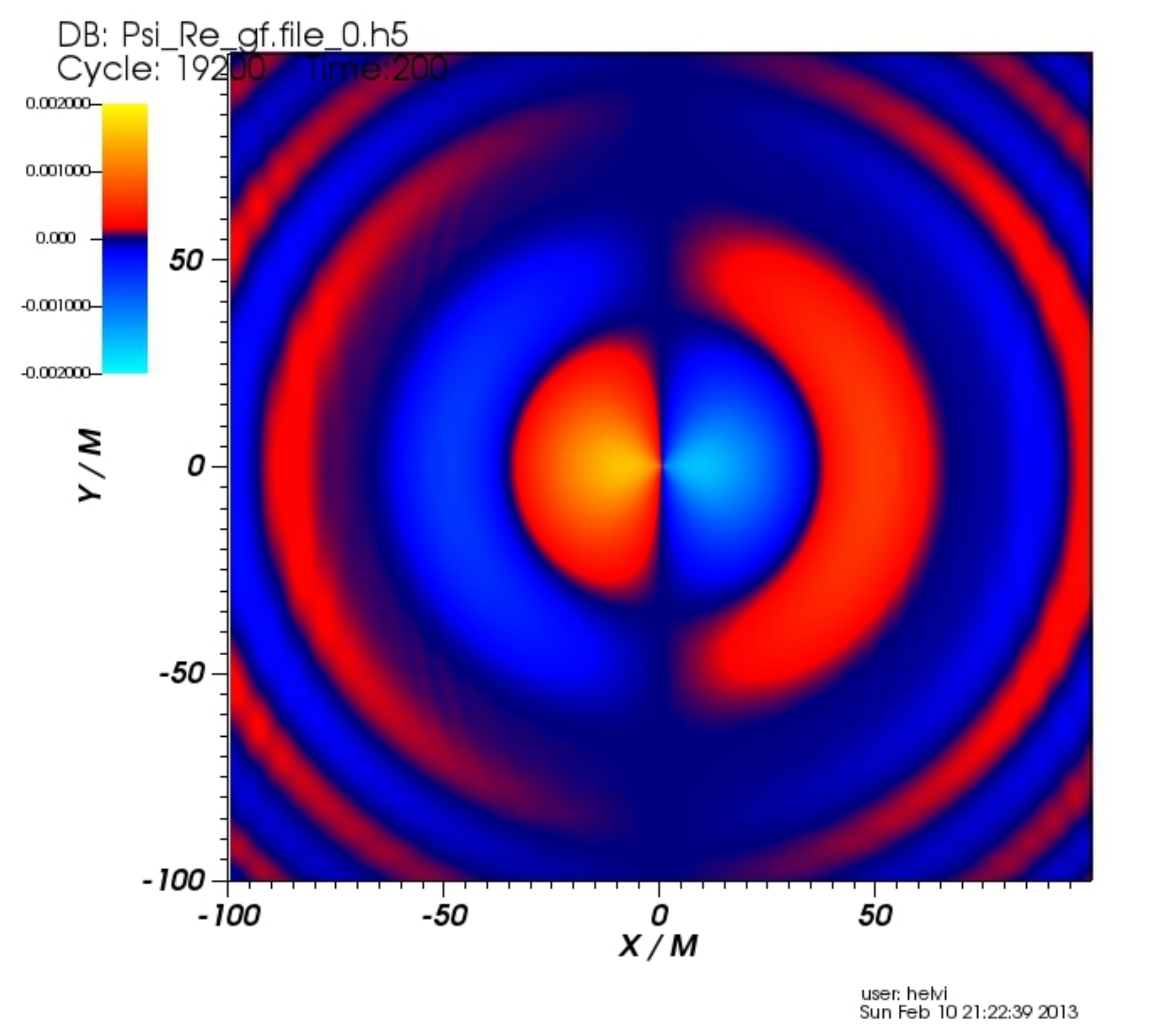}}
 \subfloat[$t=1088.36M_0$]{%\label{fig:SchMasslessWaveforms}
 \includegraphics[width=0.25\textwidth,clip]{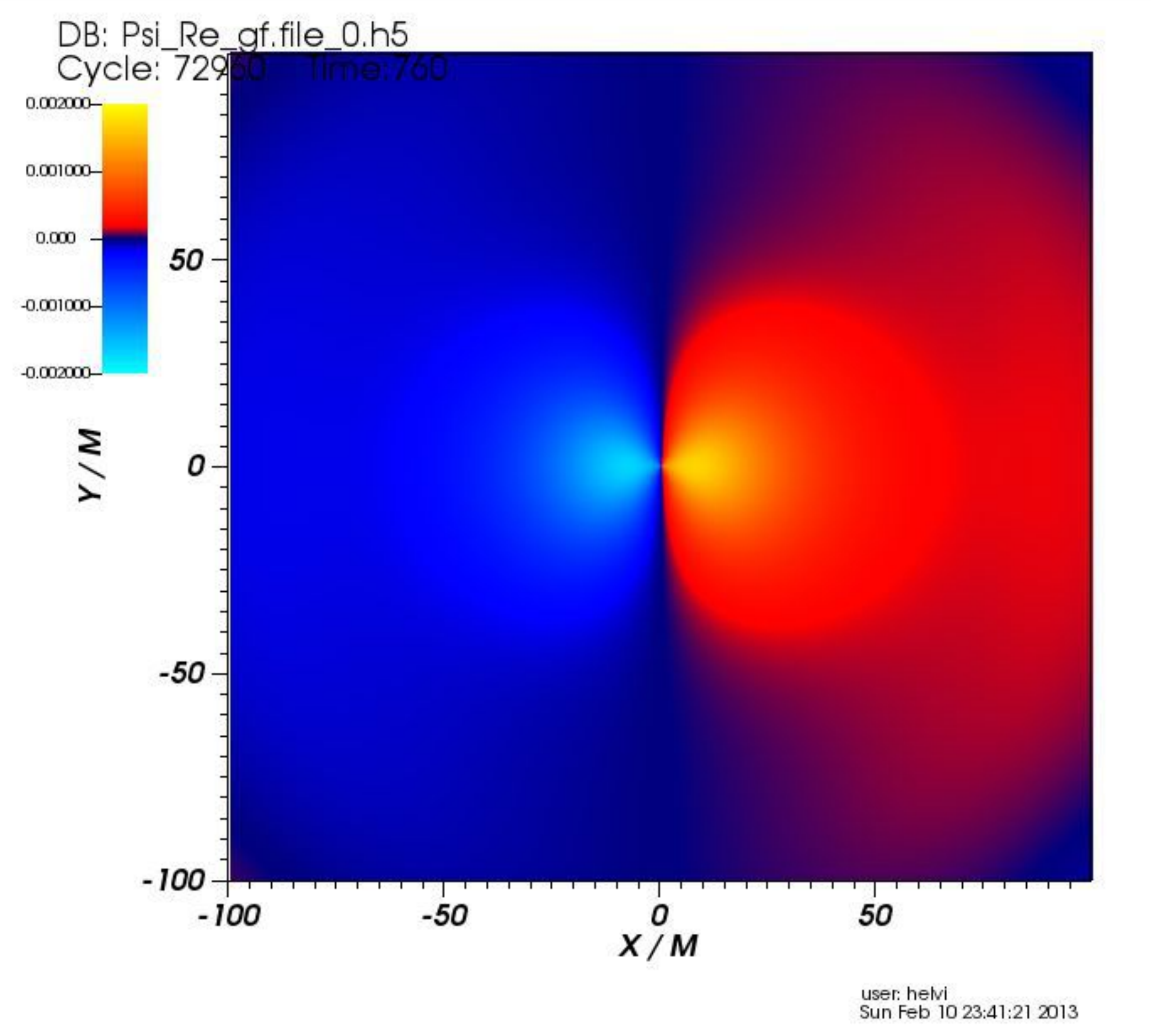}}
 \\
 \subfloat[$t=1718.46M_0$]{%\label{fig:SchMasslessWaveforms}
 \includegraphics[width=0.25\textwidth,clip]{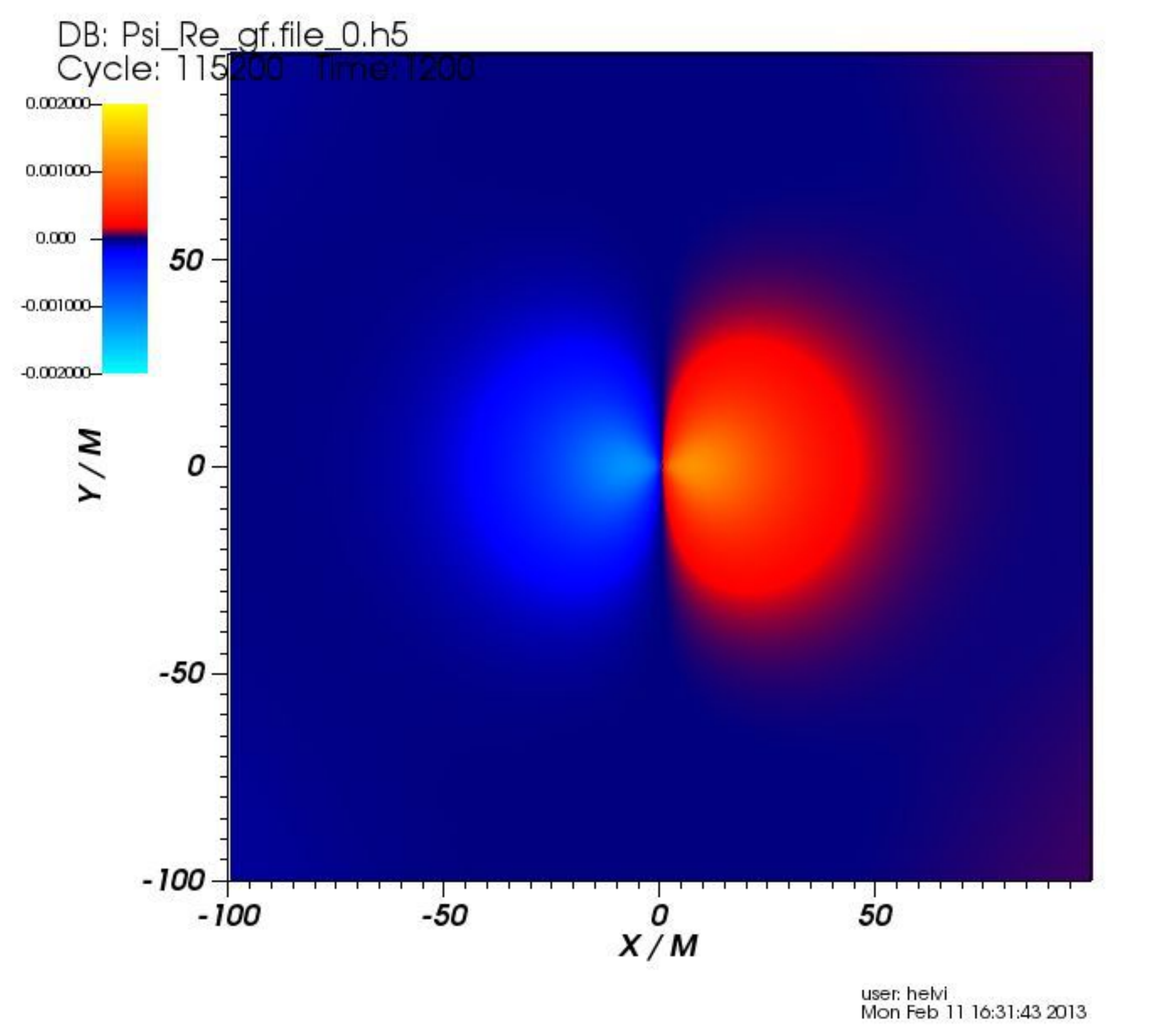}}
 \subfloat[$t=2076.47M_0$]{%\label{fig:SchMasslessWaveforms}
 \includegraphics[width=0.25\textwidth,clip]{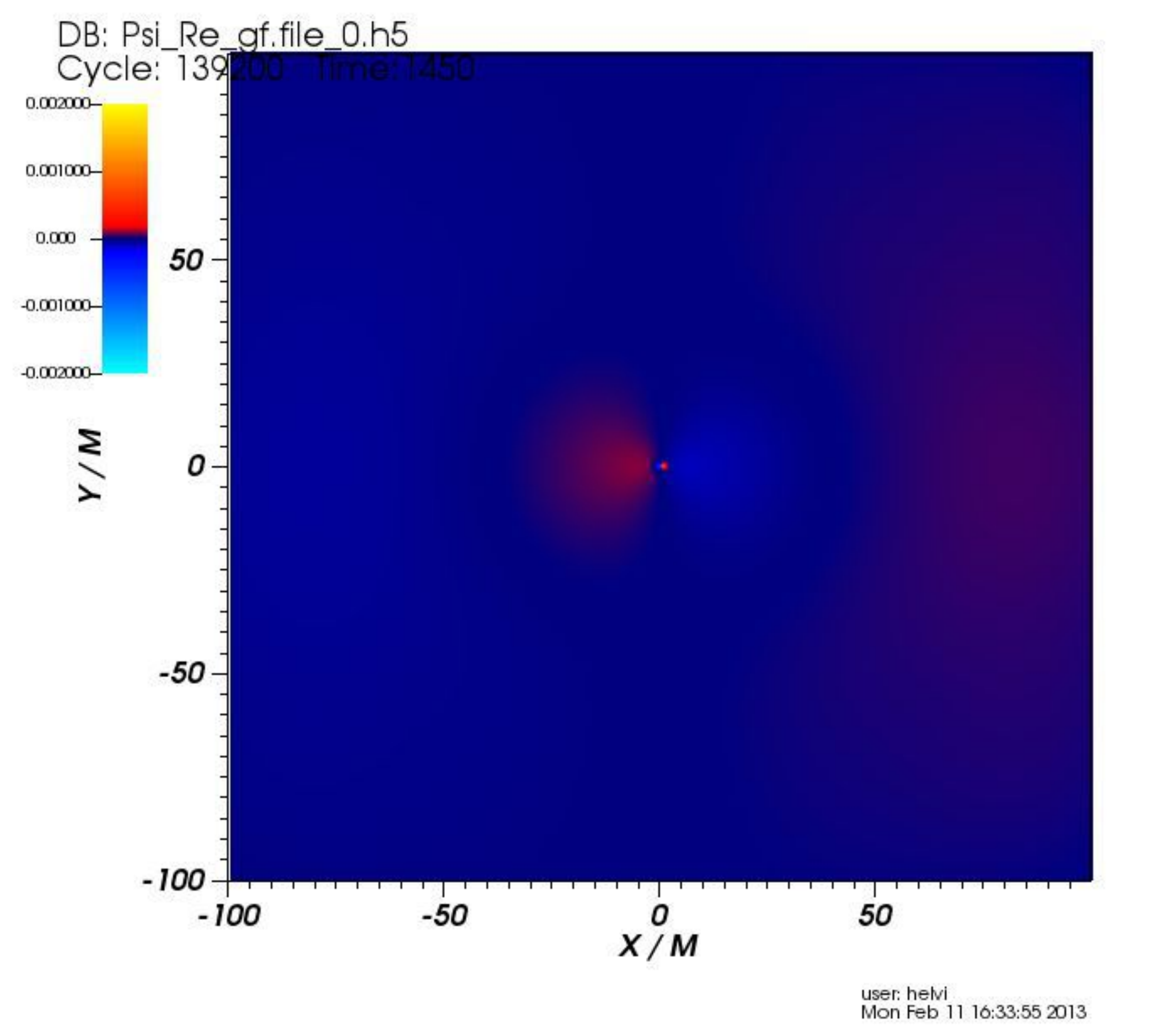}}
 \subfloat[$t=2465.89M_0$]{%\label{fig:SchMasslessWaveforms}
 \includegraphics[width=0.25\textwidth,clip]{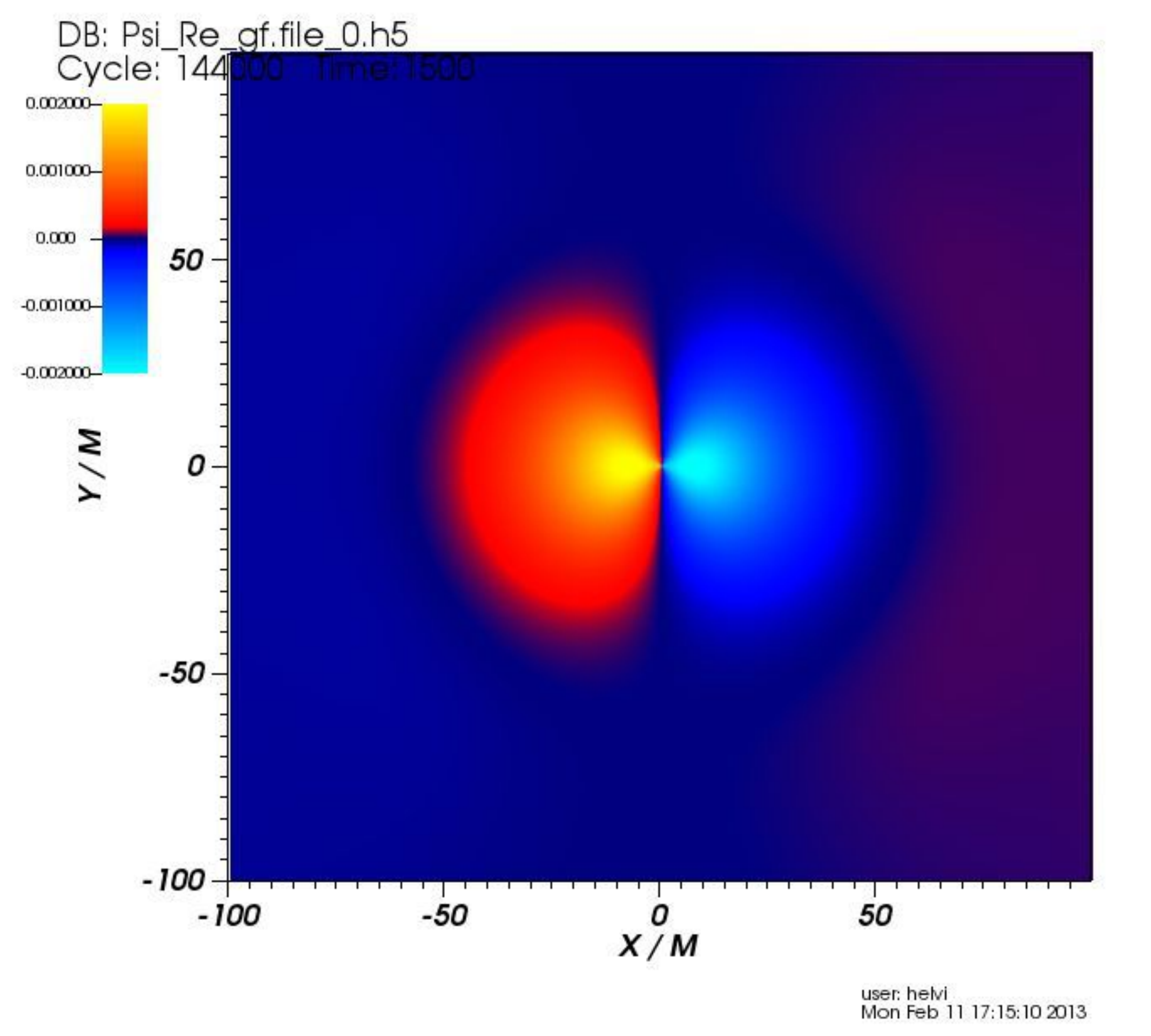}}
 \subfloat[$t=2577.69M_0$]{%\label{fig:SchMasslessWaveforms}
 \includegraphics[width=0.25\textwidth,clip]{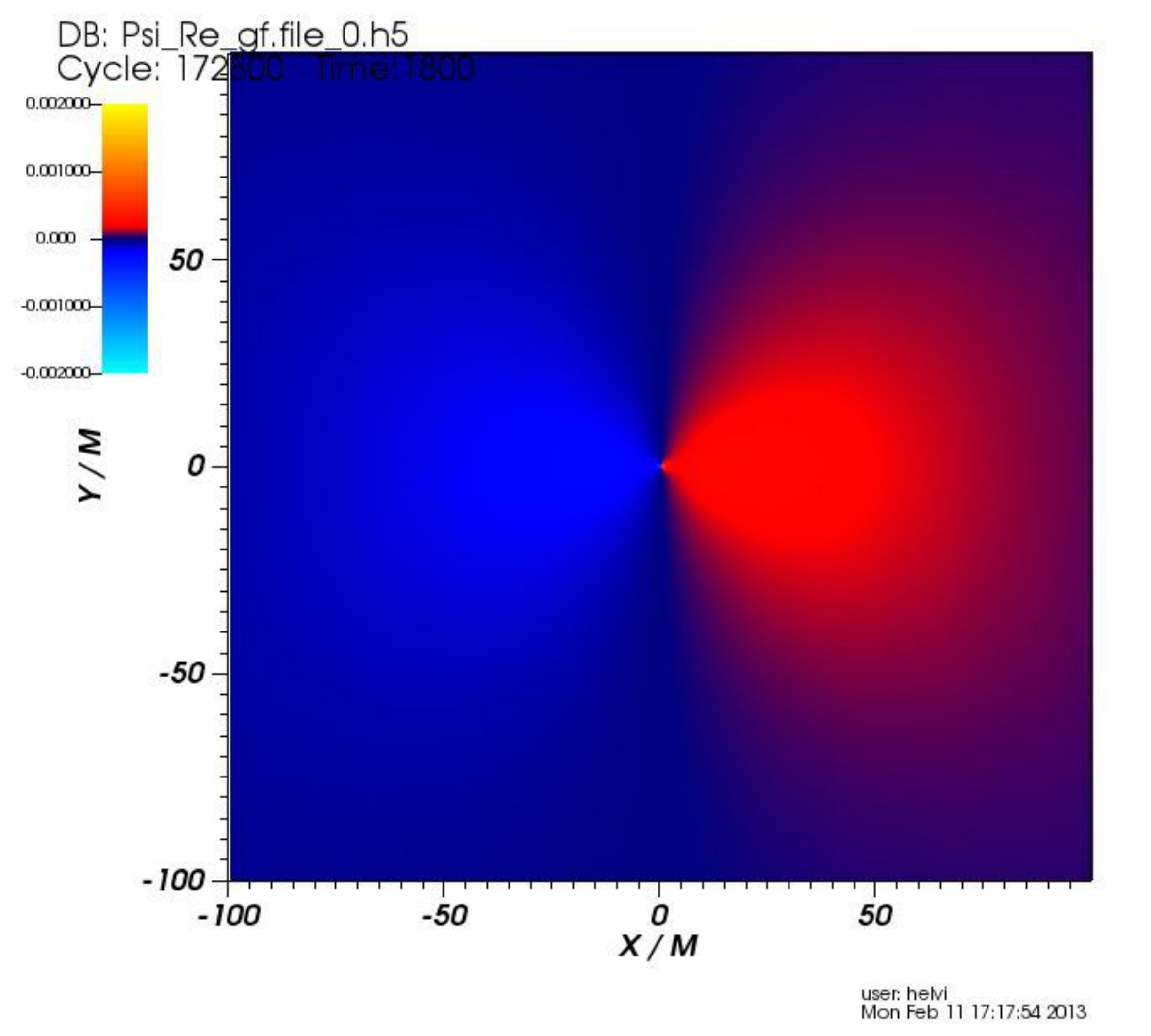}}
  \caption{\label{fig:SnapshotsSchwarzschild}
 Snapshots of a massive scalar cloud with $M_0\mu_S=0.29$ around a Schwarzschild BH
 with $M_0=0.6983$.
 We present a slice of the equatorial plane at different time steps.
 The infall of the scalar field triggers the excitation and ringdown of the BH, 
 which is followed by the formation of a flashing scalar could.
 }
 \end{center}
 \end{figure*}

\begin{figure*}[htpb!]
 \begin{center}
 \subfloat[$t=10M_0$]{%\label{fig:SchMasslessBHprop}
 \includegraphics[width=0.25\textwidth,clip]{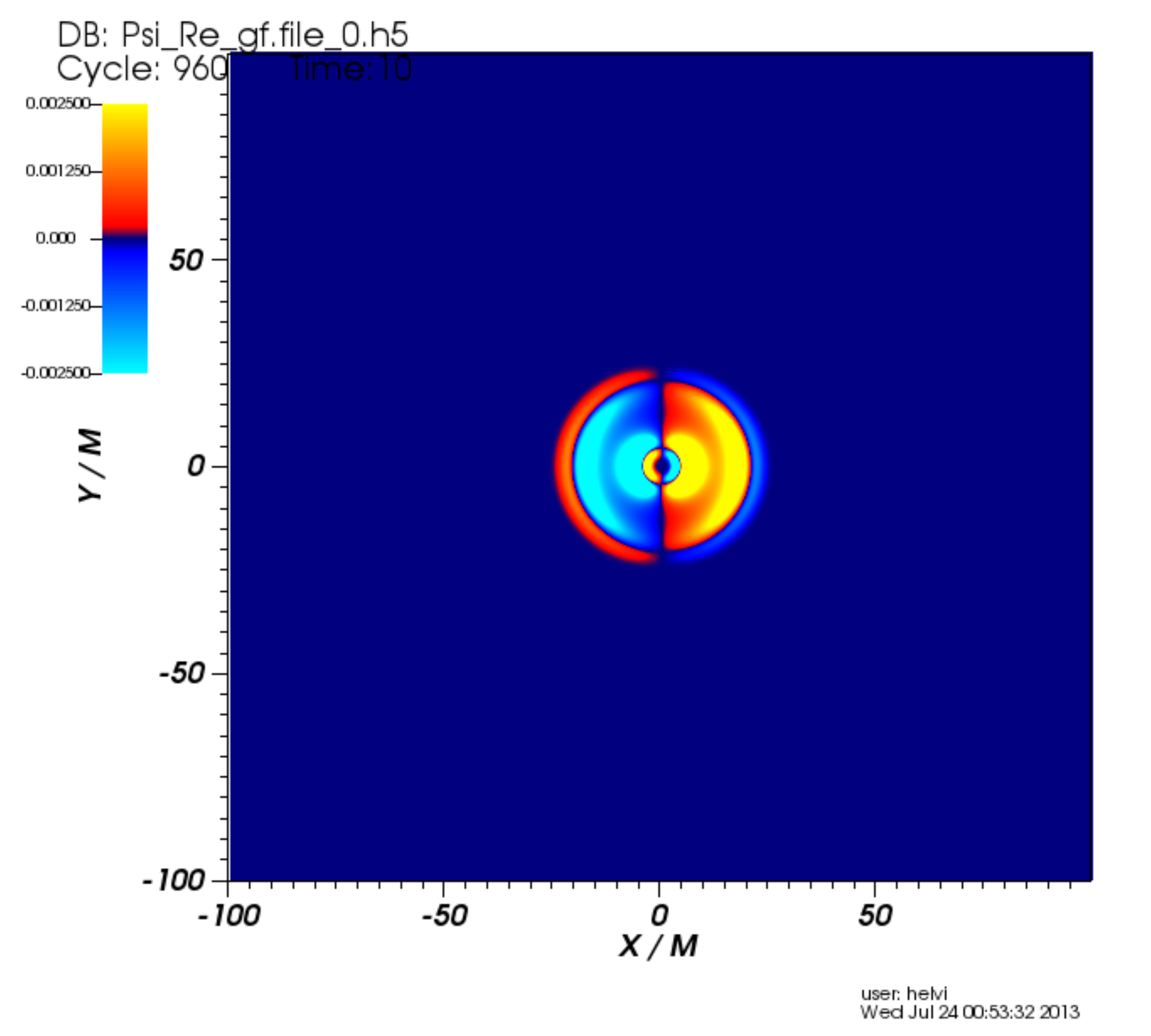}}
 \subfloat[$t=120M_0$]{%\label{fig:SchMasslessWaveforms}
 \includegraphics[width=0.25\textwidth,clip]{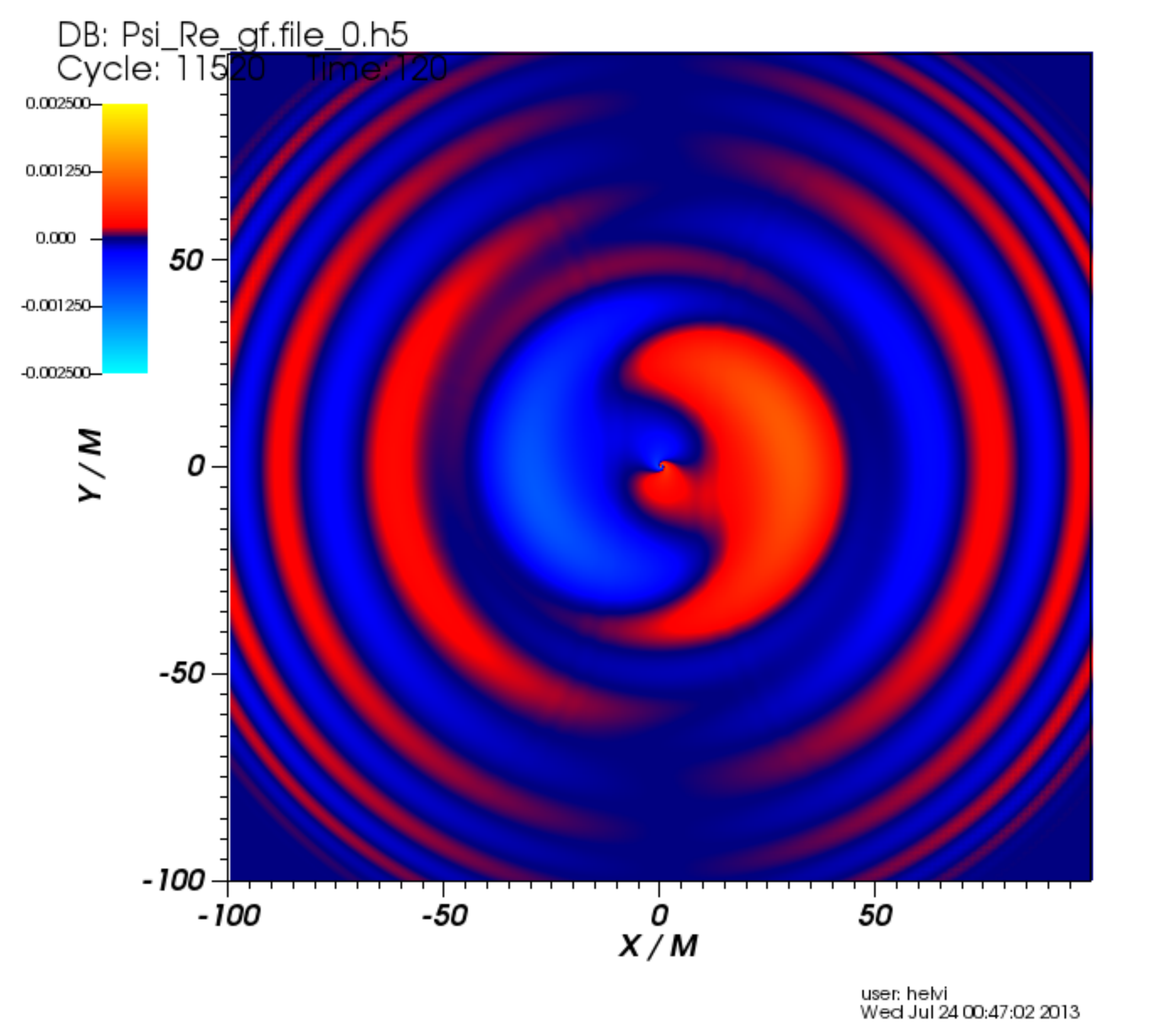}}
 \subfloat[$t=250M_0$]{%\label{fig:SchMasslessWaveforms}
 \includegraphics[width=0.25\textwidth,clip]{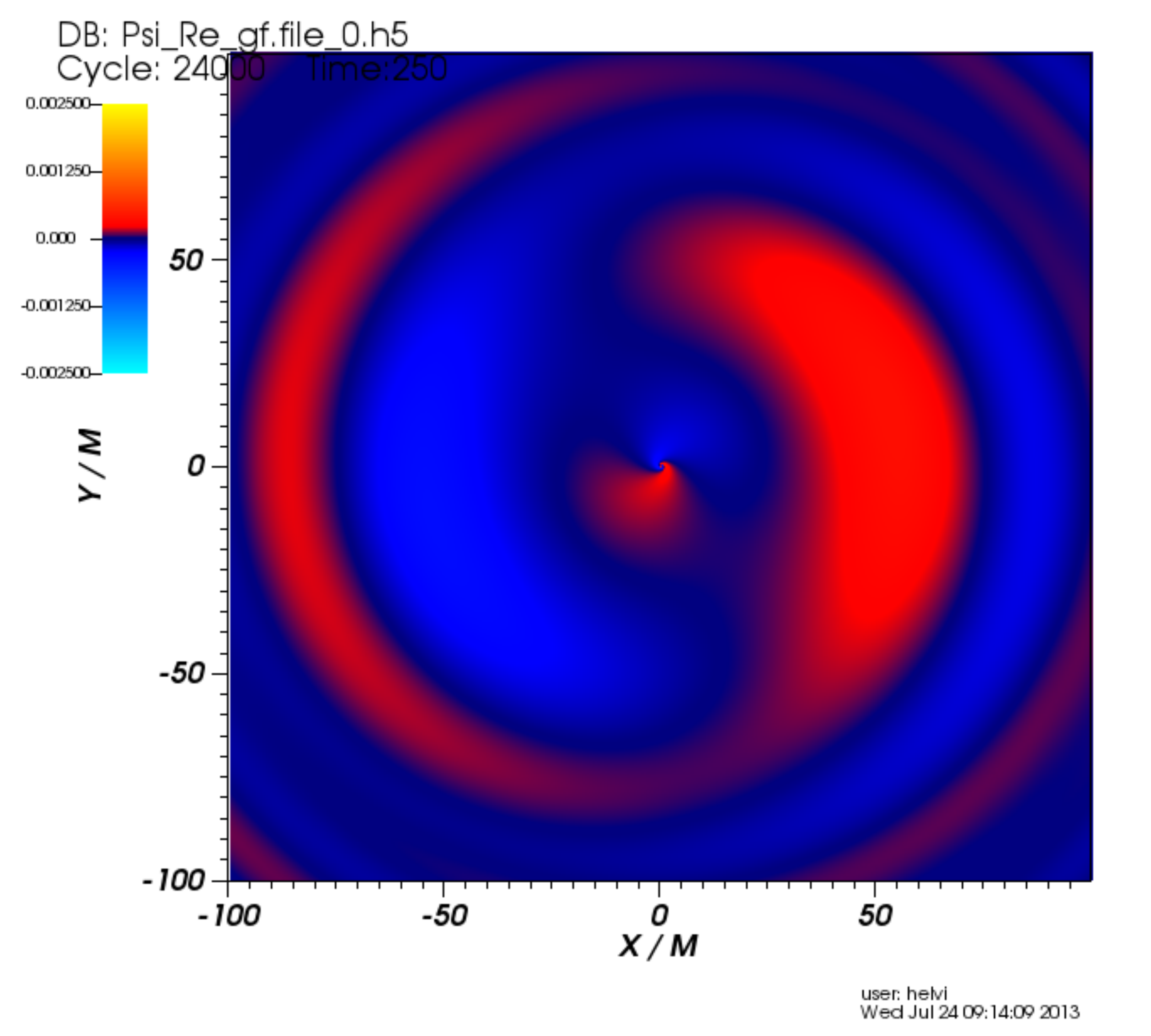}}
 \subfloat[$t=400M_0$]{%\label{fig:SchMasslessWaveforms}
 \includegraphics[width=0.25\textwidth,clip]{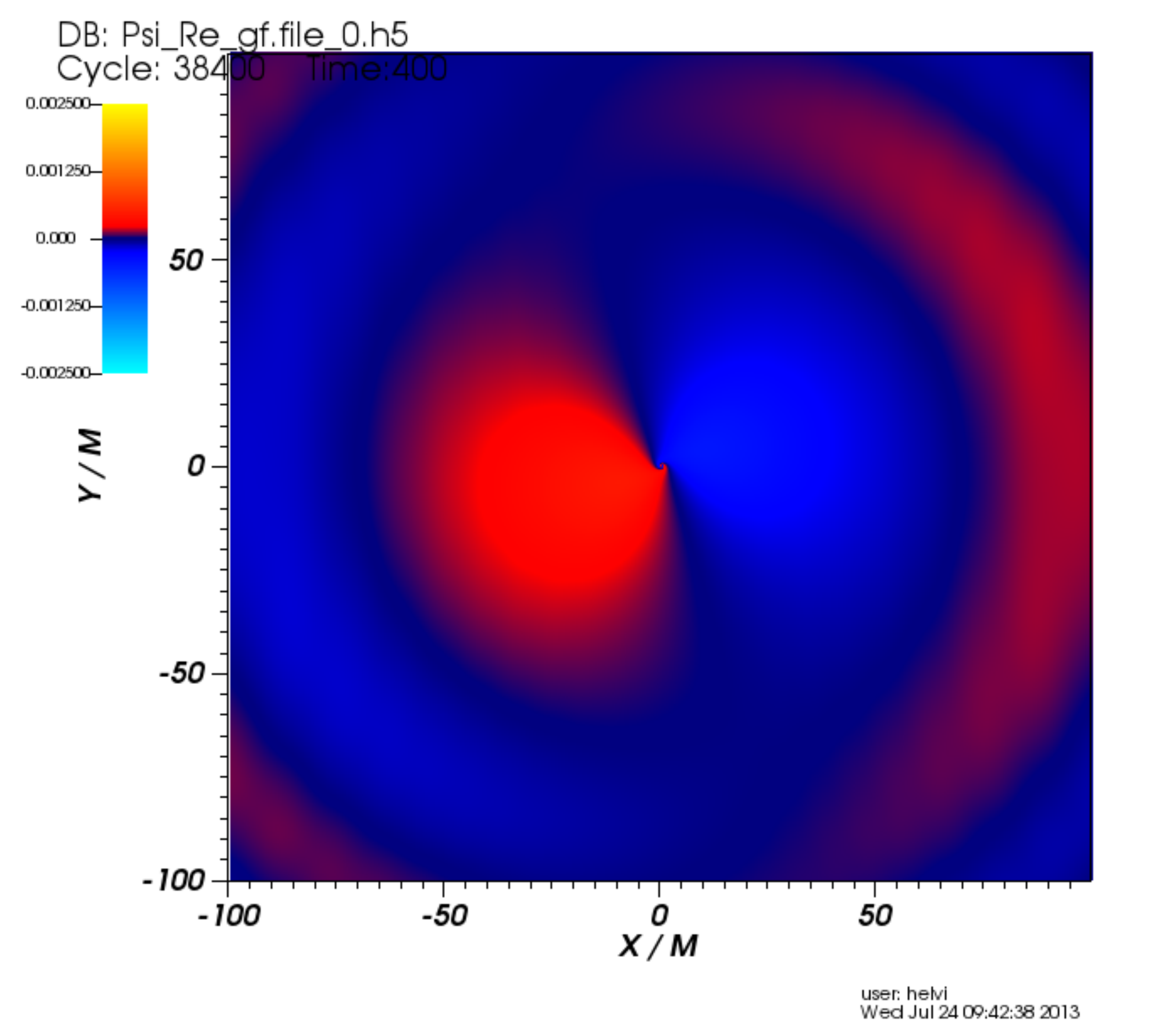}}
 \\
 \subfloat[$t=500M_0$]{%\label{fig:SchMasslessWaveforms}
 \includegraphics[width=0.25\textwidth,clip]{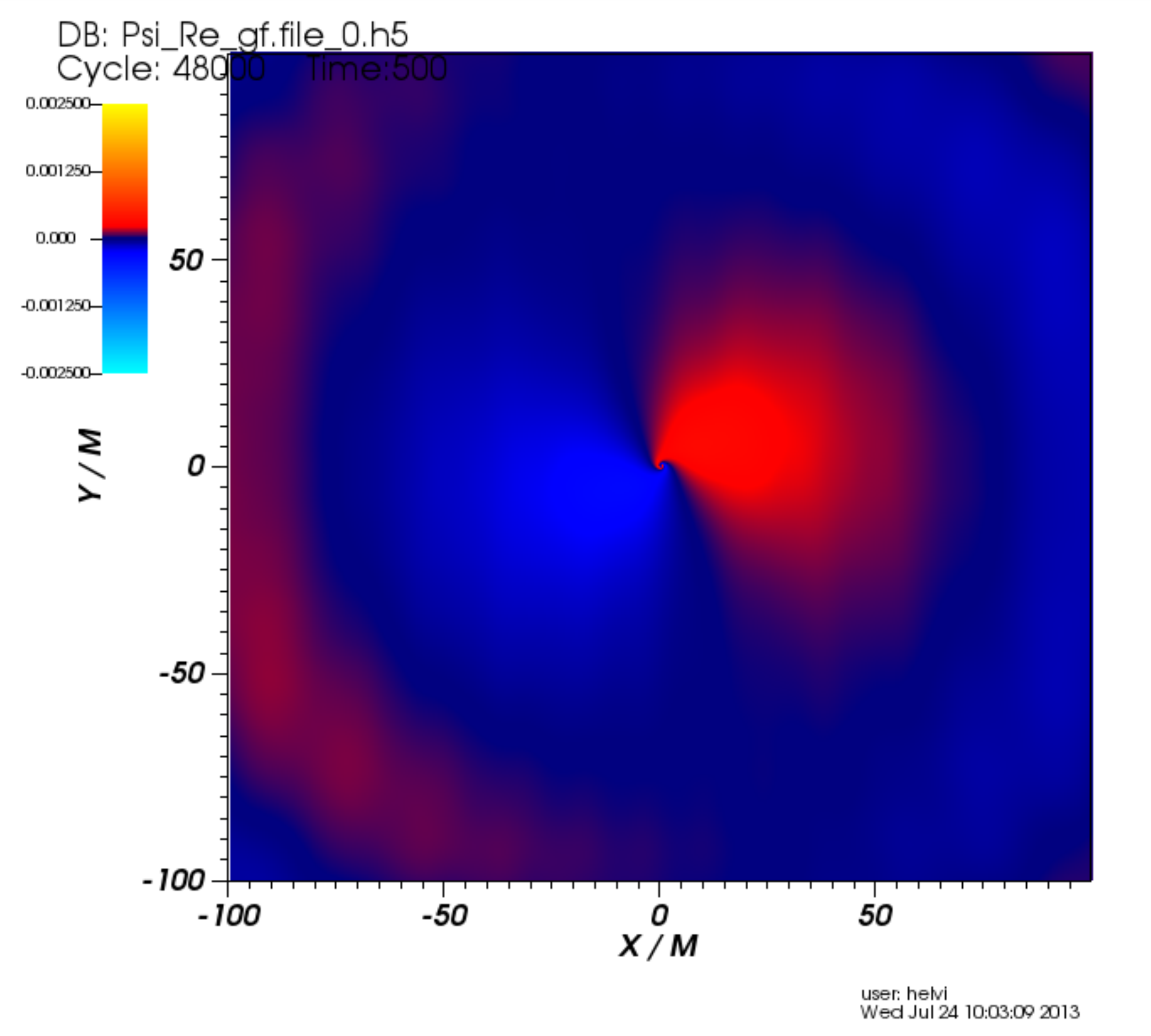}}
 \subfloat[$t=600M_0$]{%\label{fig:SchMasslessWaveforms}
 \includegraphics[width=0.25\textwidth,clip]{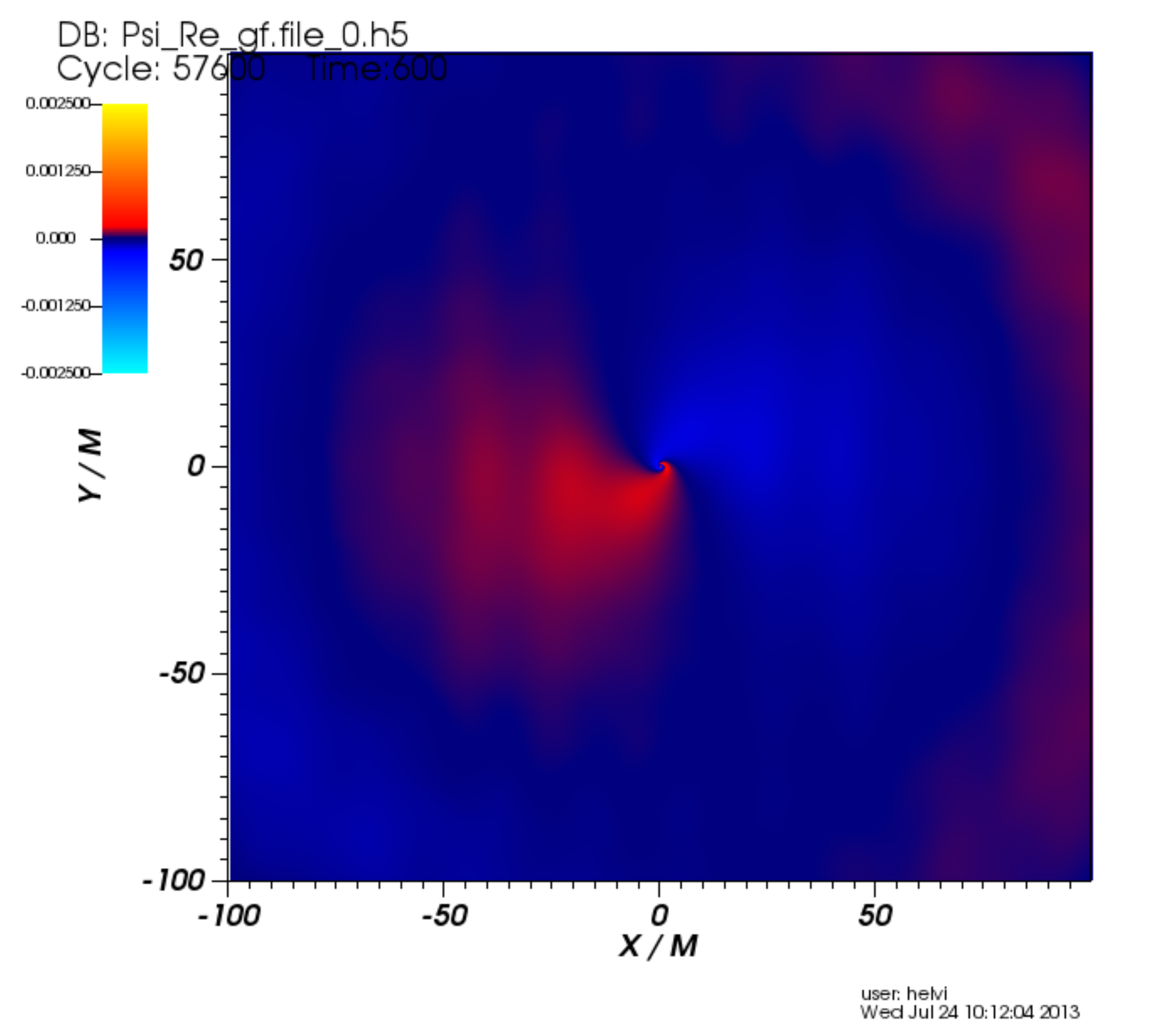}}
 \subfloat[$t=750M_0$]{%\label{fig:SchMasslessWaveforms}
 \includegraphics[width=0.25\textwidth,clip]{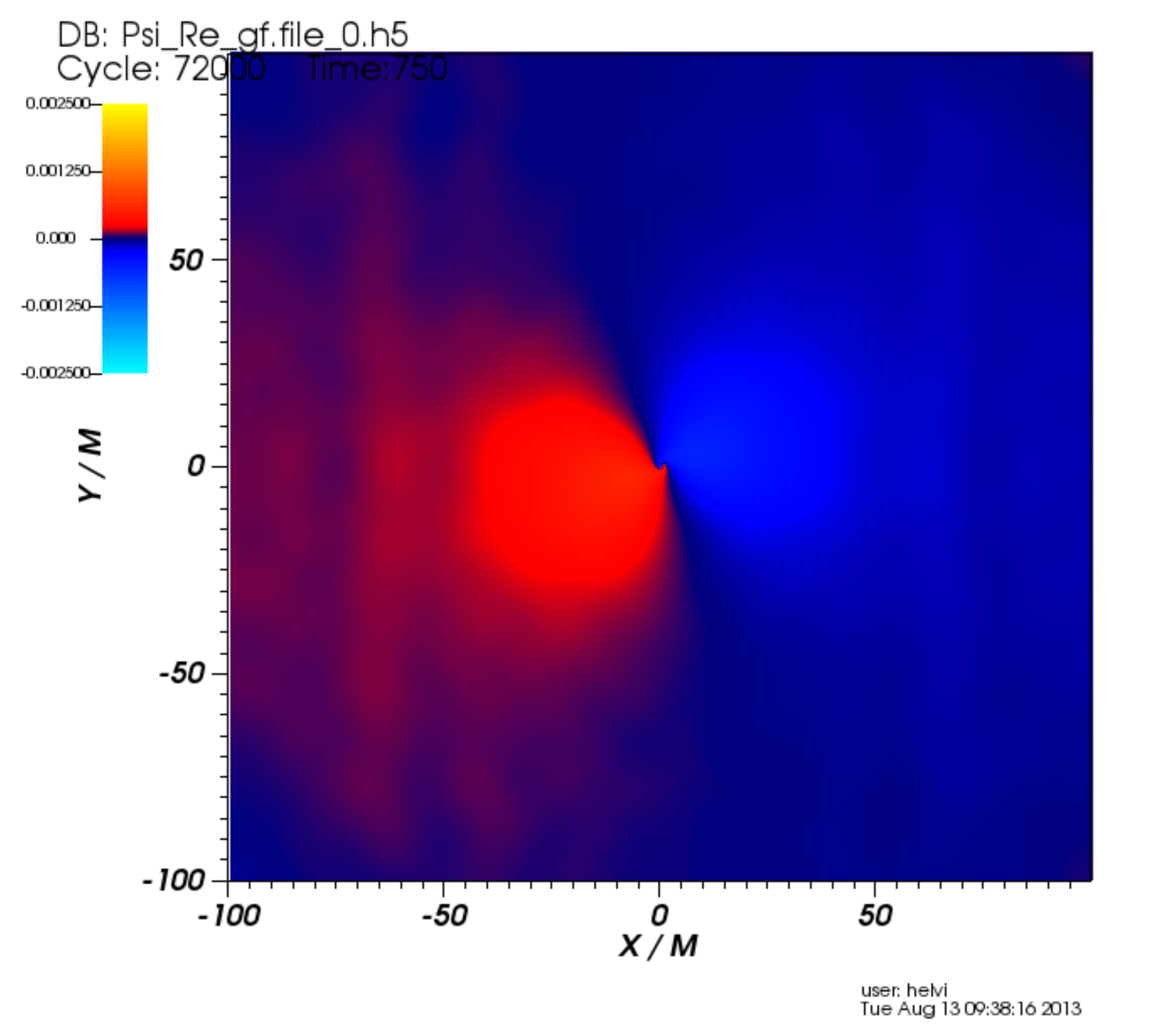}}
 \subfloat[$t=850M_0$]{%\label{fig:SchMasslessWaveforms}
 \includegraphics[width=0.25\textwidth,clip]{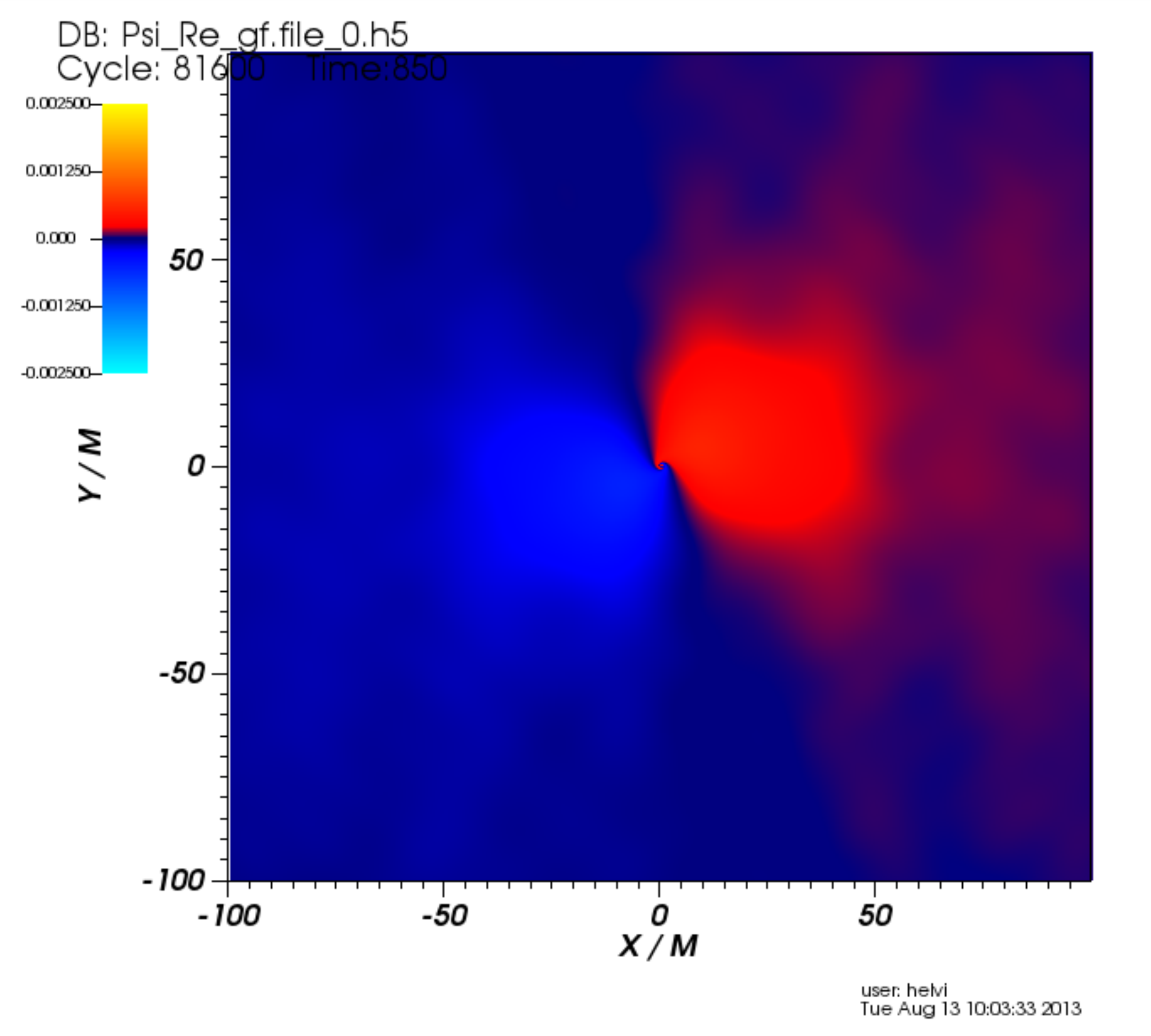}}
  \caption{\label{fig:SnapshotsKerr}
 Snapshots of a massive scalar cloud with $M_0\mu_S=0.29$ around a Kerr BH
 with $a_0/M=0.95$.
 We present a slice of the equatorial plane at different time steps.
 The infall of the scalar field induces a burst of gravitational and scalar waves
 succeeded by the formation of a flashing scalar could.
 Due to frame-dragging effects the cloud follows the rotation of the BH.
 }
 \end{center}
 \end{figure*}

%%%%%%%%%%%%%%%%%%%%%%%%%%%%%%%%%%%%%%%%%%%%%%%%%%%%%%%%%%%%%%%%%%%%%%%%%%%
In order to illustrate the dynamical behaviour of a massive scalar field coupled 
to a BH spacetime we present snapshots of its evolutions in Figs.~\ref{fig:SnapshotsSchwarzschild}
and~\ref{fig:SnapshotsKerr} for the case of a Schwarzschild and 
a Kerr BH with initial spin of $a_0/M=0.95$. 
The complete animations can be found at the website~\cite{DyBHo:web}.
Specifically we show the evolution of a dipole scalar field in the equatorial plane at different time steps.
The general development is quite similar in both cases:
The first plot depicts the system at early times of the evolution when the scalar field starts 
to be sucked into the BH. Its accretion onto the BH triggers a burst of gravitational 
and scalar radiation. The latter is shown in the second plot of the time series.
Picture (c) and (d) of Figs.~\ref{fig:SnapshotsSchwarzschild} and~\ref{fig:SnapshotsKerr}
illustrate the transition from the ringdown of the system towards the formation of 
a scalar cloud.
In the second row of Figs.~\ref{fig:SnapshotsSchwarzschild} and~\ref{fig:SnapshotsKerr}
we observe a dimming and brighening up of the massive scalar hair due to beating effects.
In case of the Kerr BH, Fig.~\ref{fig:SnapshotsKerr}, the flashing of the scalar field
is accompanied by a dragging of the scalar field along with the BH's rotation.

% %

%%%%%%%%%%%%%%%%%%%%%%%%%%%%%%%%%%%%%%%%%%%%%%%%%%%%%%%%%%%%%%%%%%%%%%%%%%%
%\clearpage
\bibliographystyle{h-physrev4}
\bibliography{NonScalar}

\begin{thebibliography}{100}

\bibitem{Belczynski:2006br}
K.~Belczynski {\em et~al.},
\newblock Astrophys.J. {\bf 648}, 1110 (2006), [astro-ph/0601458].
%%CITATION = ASTRO-PH/0601458;%%

\bibitem{McClintock:2009as}
J.~E. McClintock and R.~A. Remillard,
\newblock 0902.3488.
%%CITATION = ARXIV:0902.3488;%%

\bibitem{Ivanova:2010ia}
N.~Ivanova {\em et~al.},
\newblock Astrophys.J. {\bf 717}, 948 (2010), [1001.1767].
%%CITATION = ARXIV:1001.1767;%%

\bibitem{Nissanke:2012eh}
S.~Nissanke, M.~Vallisneri, G.~Nelemans and T.~A. Prince,
\newblock Astrophys.J. {\bf 758}, 131 (2012), [1201.4613].
%%CITATION = ARXIV:1201.4613;%%

\bibitem{Belczynski:2008nh}
K.~Belczynski, M.~Benacquista and T.~Bulik,
\newblock Astrophys.J. {\bf 725}, 816 (2010), [0811.1602].
%%CITATION = ARXIV:0811.1602;%%

\bibitem{Heinke:2013ela}
C.~Heinke {\em et~al.},
\newblock Astrophys.J. {\bf 768}, 184 (2013), [1303.5864].
%%CITATION = ARXIV:1303.5864;%%

\bibitem{Seoane:2013qna}
eLISA Collaboration, P.~A. Seoane {\em et~al.},
\newblock 1305.5720.
%%CITATION = ARXIV:1305.5720;%%

\bibitem{Reynolds:2013qqa}
C.~S. Reynolds,
\newblock 1302.3260.
%%CITATION = ARXIV:1302.3260;%%

\bibitem{Begelman:1980vb}
M.~Begelman, R.~Blandford and M.~Rees,
\newblock Nature {\bf 287}, 307 (1980).
%%CITATION = NATUA,287,307;%%

\bibitem{Rees:1984si}
M.~J. Rees,
\newblock Ann.Rev.Astron.Astrophys. {\bf 22}, 471 (1984).
%%CITATION = ARAAA,22,471;%%

\bibitem{Ferrarese:2004qr}
L.~Ferrarese and H.~Ford,
\newblock Space Sci.Rev. {\bf 116}, 523 (2005), [astro-ph/0411247].
%%CITATION = ASTRO-PH/0411247;%%

\bibitem{Alexander:2005jz}
T.~Alexander,
\newblock Phys.Rept. {\bf 419}, 65 (2005), [astro-ph/0508106].
%%CITATION = ASTRO-PH/0508106;%%

\bibitem{Ferrarese:2006fd}
L.~Ferrarese {\em et~al.},
\newblock Astrophys.J. {\bf 644}, L21 (2006), [astro-ph/0603840].
%%CITATION = ASTRO-PH/0603840;%%

\bibitem{Denney:2010cn}
K.~Denney {\em et~al.},
\newblock Astrophys.J. {\bf 721}, 715 (2010), [1006.4160].
%%CITATION = ARXIV:1006.4160;%%

\bibitem{Volonteri:2012yn}
M.~Volonteri, M.~Sikora, J.-P. Lasota and A.~Merloni,
\newblock 1210.1025.
%%CITATION = ARXIV:1210.1025;%%

\bibitem{Wang:2013oga}
J.~Wang, X.~Zhou and J.~Y. Wei,
\newblock Astrophys.J. {\bf 768}, 176 (2013), [1303.5495].
%%CITATION = ARXIV:1303.5495;%%

\bibitem{Reynolds:2013rva}
C.~S. Reynolds,
\newblock 1307.3246.
%%CITATION = ARXIV:1307.3246;%%

\bibitem{Abbott:2007kv}
LIGO Scientific Collaboration, B.~Abbott {\em et~al.},
\newblock Rept.Prog.Phys. {\bf 72}, 076901 (2009), [0711.3041].
%%CITATION = ARXIV:0711.3041;%%

\bibitem{Acernese:2008zzf}
F.~Acernese {\em et~al.},
\newblock Class.Quant.Grav. {\bf 25}, 184001 (2008).
%%CITATION = CQGRD,25,184001;%%

\bibitem{Abadie:2011kd}
LIGO Scientific Collaboration, Virgo Collaboration, J.~Abadie {\em et~al.},
\newblock Phys.Rev. {\bf D83}, 122005 (2011), [1102.3781].
%%CITATION = ARXIV:1102.3781;%%

\bibitem{Aasi:2013wya}
LIGO Scientific Collaboration, Virgo Collaboration, J.~Aasi {\em et~al.},
\newblock 1304.0670.
%%CITATION = ARXIV:1304.0670;%%

\bibitem{aLIGO}
Advanced {LIGO} webpage,
\newblock \url{https://www.advancedligo.mit.edu/}.

\bibitem{aLIGOind}
Ind{IGO} website,
\newblock \url{http://www.gw-indigo.org/}.

\bibitem{Aso:2013eba}
Y.~Aso {\em et~al.},
\newblock 1306.6747.
%%CITATION = ARXIV:1306.6747;%%

\bibitem{Somiya:2011np}
KAGRA Collaboration, K.~Somiya,
\newblock Class.Quant.Grav. {\bf 29}, 124007 (2012), [1111.7185].
%%CITATION = ARXIV:1111.7185;%%

\bibitem{AmaroSeoane:2012km}
P.~Amaro-Seoane {\em et~al.},
\newblock 1201.3621.
%%CITATION = ARXIV:1201.3621;%%

\bibitem{Eda:2013gg}
K.~Eda, Y.~Itoh, S.~Kuroyanagi and J.~Silk,
\newblock Phys.Rev.Lett. {\bf 110}, 221101 (2013), [1301.5971].
%%CITATION = ARXIV:1301.5971;%%

\bibitem{Macedo:2013qea}
C.~F. Macedo, P.~Pani, V.~Cardoso and L.~C. Crispino,
\newblock Astrophys.J. {\bf 774}, 48 (2013), [1302.2646].
%%CITATION = ARXIV:1302.2646;%%

\bibitem{Arvanitaki:2010sy}
A.~Arvanitaki and S.~Dubovsky,
\newblock Phys.Rev. {\bf D83}, 044026 (2011), [1004.3558].
%%CITATION = ARXIV:1004.3558;%%

\bibitem{zeldovich1}
Y.~B. Zel'dovich,
\newblock Pis'ma Zh. Eksp. Teor. Fiz. {\bf 14}, 270 (1971).

\bibitem{zeldovich2}
Y.~B. Zel'dovich,
\newblock Zh. Eksp. Teor. Fiz {\bf 62}, 2076 (1972).

\bibitem{East:2013mfa}
W.~E. East, F.~M. Ramazanoglu and F.~Pretorius,
\newblock 1312.4529.
%%CITATION = ARXIV:1312.4529;%%

\bibitem{Press:1972zz}
W.~H. Press and S.~A. Teukolsky,
\newblock Nature {\bf 238}, 211 (1972).
%%CITATION = NATUA,238,211;%%

\bibitem{Cardoso:2004nk}
V.~Cardoso, O.~J. Dias, J.~P. Lemos and S.~Yoshida,
\newblock Phys.Rev. {\bf D70}, 044039 (2004), [hep-th/0404096].
%%CITATION = HEP-TH/0404096;%%

\bibitem{Hod:2009cp}
S.~Hod and O.~Hod,
\newblock Phys.Rev. {\bf D81}, 061502 (2010), [0910.0734].
%%CITATION = ARXIV:0910.0734;%%

\bibitem{Rosa:2009ei}
J.~Rosa,
\newblock JHEP {\bf 1006}, 015 (2010), [0912.1780].
%%CITATION = ARXIV:0912.1780;%%

\bibitem{Witek:2010qc}
H.~Witek {\em et~al.},
\newblock Phys.Rev. {\bf D82}, 104037 (2010), [1004.4633].
%%CITATION = ARXIV:1004.4633;%%

\bibitem{Dolan:2012yt}
S.~R. Dolan,
\newblock Phys. Rev. D 87, {\bf 124026} (2013), [1212.1477].
%%CITATION = ARXIV:1212.1477;%%

\bibitem{Peccei:1977hh}
R.~Peccei and H.~R. Quinn,
\newblock Phys.Rev.Lett. {\bf 38}, 1440 (1977).
%%CITATION = PRLTA,38,1440;%%

\bibitem{Arvanitaki:2009fg}
A.~Arvanitaki, S.~Dimopoulos, S.~Dubovsky, N.~Kaloper and J.~March-Russell,
\newblock Phys.Rev. {\bf D81}, 123530 (2010), [0905.4720].
%%CITATION = ARXIV:0905.4720;%%

\bibitem{Hawking:1999dp}
S.~Hawking and H.~Reall,
\newblock Phys.Rev. {\bf D61}, 024014 (2000), [hep-th/9908109].
%%CITATION = HEP-TH/9908109;%%

\bibitem{Cardoso:2004hs}
V.~Cardoso and O.~J. Dias,
\newblock Phys.Rev. {\bf D70}, 084011 (2004), [hep-th/0405006].
%%CITATION = HEP-TH/0405006;%%

\bibitem{Cardoso:2006wa}
V.~Cardoso, O.~J. Dias and S.~Yoshida,
\newblock Phys.Rev. {\bf D74}, 044008 (2006), [hep-th/0607162].
%%CITATION = HEP-TH/0607162;%%

\bibitem{Cardoso:2013pza}
V.~Cardoso, O.~J.~C. Dias, G.~S. Hartnett, L.~Lehner and J.~E. Santos,
\newblock 1312.5323.
%%CITATION = ARXIV:1312.5323;%%

\bibitem{Damour:1976}
T.~Damour, N.~Deruelle and R.~Ruffini,
\newblock Nuovo Cimento Lettere {\bf 15}, 257 (1976).

\bibitem{Detweiler:1980uk}
S.~L. Detweiler,
\newblock Phys.Rev. {\bf D22}, 2323 (1980).
%%CITATION = PHRVA,D22,2323;%%

\bibitem{Zouros:1979iw}
T.~Zouros and D.~Eardley,
\newblock Annals Phys. {\bf 118}, 139 (1979).
%%CITATION = APNYA,118,139;%%

\bibitem{Cardoso:2011xi}
V.~Cardoso, S.~Chakrabarti, P.~Pani, E.~Berti and L.~Gualtieri,
\newblock Phys.Rev.Lett. {\bf 107}, 241101 (2011), [1109.6021].
%%CITATION = ARXIV:1109.6021;%%

\bibitem{Pani:2012bp}
P.~Pani, V.~Cardoso, L.~Gualtieri, E.~Berti and A.~Ishibashi,
\newblock Phys.Rev. {\bf D86}, 104017 (2012), [1209.0773].
%%CITATION = ARXIV:1209.0773;%%

\bibitem{Cardoso:2005vk}
V.~Cardoso and S.~Yoshida,
\newblock JHEP {\bf 0507}, 009 (2005), [hep-th/0502206].
%%CITATION = HEP-TH/0502206;%%

\bibitem{Dolan:2007mj}
S.~R. Dolan,
\newblock Phys.Rev. {\bf D76}, 084001 (2007), [0705.2880].
%%CITATION = ARXIV:0705.2880;%%

\bibitem{Berti:2009kk}
E.~Berti, V.~Cardoso and A.~O. Starinets,
\newblock Class.Quant.Grav. {\bf 26}, 163001 (2009), [0905.2975].
%%CITATION = ARXIV:0905.2975;%%

\bibitem{Burt:2011pv}
J.~Barranco {\em et~al.},
\newblock Phys.Rev. {\bf D84}, 083008 (2011), [1108.0931].
%%CITATION = ARXIV:1108.0931;%%

\bibitem{Barranco:2013rua}
J.~Barranco {\em et~al.},
\newblock 1312.5808.
%%CITATION = ARXIV:1312.5808;%%

\bibitem{Strafuss:2004qc}
M.~J. Strafuss and G.~Khanna,
\newblock Phys.Rev. {\bf D71}, 024034 (2005), [gr-qc/0412023].
%%CITATION = GR-QC/0412023;%%

\bibitem{Barranco:2012qs}
J.~Barranco {\em et~al.},
\newblock Phys.Rev.Lett. {\bf 109}, 081102 (2012), [1207.2153].
%%CITATION = ARXIV:1207.2153;%%

\bibitem{Yoshino:2012kn}
H.~Yoshino and H.~Kodama,
\newblock Prog.Theor.Phys. {\bf 128}, 153 (2012), [1203.5070].
%%CITATION = ARXIV:1203.5070;%%

\bibitem{Witek:2012tr}
H.~Witek, V.~Cardoso, A.~Ishibashi and U.~Sperhake,
\newblock Phys.Rev. {\bf D87}, 043513 (2013), [1212.0551].
%%CITATION = ARXIV:1212.0551;%%

\bibitem{Damour:1993hw}
T.~Damour and G.~Esposito-Farese,
\newblock Phys.Rev.Lett. {\bf 70}, 2220 (1993).
%%CITATION = PRLTA,70,2220;%%

\bibitem{Yunes:2013dva}
N.~Yunes and X.~Siemens,
\newblock 1304.3473.
%%CITATION = ARXIV:1304.3473;%%

\bibitem{Cardoso:2013fwa}
V.~Cardoso, I.~P. Carucci, P.~Pani and T.~P. Sotiriou,
\newblock Phys. Rev. Lett. 111, {\bf 111101} (2013), [1308.6587].
%%CITATION = ARXIV:1308.6587;%%

\bibitem{Stein:2013wza}
L.~C. Stein and K.~Yagi,
\newblock 1310.6743.
%%CITATION = ARXIV:1310.6743;%%

\bibitem{Kodama:2011zc}
H.~Kodama and H.~Yoshino,
\newblock Int.J.Mod.Phys.Conf.Ser. {\bf 7}, 84 (2012), [1108.1365].
%%CITATION = ARXIV:1108.1365;%%

\bibitem{Pani:2012vp}
P.~Pani, V.~Cardoso, L.~Gualtieri, E.~Berti and A.~Ishibashi,
\newblock Phys.Rev.Lett. {\bf 109}, 131102 (2012), [1209.0465].
%%CITATION = ARXIV:1209.0465;%%

\bibitem{Brito:2013wya}
R.~Brito, V.~Cardoso and P.~Pani,
\newblock Phys. Rev. {\bf D88}, 023514 (2013), [1304.6725].
%%CITATION = ARXIV:1304.6725;%%

\bibitem{Brito:2013yxa}
R.~Brito, V.~Cardoso and P.~Pani,
\newblock Phys. Rev. {\bf D87}, 124024 (2013), [1306.0908].
%%CITATION = ARXIV:1306.0908;%%

\bibitem{Pani:2013hpa}
P.~Pani and A.~Loeb,
\newblock Phys.Rev. {\bf D88}, 041301 (2013), [1307.5176].
%%CITATION = ARXIV:1307.5176;%%

\bibitem{Cardoso:2013opa}
V.~Cardoso, I.~P. Carucci, P.~Pani and T.~P. Sotiriou,
\newblock 1305.6936.
%%CITATION = ARXIV:1305.6936;%%

\bibitem{Degollado:2013eqa}
J.~C. Degollado and C.~A.~R. Herdeiro,
\newblock 1303.2392.
%%CITATION = ARXIV:1303.2392;%%

\bibitem{Herdeiro:2013pia}
C.~A.~R. Herdeiro, J.~C. Degollado and H.~F. Rúnarsson,
\newblock 1305.5513.
%%CITATION = ARXIV:1305.5513;%%

\bibitem{Degollado:2013bha}
J.~C. Degollado and C.~A.~R. Herdeiro,
\newblock 1312.4579.
%%CITATION = ARXIV:1312.4579;%%

\bibitem{Hod:2013fvl}
S.~Hod,
\newblock Physical Review D 88, {\bf 064055} (2013), [1310.6101].
%%CITATION = ARXIV:1310.6101;%%

\bibitem{Pani:2013ija}
P.~Pani, E.~Berti and L.~Gualtieri,
\newblock Phys.Rev.Lett. {\bf 110}, 241103 (2013), [1304.1160].
%%CITATION = ARXIV:1304.1160;%%

\bibitem{Pani:2013wsa}
P.~Pani, E.~Berti and L.~Gualtieri,
\newblock 1307.7315.
%%CITATION = ARXIV:1307.7315;%%

\bibitem{Rosa:2011my}
J.~G. Rosa and S.~R. Dolan,
\newblock Phys.Rev. {\bf D85}, 044043 (2012), [1110.4494].
%%CITATION = ARXIV:1110.4494;%%

\bibitem{Cardoso:2013krh}
V.~Cardoso,
\newblock Gen.Rel.Grav. {\bf 45}, 2079 (2013), [1307.0038].
%%CITATION = ARXIV:1307.0038;%%

\bibitem{Berti:2013uda}
E.~Berti,
\newblock Braz.J.Phys. {\bf 43}, 341 (2013), [1302.5702].
%%CITATION = ARXIV:1302.5702;%%

\bibitem{Herdeiro:2014goa}
C.~A.~R. Herdeiro and E.~Radu,
\newblock 1403.2757.
%%CITATION = ARXIV:1403.2757;%%

\bibitem{Yoshino:2013ofa}
H.~Yoshino and H.~Kodama,
\newblock 1312.2326.
%%CITATION = ARXIV:1312.2326;%%

\bibitem{Mocanu:2012fd}
G.~Mocanu and D.~Grumiller,
\newblock Phys.Rev. {\bf D85}, 105022 (2012), [1203.4681].
%%CITATION = ARXIV:1203.4681;%%

\bibitem{Pani:2010jz}
P.~Pani, E.~Barausse, E.~Berti and V.~Cardoso,
\newblock Phys.Rev. {\bf D82}, 044009 (2010), [1006.1863].
%%CITATION = ARXIV:1006.1863;%%

\bibitem{Zhang:2013ksa}
Z.~Zhang, E.~Berti and V.~Cardoso,
\newblock 1305.4306.
%%CITATION = ARXIV:1305.4306;%%

\bibitem{Liu:2009al}
Y.~T. Liu, Z.~B. Etienne and S.~L. Shapiro,
\newblock Phys.Rev. {\bf D80}, 121503 (2009), [1001.4077].
%%CITATION = ARXIV:1001.4077;%%

\bibitem{1975problembook}
A.~P. {Lightman}, W.~H. {Press}, R.~H. {Price} and S.~A. {Teukolsky},
\newblock {\em {Problem book in relativity and gravitation}} (Princeton
  University Press, 1975).

\bibitem{Wald:1984rg}
R.~M. Wald,
\newblock (1984).
%%CITATION = INSPIRE-209356;%%

\bibitem{Alcubierre:2008}
M.~Alcubierre,
\newblock {\em {Introduction to 3+1 numerical relativity }} (Oxford Univ.
  Press, Oxford, 2008).

\bibitem{Baumgarte2010}
T.~W. Baumgarte and S.~L. Shapiro,
\newblock {\em {Numerical Relativity}} (Cambridge University Press, 2010).

\bibitem{York:1979}
J.~W. {York}, Jr.,
\newblock {Kinematics and dynamics of general relativity},
\newblock in {\em Sources of Gravitational Radiation}, edited by {L.~L.~Smarr},
  pp. 83--126, 1979.

\bibitem{Gourgoulhon:2007ue}
E.~Gourgoulhon,
\newblock gr-qc/0703035.
%%CITATION = GR-QC/0703035;%%

\bibitem{Centrella:2010mx}
J.~Centrella, J.~G. Baker, B.~J. Kelly and J.~R. van Meter,
\newblock Rev.Mod.Phys. {\bf 82}, 3069 (2010), [1010.5260].
%%CITATION = ARXIV:1010.5260;%%

\bibitem{Hinder:2010vn}
I.~Hinder,
\newblock Class.Quant.Grav. {\bf 27}, 114004 (2010), [1001.5161].
%%CITATION = ARXIV:1001.5161;%%

\bibitem{Baumgarte:2002jm}
T.~Baumgarte and S.~Shapiro,
\newblock Phys.Rept. {\bf 376}, 41 (2003), [gr-qc/0211028].
%%CITATION = GR-QC/0211028;%%

\bibitem{Arnowitt:1962hi}
R.~L. Arnowitt, S.~Deser and C.~W. Misner,
\newblock gr-qc/0405109.
%%CITATION = GR-QC/0405109;%%

\bibitem{Sarbach:2012pr}
O.~Sarbach and M.~Tiglio,
\newblock Living Rev.Rel. {\bf 15}, 9 (2012), [1203.6443].
%%CITATION = ARXIV:1203.6443;%%

\bibitem{Hilditch:2013sba}
D.~Hilditch,
\newblock Int.J.Mod.Phys. {\bf A28}, 1340015 (2013), [1309.2012].
%%CITATION = ARXIV:1309.2012;%%

\bibitem{Baumgarte:1998te}
T.~W. Baumgarte and S.~L. Shapiro,
\newblock Phys.Rev. {\bf D59}, 024007 (1999), [gr-qc/9810065].
%%CITATION = GR-QC/9810065;%%

\bibitem{Shibata:1995we}
M.~Shibata and T.~Nakamura,
\newblock Phys.Rev. {\bf D52}, 5428 (1995).
%%CITATION = PHRVA,D52,5428;%%

\bibitem{Witek:2013ora}
H.~Witek,
\newblock Int.J.Mod.Phys. {\bf A28}, 1340017 (2013), [1308.1686].
%%CITATION = ARXIV:1308.1686;%%

\bibitem{Campanelli:2005dd}
M.~Campanelli, C.~Lousto, P.~Marronetti and Y.~Zlochower,
\newblock Phys.Rev.Lett. {\bf 96}, 111101 (2006), [gr-qc/0511048].
%%CITATION = GR-QC/0511048;%%

\bibitem{Baker:2005vv}
J.~G. Baker, J.~Centrella, D.-I. Choi, M.~Koppitz and J.~van Meter,
\newblock Phys.Rev.Lett. {\bf 96}, 111102 (2006), [gr-qc/0511103].
%%CITATION = GR-QC/0511103;%%

\bibitem{vanMeter:2006vi}
J.~R. van Meter, J.~G. Baker, M.~Koppitz and D.-I. Choi,
\newblock Phys.Rev. {\bf D73}, 124011 (2006), [gr-qc/0605030].
%%CITATION = GR-QC/0605030;%%

\bibitem{Friedrich:1996hq}
H.~Friedrich,
\newblock Class. Quant. Grav. {\bf 13}, 1451 (1996).
%%CITATION = CQGRD,13,1451;%%

\bibitem{Teukolsky:1973ha}
S.~A. Teukolsky,
\newblock Astrophys. J. {\bf 185}, 635 (1973).
%%CITATION = ASJOA,185,635;%%

\bibitem{Thornburg:1995cp}
J.~Thornburg,
\newblock Phys.Rev. {\bf D54}, 4899 (1996), [gr-qc/9508014].
%%CITATION = GR-QC/9508014;%%

\bibitem{Thornburg:2003sf}
J.~Thornburg,
\newblock Class.Quant.Grav. {\bf 21}, 743 (2004), [gr-qc/0306056].
%%CITATION = GR-QC/0306056;%%

\bibitem{Christodoulou:1970wf}
D.~Christodoulou,
\newblock Phys.Rev.Lett. {\bf 25}, 1596 (1970).
%%CITATION = PRLTA,25,1596;%%

\bibitem{Goodale:2002a}
T.~Goodale {\em et~al.},
\newblock The {Cactus} framework and toolkit: Design and applications,
\newblock Berlin, 2003, Springer.

\bibitem{Cactuscode:web}
{Cactus Computational Toolkit}.

\bibitem{Loffler:2011ay}
F.~Loffler {\em et~al.},
\newblock Class.Quant.Grav. {\bf 29}, 115001 (2012), [1111.3344].
%%CITATION = ARXIV:1111.3344;%%

\bibitem{EinsteinToolkit:web}
{Einstein Toolkit}: Open software for relativistic astrophysics.

\bibitem{Sperhake:2006cy}
U.~Sperhake,
\newblock Phys.Rev. {\bf D76}, 104015 (2007), [gr-qc/0606079].
%%CITATION = GR-QC/0606079;%%

\bibitem{Ansorg:2004ds}
M.~Ansorg, B.~Bruegmann and W.~Tichy,
\newblock Phys.Rev. {\bf D70}, 064011 (2004), [gr-qc/0404056].
%%CITATION = GR-QC/0404056;%%

\bibitem{Schnetter:2003rb}
E.~Schnetter, S.~H. Hawley and I.~Hawke,
\newblock Class.Quant.Grav. {\bf 21}, 1465 (2004), [gr-qc/0310042].
%%CITATION = GR-QC/0310042;%%

\bibitem{CarpetCode:web}
{Carpet}: Adaptive Mesh Refinement for the {Cactus} Framework.

\bibitem{Yamamoto:2008js}
T.~Yamamoto, M.~Shibata and K.~Taniguchi,
\newblock Phys.Rev. {\bf D78}, 064054 (2008), [0806.4007].
%%CITATION = ARXIV:0806.4007;%%

\bibitem{Okawa:2013afa}
H.~Okawa,
\newblock International Journal of Modern Physics A  (2013), [1308.3502].
%%CITATION = ARXIV:1308.3502;%%

\bibitem{Shibata:1997nc}
M.~Shibata,
\newblock Phys.Rev. {\bf D55}, 2002 (1997).
%%CITATION = PHRVA,D55,2002;%%

\bibitem{Shibata:2000nw}
M.~Shibata and K.~Uryu,
\newblock Phys.Rev. {\bf D62}, 087501 (2000).
%%CITATION = PHRVA,D62,087501;%%

\bibitem{Cook:2000vr}
G.~B. Cook,
\newblock Living Rev.Rel. {\bf 3}, 5 (2000), [gr-qc/0007085].
%%CITATION = GR-QC/0007085;%%

\bibitem{Healy:2011ef}
J.~Healy {\em et~al.},
\newblock 1112.3928.
%%CITATION = ARXIV:1112.3928;%%

\bibitem{Berti:2013gfa}
E.~Berti, V.~Cardoso, L.~Gualtieri, M.~Horbatsch and U.~Sperhake,
\newblock Phys.Rev. {\bf D87}, 124020 (2013), [1304.2836].
%%CITATION = ARXIV:1304.2836;%%

\bibitem{Okawa:2013jba}
H.~Okawa, V.~Cardoso and P.~Pani,
\newblock Phys.Rev. {\bf D89}, 041502 (2014), [1311.1235].
%%CITATION = ARXIV:1311.1235;%%

\bibitem{Brill:1963yv}
D.~R. Brill and R.~W. Lindquist,
\newblock Phys.Rev. {\bf 131}, 471 (1963).
%%CITATION = PHRVA,131,471;%%

\bibitem{Lindquist1963}
R.~W. Lindquist,
\newblock J. Math. Phys. {\bf 4}, 938 (1963).

\bibitem{Brandt:1994ee}
S.~R. Brandt and E.~Seidel,
\newblock Phys.Rev. {\bf D52}, 856 (1995), [gr-qc/9412072].
%%CITATION = GR-QC/9412072;%%

\bibitem{Brandt:1996si}
S.~R. Brandt and E.~Seidel,
\newblock Phys.Rev. {\bf D54}, 1403 (1996), [gr-qc/9601010].
%%CITATION = GR-QC/9601010;%%

\bibitem{Konoplya:2006br}
R.~Konoplya and A.~Zhidenko,
\newblock Phys.Rev. {\bf D73}, 124040 (2006), [gr-qc/0605013].
%%CITATION = GR-QC/0605013;%%

\bibitem{Price:1971fb}
R.~H. Price,
\newblock Phys.Rev. {\bf D5}, 2419 (1972).
%%CITATION = PHRVA,D5,2419;%%

\bibitem{Ching:1995tj}
E.~Ching, P.~Leung, W.~Suen and K.~Young,
\newblock Phys.Rev. {\bf D52}, 2118 (1995), [gr-qc/9507035].
%%CITATION = GR-QC/9507035;%%

\bibitem{Gundlach:1993tn}
C.~Gundlach, R.~H. Price and J.~Pullin,
\newblock Phys.Rev. {\bf D49}, 890 (1994), [gr-qc/9307010].
%%CITATION = GR-QC/9307010;%%

\bibitem{DyBHo:web}
Gravity group {CENTRA/IST} {Lisbon}.

\bibitem{Hod:1998ra}
S.~Hod and T.~Piran,
\newblock Phys.Rev. {\bf D58}, 044018 (1998), [gr-qc/9801059].
%%CITATION = GR-QC/9801059;%%

\bibitem{Koyama:2001ee}
H.~Koyama and A.~Tomimatsu,
\newblock Phys.Rev. {\bf D64}, 044014 (2001), [gr-qc/0103086].
%%CITATION = GR-QC/0103086;%%

\bibitem{Koyama:2001qw}
H.~Koyama and A.~Tomimatsu,
\newblock Phys.Rev. {\bf D65}, 084031 (2002), [gr-qc/0112075].
%%CITATION = GR-QC/0112075;%%

\bibitem{Burko:2004jn}
L.~M. Burko and G.~Khanna,
\newblock Phys.Rev. {\bf D70}, 044018 (2004), [gr-qc/0403018].
%%CITATION = GR-QC/0403018;%%

\bibitem{Bizon:2011gg}
P.~Bizon and A.~Rostworowski,
\newblock Phys.Rev.Lett. {\bf 107}, 031102 (2011), [1104.3702].
%%CITATION = ARXIV:1104.3702;%%

\bibitem{Maliborski:2012gx}
M.~Maliborski,
\newblock Phys.Rev.Lett. {\bf 109}, 221101 (2012), [1208.2934].
%%CITATION = ARXIV:1208.2934;%%

\bibitem{Buchel:2013uba}
A.~Buchel, S.~L. Liebling and L.~Lehner,
\newblock Phys.Rev. {\bf D87}, 123006 (2013), [1304.4166].
%%CITATION = ARXIV:1304.4166;%%

\bibitem{Adams:2013vsa}
A.~Adams, P.~M. Chesler and H.~Liu,
\newblock 1307.7267.
%%CITATION = ARXIV:1307.7267;%%

\bibitem{Yang:2014tla}
H.~Yang, A.~Zimmerman and L.~Lehner,
\newblock 1402.4859.
%%CITATION = ARXIV:1402.4859;%%

\bibitem{Bardeen:1972fi}
J.~M. Bardeen, W.~H. Press and S.~A. Teukolsky,
\newblock Astrophys.J. {\bf 178}, 347 (1972).
%%CITATION = ASJOA,178,347;%%

\bibitem{Teukolsky:1974yv}
S.~Teukolsky and W.~Press,
\newblock Astrophys.J. {\bf 193}, 443 (1974).
%%CITATION = ASJOA,193,443;%%

\end{thebibliography}

\end{document}